\documentclass[twocolumn,superscriptaddress,floatfix,prb,aps,showpacs]{revtex4}
\usepackage{graphicx}
\begin{document}
\bibliographystyle{apsrev}
\def\sa{\section}
\def\sb{\subsection}
\def\sc{\subsubsection}

\def\ind{\ \ \ \ }

\def\be{\begin{equation}}
\def\ee{\end{equation}}

\def\bea{\begin{eqnarray}}
\def\eea{\end{eqnarray}}

\def\ba{\begin{array}}
\def\ea{\end{array}}

\def\nn{\nonumber}

\def\ben{\begin{enumerate}}
\def\een{\end{enumerate}}

\def\fn{\footnote}

\def\rd{\partial}
\def\rot{\nabla\times}

\def\r{\right}
\def\l{\left}

\def\gt{\rightarrow}
\def\cf{\leftarrow}
\def\bw{\leftrightarrow}

\def\ra{\rangle}
\def\la{\langle} 
\def\bla{\big\langle}
\def\bra{\big\rangle}
\def\bbla{\bigg\langle}
\def\bbra{\bigg\rangle}
\def\Bla{\Big\langle}
\def\Bra{\Big\rangle}
\def\BBla{\Bigg\langle}
\def\BBra{\Bigg\rangle}

\def\ddt{{d\over dt}}

\def\rdt{{\rd\over\rd t}}
\def\rdx{{\rd\over\rd x}}

\def\bb{\begin{thebibliography}{99}}
\def\eb{\end{thebibliography}}
\def\bit{\bibitem}

\def\bc{\begin{center}}
\def\ec{\end{center}}

\title{Gradient expansion approach to multiple-band Fermi liquids}
\author{Ryuichi Shindou}
 \altaffiliation[Present address: ]{Division of Materials Physics, Department of Physical Science,
 Graduate School of Engineering Science, Osaka University} 
 \email{rshindou@postman.riken.jp}
\author{Leon Balents}
\affiliation{Department of Physics, University of California, 
Santa Barbara, California 93106, USA}
\date{today}
\begin{abstract}
 Promoted by the recent progress of Berry phase physics in spin galvanomagnetic 
 communities, we develop a systematic derivation of the reduced Keldysh equation (RKE) 
 which captures the low-energy dynamics of quasi-particles constrained within 
 doubly degenerate bands forming a single Fermi surface.  
 The derivation begins with the Keldysh equation for a quite general multiple-band 
 interacting Fermi systems, which is originally an $N_b$ by $N_b$ matrix-formed integral 
 (or infinite-order differential) equation, with $N_b$ being the total number of bands. 
 To derive the RKE for quasi-particle on a Fermi surface in question, we  
 project out the fully occupied/empty band degrees  
 of freedom perturbatively in the gradient expansion, whose coupling constant 
 measures how a system is disequilibrated. As for the 
 electron-electron interactions, however, we only employ the so-called adiabatic assumption 
 of the Fermi liquid theory, so that the electron correlations effect onto the 
 adiabatic transport of quasi-particles, i.e. the hermitian (real) part of the 
 self-energy, is taken into account in an unbiased manner.  
 The RKE thus derived becomes an $SU(2)$ {\it covariant} differential equation and 
 treats the spin and charge degrees of freedom on an equal footing. Namely,  
 the quasi-particle spin precessions due to the non-abelian gauge fields are 
 automatically encoded into its covariant derivatives. When further solved in favor 
 of spectral functions, this covariant differential equation also suggests that quasi-particles 
 on a doubly degenerate Fermi surface acquire {\it spin-selective} Berry curvature 
 corrections under the applied electromagnetic fields. This theoretical observation 
 gives us some hints of possible experimental methodology for  
 measuring the $SU(2)$ Berry's curvatures by spin-resolved photoemission experiments. 
 Due to the non-trivial frequency dependence of (the hermitian part of) self energy, 
 our RKE is composed of Berry's curvatures in the $d+1$ dual space, i.e. $k$-$\omega$ space, 
 so that the dual electric field is already introduced. To provide a simple way to 
 understand this ``temporal'' component of the $U(1)$ Berry's curvature, 
 we also provide the dual analogue of the Ampere  
 law, where the ``spatial'' rotation of the electric field in combination  
 with the ``temporal'' derivative of the well-known magnetic component  
 are determined by the $U(1)$ magnetic monopole ``current''.            
\end{abstract} 
\maketitle 
\section{Introduction}
Gauge fields 
often appear in effective low energy theories in condensed matter
physics, whenever the system's low energy manifold is restricted by some
``local'' constraints \cite{dimer}. Classic examples include doped
Mott-Hubbard insulators, two dimensional electron gases in the
fractional quantum Hall regime, low-dimensional quantum spin systems and
highly frustrated magnets.

In these ``strong-coupling'' problems, the constraints are implemented 
locally in {\sl real space}.  For example, the on-site Coulomb 
interaction in the Hubbard model or t-J model forbids double occupancy 
on {\sl every} site. 
In strong magnetic fields, the wavefunction for electrons in a two
dimensional electron gas must be annihilated by the Landau level
lowering operator, which is a differential operator defined locally for
{\sl each} value of the two-dimensional coordinate.  In the quantum
dimer model, spins are presumed to be bound into singlets which cover {\sl
  each} site exactly once.  In many cases, these local
constraints 
take the form of a version of ``Gauss' law'' 
imposed on suitably introduced ``electromagnetic field'' and ``charge''
variables.  Due to the Gauss' law constraint, the low energy effective
theory for such systems becomes mathematically equivalent to a form of
(usually compact) quantum electro-dynamics\cite{corner,cubic,dimer}. 

Another kind of constrained system occurs in the weak coupling region, 
such as in 
a Fermi liquid (FL), where 
the states at the fermi surface are supposed to be 
well separated from the other fully occupied/empty bands by 
a sufficiently large (direct) band gap.  The low energy manifold  
is thus spanned only by the Bloch states of those conduction bands forming this
Fermi surface. 
Even in the presence of strong electron correlations, 
we may still assume that the low energy Hilbert 
space is spanned only by those quasi-particle excitations constrained 
within a Fermi surface,  as far as electron-electron interactions 
can be introduced adiabatically in comparison with the direct band gap.

In most of the literature, Fermi liquids 
are assumed to be 
well described, by ignoring electron correlations completely.  
In such a non-interacting case, the projection to a single
low-energy band or degenerate bands has been studied extensively
\cite{sn, mnz, omn, cn, bb, si}, and yields an effective equation of motion
(EOM) for the conduction electrons in this 
band, say the $\alpha$-th band, moving under the
influence of {\sl external} electric field $\bf e$ 
and magnetic field $\bf b$.  Such an EOM contains an {\sl
  dual magnetic field} ${\cal B}^{\alpha}$, which acts on a
quasi-particle like a {\sl Lorentz force in $k$-space}:
\begin{eqnarray} 
\frac{dR}{dT}&=&  {\bf z}^{\dagger}     
\Big\{\hat{\bf v}_{\alpha} + {\cal B}^{\alpha}  
\times \frac{dk}{dT}\Big\}{\bf z},\nonumber \\
\frac{dk}{dT} &=& - {\bf e} + 
{\bf b} \times \frac{dR}{dT}, \label{1-1} \\
i\frac{d{\bf z}}{dT} &=& \Big\{ {\bf M}_{\alpha}\cdot {\bf b} + \sum_{i}^{d} 
{\cal A}^{\alpha}\cdot\frac{dk}{dT}\Big\} {\bf z}. \nonumber  
\end{eqnarray} 
${\bf z}$ is a ${\rm CP}^{N-1}$ vector in FLs with $N$-fold
degenerate conduction bands, i.e.  ${\bf z}=({\rm z}_1,{\rm
  z}_2,\cdots,{\rm z}_N)$ with ${\bf z}^{\dagger}{\bf z} = 1$.  This
complex-valued vector describes the internal degrees of freedom associated with
the degeneracy at each $k$-point.  Correspondingly, the dual 
magnetic field, magnetic gauge field ${\cal A}^{\alpha}$ and ``orbital
magnetization'' ${\bf M}_{\alpha}$ are all $N$ by $N$ Hermitian matrices.  In a
non-interacting fermi gas, they are defined solely in terms of the periodic
part of Bloch wavefunctions of conduction bands 
$|u_{\sigma}\ra \ (\sigma=1,\cdots,N \in \alpha)$;  
\begin{eqnarray} 
{\cal B}^{\alpha}_{i}   
& \equiv & i\epsilon_{ijm}\partial_{k_j}{\cal A}^{\alpha}_{m} +  
i\epsilon_{ijm}{\cal A}^{\alpha}_{j}{\cal A}^{\alpha}_{m}, 
\label{1-2} \\ 
\big[{\cal A}^{\alpha}_j\big]_{(\sigma|\sigma')} 
&\equiv & \bla u^{\alpha\sigma}\big|
\partial_{k_j}u^{\alpha\sigma'}\bra, \label{1-2-a} \\ 
\big[{\bf M}_{\alpha,m}]_{(\sigma|\sigma')}
&\equiv& \frac{i\epsilon_{mnl}}{2}\bla\partial_{k_n} 
u^{\alpha\sigma}\big|\hat{H}-E_{\alpha}\big|\partial_{k_l}
u^{\alpha\sigma'}\bra,  \label{1-2m}
\end{eqnarray} 
with $E_\alpha$ being the energy dispersion for the $\alpha$-th band. 
Especially, the third EOM dictates that the 
$CP^{N-1}$ vector $\bf z$ precesses due to the nontrivial matrix 
structures of $[{\bf M}_{\alpha}]$ (Zeeman field) 
and $[{\cal A}^{\alpha}_j]$ (Wilczek-Zee phase) 
respectively.

One purpose of this paper is to enlarge the regime of validity of this
effective EOM, so as to include metals for which electron-electron
interactions are significant, i.e. Fermi {\sl liquids} rather than Fermi
{\sl gases}.  The $SU(N)$ effective EOM mentioned above is clearly valid
{\it only} for a {\it non-interacting} Fermi gas, or within a mean-field
description of ordered states such as (ferro)magnetic metals {\it in 
which the ordered moment does not fluctuate at all}. The fundamental
framework of Fermi liquid theory, however, implies that these effective
EOMs should be properly generalized into a realistic metal, where the
electron-electron interaction/magnetic fluctuations are not
weak. 
In fact, Haldane has recently argued that the ``renormalized'' Bloch
wavefunction for a quasi-particle should be taken as the eigenvector of
 spectral function.~\cite{haldane1,haldane3}  
In this article, we provide a quite 
general derivation of the effective EOM for quasi-particles based on the 
Keldysh formalism, which confirms in part that this notion is valid in an
arbitrary $U(1)$ Fermi liquid. 

For $N=1$ case, the {\sl form} of this EOM is identical 
to Eq.(\ref{1-1}).  However, 
in passing from the non-interacting version of  
gauge field definition, i.e. Eq.(\ref{1-2-a}), to that of the many-body 
case, we encounter an additional complication.  Because  
of the energy dependence of the self-energy, the renormalized Bloch   
wavefunction for quasi-particles -- defined from the Green's function --   
depends on the energy or frequency $\omega$, in addition to crystal   
momentum $k$, i.e. $|u^{\alpha}(k,\omega)\rangle$.  Accordingly, we 
are naturally led to introduce a sort of 
{\it dual version of the electric field and 
associated electrostatic potential}:     
\begin{eqnarray} 
  {\cal E}^{\alpha}_{j} & \equiv & 
  i\big(\partial_{\omega}{\cal A}^{\alpha}_{j}  
  - \partial_{k_j}{\cal A}^{\alpha}_{0}\big) +   
  i\big[{\cal A}^{\alpha}_{0},{\cal A}^{\alpha}_{j}\big], \label{1-4} \\
  \big[ {\cal A}^{\alpha}_{0} \big]_{(\sigma|\sigma')}
  & \equiv & \bla u^{\alpha\sigma}\big|
  \partial_{\omega} u^{\alpha\sigma'}\bra,  \label{1-4-a} 
\end{eqnarray} 
which, as well as the magnetic component, generally 
have non-trivial structures in the $(d+1)$-dimensional dual 
space.  

As will be shown in this paper, the {\sl renormalized} 
$k$-space Lorentz force appearing in the 
effective EOM for quasi-particles    
is composed both of these magnetic and electric 
component estimated on shell:
\begin{eqnarray}
\tilde{\cal B}^{\alpha} &\equiv& 
\bar{\cal B}^{\alpha} - \bar{\cal E}^{\alpha} 
\times {\bf v}_{\alpha}, \label{eq:1} \\
\bar{\cal B}^{\alpha} &\equiv& \big({\cal B}^{\alpha}\big)_{|\omega=\epsilon_{\alpha,k}},   
\bar{\cal E}^{\alpha} \equiv \big({\cal E}^{\alpha}\big)_{|\omega=\epsilon_{\alpha,k}},  
{\bf v}_{\alpha,j} = \frac{\partial \epsilon_{\alpha,k}}{\partial k_j}.
\label{eq:1-1}  
\end{eqnarray}  
$\epsilon_{\alpha,k}$ is renormalized energy dispersion 
for a quasi-particle in question, which is defined as 
the pole of the single-point Green 
function with respect to $\omega$ (see Eq.(\ref{2-2})).  
In this sense, this newly introduced electric component 
{\it provides another source of the Lorentz force} in the 
$k$ space, particular only to {\sl interacting} Fermi 
systems. 

\begin{figure}
\begin{center}
\includegraphics[width=0.40\textwidth]{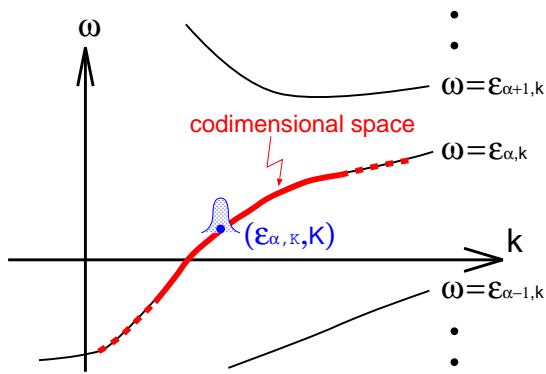}
\end{center}
\caption{$\omega$-$k$ Euclidean space and  
its codimensional subspace associated with 
$\omega=\epsilon_{\alpha,k}$ (Red line). $\tilde{\cal B}^{\alpha}$ 
can be viewed as just the ``magnetic component'' of Berry's curvature 
defined in latter subspace.}
\label{fig1}
\end{figure}

From the viewpoint of this EOM alone, however, the significance of the 
temporal component of the Berry's curvature is not clear.  Namely, 
the {\it renormalized Lorentz field} $\tilde{\cal B}^{\alpha}$ 
can be also viewed as just the ``magnetic component'' {\it defined 
in the codimensional space} associated with the quasi-particle energy 
dispersion $\omega=\epsilon_{\alpha,k}$;
\begin{eqnarray}
\tilde{\cal B}^{\alpha}_{j}&\equiv& i\epsilon_{jml}\partial_{k_m}
\tilde{\cal A}^{\alpha}_{k_l} + i\epsilon_{jml}\tilde{\cal A}^{\alpha}_{k_m}
\tilde{\cal A}^{\alpha}_{k_l}, \label{codim1} \\ 
\tilde{\cal A}^{\alpha}_{k_m}&\equiv& \bla \tilde{u}^{\alpha}\big|
\partial_{k_m} 
\tilde{u}^{\alpha} \bra, \label{codim2} \\
|\tilde{u}^{\alpha}(k)\rangle &\equiv &
|u^{\alpha}(k,\omega)\rangle_{|\omega=\epsilon_{\alpha,k}},  \nonumber
\end{eqnarray}        
(see Fig.\ref{fig1}). Therefore, it is tempting to regard that 
the separate definition of the 
electric and magnetic fields, i.e. eqs. (\ref{1-2},\ref{1-4}), 
is a sort of redundancy.  Physically, this is somewhat sensible, 
since ``sharply-peaked'' 
quasiparticles (``infinite'' life time with a definite dispersion relation 
$\omega=\epsilon_{\alpha,k}$) should be linked with those wavefunctions which are 
defined only in this codimensional subspace of the $\omega-k$ Euclidean space.  
Such thinking suggests that only 
derivatives along this subspace are meaningful, which thereby involve  
appropriate linear combinations of $\omega$ and $k$ derivatives.

Nevertheless, it can be shown that the dual electric and magnetic 
fields defined in the $\omega$-$k$ Euclidean space 
{\sl do} have distinct physical meaning.  Specifically, 
we will show that 
these two fields enter into the linear response of the 
spectral {\it weight} to the applied electro and magnetic fields 
{\it respectively} (see eqs.~(\ref{4-6},\ref{4-7a}))~\cite{Xiao}.  
This conclusion is obtained by {\it solving the quasi-particle spectral function}   
perturbatively with respect to the gradient expansion.   
Therein,  
we observe, in both $U(1)$ and $SU(2)$ case, that these Berry's curvatures 
characterize the 1st order correction to the weight  
of the quasi-particle spectral function. Since the spectral weight in 
principle can be detected {\it in a momentum resolved way}, one may at 
least envision several photoemission experiments as 
candidate tools to make a ``contour map'' of the dual electro 
and magnetic fields separately (see sec.~IV). 

A momentum-resolved measurement of the dual electromagnetic fields 
is of interest not only as a matter of principle, but also potentially 
in diverse experimentally active areas such as anomalous and spin Hall 
effects in metals and semiconductors, and unconventional 
superconductivity in some ferromagnetic metals.  Recent experimental
activities in metals\cite{princeton,Harvard,TEXAS1-1} and semiconductors 
\cite{UCSB,TEXAS1} have focused on novel observations of anomalous and
spin Hall effects, leading to controversy over the origin of these
effects.  On the one hand are a variety of proposals of {\sl intrinsic}
effects, related to Berry phases and the above-defined dual 
electromagnetic fields in the clean band theory limit. \cite{mnz,TEXAS1}
However, such Hall effects could also be of an {\sl extrinsic} origin, 
i.e. a result of spin-dependent scattering of electrons from the 
spin-orbit potential of {\sl impurity} atoms \cite{KITP}.   
Currently, the discrimination between these
scenarios is rather indirect and based primarily upon the comparison to
diverse theoretical model calculations and approximations whose physical
applicability is difficult to judge \cite{sds}.  If one could somehow
{\it experimentally} ``visualize'' the distribution of these dual 
electromagnetic fields in a momentum space, 
independently from Hall/galvanomagnetic measurements, 
experimentalists could readily judge {\sl for themselves} whether
intrinsic or extrinsic contributions are dominant in their material
sample, by comparing with their transport measurements.  

A second potential use for a measurement of dual electromagnetic
fields stems for a recent proposal by Shi and Niu \cite{ShN} of a new
formulation of the many-body problem based on the 
non-commutative quantum mechanics.  They begin with 
a quantization of the EOM given in Eq.(\ref{1-1}), 
leading to a non-zero commutator between position operators,
$\big[\hat{R}_j,\hat{R}_m\big]=i\epsilon_{jml}{\cal B}_l(\hat{k})$~\cite{mnz}, 
in addition to the usual commutators of momenta and position, i.e. 
$\big[\hat{k}_j,\hat{k}_m\big]=i\epsilon_{jml}{\bf b}_l(\hat{R})$ and 
$[\hat{R}_j,\hat{k}_m]=i\delta_{jm}$.   
As is the case with the momentum in the presence of a real magnetic
field ${\bf b}$, one can also introduce the {\it canonical} position operator
$\hat{R}'_{j}\equiv \hat{R}_j - i{\cal A}_{j}(\hat{k})$ such that their
commutators become free from the dual magnetic field; 
$[\hat{R}'_j,\hat{R}'_m]=0$. At a price for this, however, the
electron-electron interaction acquires an additional phase factor, 
which can transform a repulsive electron-electron  
interaction into an attractive one. Observing this,  
Shi and Niu attributed to the $k$-space Berry phase an unique   
origin of the superconducting phases realized only within ferromagnetic
metals, such as in ${\rm UGe}_2$, ${\rm ZrZn}_2$ and 
${\rm URhGe}$.  A detailed {\it experimental} information 
constraining ${\cal B}(k)$ as a function of $k$, in
combination with a measurement of band dispersions would clearly 
provide 
these effective theoretical approaches an ``ab-initio'' model
Hamiltonian, only to aid a quantitative comparison with material physics.
  
The structure of this paper is as follows.  In section II, we introduce
a quite general multiple-band continuum model and the Keldysh
equation for the lesser (greater) Green function/spectral function.  In
section III, based on this (dissipationless) Keldysh equation, we
provide a systematic procedure for carrying out the projection into the
low-energy band, so as to obtain the effective (reduced) Keldysh equation for
those Green functions of quasi-particles forming a Fermi surface. 

Thanks to the Fermi liquid assumption, this effective Keldysh equation
can be further solved in favor of the spectral function, perturbatively
in the gradient expansion. Then, we observe in section IV that the 
Berry's curvatures, i.e. both dual electric fields and
magnetic fields, indeed enter into the 1st order correction to the
renormalization factor for quasi-particles.  Using this solution for the
spectral function, we further derive in section V the 
effective Boltzmann equation for the occupation number of 
quasi-particles in the phase space. Based on this equation, we can
finally read off the $U(1)$ effective EOM 
in interacting Fermi liquids. Section~VI is devoted to the summary
and discussions of the present paper.

A number of appendices describe other topics useful in understanding the
main text in more detail.  For completeness, we briefly review the
Keldysh formalism and our notations in the appendix A. The appendix B is
devoted to demonstrating the logical consistency between our derived
$U(1)$ effective EOM and the so-called Ishikawa-Matsuyawa-Haldane
formula \cite{IMH} within the linear response regime.  Appendix C
describes how a $U(1)$ magnetic monopole ``current'' in $\omega-k$ space (which  
extends the notion of magnetic monopole density in $k$-space  
into the $\omega$-$k$ space) determines the
distribution of dual electromagnetic fields in the $U(1)$ case.
As a specific example which gives a quantitative idea of the
significance of the many-body correction 
i.e. ${\cal E}$, we also present in appendix D 
some specific model calculations based on an electron-phonon 
coupling Hamiltonian.

\section{Keldysh Formalism} 

\subsection{effective continuum model}
We begin with a general 
semi-microscopic model in which the electronic spectrum is described by
a $k\cdot p$ type expansion about some (arbitrary) point in the
Brillouin zone.  
One may keep as many bands as are deemed close enough 
in energy to be relevant to the physics, and our arguments do not depend
upon the order in the expansion in $k$.  Familiar examples would be the
multiple-band Luttinger models, Dresselhaus models, and Rashba model 
commonly studied in semiconductors, in 
which the expansion point is chosen at the $\Gamma$ point. 
The advantage of this formulation is that we can Fourier 
transform, in 
a usual way, to a continuous real space coordinate $r$, i.e. $k
\rightarrow -i\nabla_r$.  The non-interacting  Hamiltonian is thereby 
expressed in terms of the slowly varying ``envelope fields''
$\psi_{\alpha}(r)$:
\begin{eqnarray}
  {\cal H}_{0}=\sum_{\alpha,\alpha'}  
  \int dr\psi^{\dagger}_{\alpha}(r)
  [\hat{H}_{0}(-i\nabla_{r},r,t)]_{\alpha\alpha'}\psi_{\alpha'}(r),  
  \label{non-interacting}
\end{eqnarray}
where the band index $\alpha^{(\prime)}$ runs 
from $1$ to  $N_b$.  
Specific informations about lattice, orbital and spin-orbit 
couplings are encoded into the matrix structure of $[\hat{H}_{0}]$.  
Following the standard literatures~\cite{KB}, 
we will employ a short-ranged electron-electron interaction potential;  
\begin{eqnarray}
{\cal H}_{1}&=& 
\sum\int\int V_{\alpha_1\alpha_2\alpha^{'}_2\alpha^{'}_1}(r_1,r_2) \nonumber \\
&&\hspace{1.0cm} \times\ \psi^{\dagger}_{\alpha_1}(r_1) \psi^{\dagger}_{\alpha_2}(r_2)
\psi_{\alpha^{'}_2}(r_2) \psi_{\alpha^{'}_1}(r_1), \label{interacting}
\end{eqnarray}
though the form of our results does not depend in detail 
upon this. 

\subsection{dissipationless Keldysh equation} 

Apart from the perturbation theory at equilibrium, 
the Keldysh formalism is constituted by 
the lesser and greater Green functions; 
\begin{eqnarray}
g^{>}(1|1')&\equiv& -i \bla \psi(1)\psi^{\dagger}(1') \bra, \nonumber \\
g^{<}(1|1')&\equiv& i \bla \psi^{\dagger}(1')\psi(1) \bra,  \nonumber 
\end{eqnarray}
where the time evolution of 
$\psi(1)\equiv \psi_{\alpha_1}(r_1,t_1)$ is determined by  
the interacting Hamiltonian introduced above;
\begin{eqnarray}
-i\frac{\partial \psi(1)}{\partial t_1} \equiv 
\big[{\cal H}_{0} + {\cal  H}_1,\psi(1)\big].  \nonumber 
\end{eqnarray} 
Because these Green functions are {\it not} the 
time-ordered ones, they  
acquire the translational symmetry in space and {\it time} 
at equilibrium. Furthermore, they are always 
Hermite matrices with respect to space and 
time coordinates and band indices.    

Putting its derivation aside (see the Appendix A), 
let us begin with the Keldysh equation for 
these lesser and greater Green functions;
\begin{eqnarray}
&&\big[{G_0}^{-1} - \Sigma^{\rm HF} - \sigma , {\sf g}^{<(>)} \big]_{\otimes,-} 
-\big[\Sigma_{\rm c}^{<(>)} , {\sf b} \big]_{\otimes,-} \nonumber \\
&&\ \ = \ \frac{1}{2} \big[\ {\Sigma}_{\rm c}^{>} , {\sf g}^{<} \big]_{\otimes,+} 
-\frac{1}{2} \big[\ {\Sigma}_{\rm c}^{<} , {\sf g}^{>} \big]_{\otimes,+}. 
\label{1-6} 
\end{eqnarray} 
The (anti-)commutator here is defined by the convolution with 
respect to time $t$, space $r$ and band index $\alpha$; 
\begin{eqnarray}
[\hat{B},\hat{C}]_{\otimes,\pm}(1,1') 
&\equiv& \int d\bar{1} \hat{B}(1|\bar{1})\cdot \hat{C}(\bar{1}|1') \nonumber \\
&&\ \ \ \pm \ \int d\bar{1} \hat{C}(1|\bar{1})\cdot \hat{B}(\bar{1}|1'), \nonumber 
\end{eqnarray}
with $\int d1\equiv \sum_{\alpha_1}\int dr_1dt_1$.
We denoted the bare Green function as $\hat{G}_{0}$ which 
is composed of the quadratic part of the Hamiltonian; 
\begin{eqnarray}
&&\hat{G}^{-1}_{0}(1|1')\ \equiv \nonumber \\ 
&&\ [i\partial_{t_1}\hat{1} -
{\hat{H}_{0}(-i\nabla_{r_1},r_1,t_1)}]_{\alpha_1\alpha^{'}_1}\delta(t_1-t^{'}_1)
\delta^{d}(r_1-r^{'}_1). \nonumber 
\end{eqnarray}
``$d$'' above represents the spatial dimension of our system.

Electron-electron interaction effects, on the other hand, are 
encoded in the self-energy such as $\hat{\Sigma}^{\rm HF}(1|1')$   
and $\hat{\Sigma}^{<(>)}_{\rm c}(1|1')$ 
, which are self-consistently 
given in terms of lesser/greater Green functions (see 
eqs.~(\ref{6-5}-\ref{6-6a})
in the Appendix A). The former self-energy is the  
Hartree-Fock part, which is temporally instantaneous, i.e. 
$\Sigma^{\rm HF}(1|1')\sim \delta(t_1-t_{1'})$.
The latter one is usually dubbed as the 
lesser (greater) collisional self-energy, 
which is at least 2nd order in   
electron-electron interactions and 
thus temporally non-instantaneous. 
The physical role of this collisional self-energy  
is two-folded, corresponding to its following  
decomposition; 
\begin{eqnarray}
\sigma(1|1')&\equiv& \frac{1}{2}\frac{t_1-t_{1'}}{|t_1-t_{1'}|} 
\big(\Sigma^{>}_{\rm c}(1|1')-\Sigma^{<}_{\rm c}(1|1')\big) \nonumber  \\
&\equiv & \frac{1}{2}\big(\Sigma^{\rm R}(1|1')
+\Sigma^{\rm A}(1|1')\big), 
\label{1-6a} \\
\Gamma(1|1')&\equiv& i\big(\Sigma^{>}_{\rm c}(1|1') 
-\Sigma^{<}_{\rm c}(1|1')\big) \nonumber \\
&\equiv &i\big(\Sigma^{\rm R}(1|1')
-\Sigma^{\rm A}(1|1')\big), \label{1-6b} 
\end{eqnarray} 
with $\sigma(1|1')^{*} \equiv \sigma(1'|1)$ and 
$\Gamma(1|1')^{*}\equiv \Gamma(1'|1)$. 
$\Sigma^{\rm R}(1|1')$ and $\Sigma^{\rm A}(1|1')$  
in the right hand side denotes the retarded and advanced 
self-energy respectively. Thus,    
$\sigma(1|1')$ stands for the  
real (Hermitian) part of the self-energy associated with 
the time-ordered Green function, which  
plays role of the renormalization of the quasi-particle 
energy and wavefunction.  On the other hand, 
$\Gamma(1|1')$ is its imaginary (anti-hermitian) part, 
bringing about a finite life-time for quasi-particles. 

In a same way, we can define the real/imaginary part 
of the Green function;
\begin{eqnarray}
{\sf b}(1|1')&\equiv& 
\frac{1}{2}\frac{t_1-t_{1'}}{|t_1-t_{1'}|}
\big({\sf g}^{>}(1|1')-{\sf g}^{<}(1|1')\big) \nonumber \\
&\equiv& \frac{1}{2}\big({\sf g}^{\rm R}(1|1')+{\sf g}^{\rm A}(1|1')\big), \label{b-def} \\
{\sf A}(1|1')&\equiv& i\big({\sf g}^{>}(1|1')-{\sf g}^{<}(1|1')\big) 
\nonumber \\ 
&\equiv& i\big({\sf g}^{\rm R}(1|1')-{\sf g}^{\rm A}(1|1')\big), \label{A-def} 
\end{eqnarray}  
with ${\sf b}(1|1')^{*}\equiv {\sf b}(1'|1)$ and 
${\sf A}(1|1')^{*}\equiv {\sf A}(1'|1)$.
Note that especially the latter one is nothing but the spectral function. 
As is clear from eq. (\ref{1-6}), 
the Keldysh equation for this spectral 
function is composed only by the real/imaginary part of the self-energy 
and Green functions introduced above;
\begin{eqnarray}
&&\big[G_0^{-1} - \Sigma^{\rm HF} - \sigma , {\sf A} \big]_{\otimes,-} 
-\big[\Gamma , {\sf b} \big]_{\otimes,-} = 0. \label{1-6c}
\end{eqnarray}  

We will dub $G_0^{-1}(1|1')-\Sigma^{\rm HF}(1|1')-\sigma(1|1')$ as 
a ``Lagrangian'' ${\sf L}(1|1')$  
in a sense that, when Fourier-transformed with respect to 
its relative coordinate, i.e. $1-1'$,  
it reduces to $\omega$ minus a ``renormalized'' Hamiltonian 
for quasi-particles (q.p.). Namely, its eigenvalues specify the 
renormalized energy dispersions for q.p. 
as their zeros with respect to $\omega$, while their corresponding 
eigenvectors constitute an orthogonal set. Latter of which 
therefore can be regarded as (a periodic part of) the renormalized 
Bloch wavefunction. Accordingly, we are led to define 
the Berry curvatures and gauge connections in the dual space   
in terms of the unitary matrix 
diagonalizing this Lagrangian (see eqs. (\ref{3-6}), 
(\ref{3-5a}) and (\ref{2-1a}) respectively).  
As is shown in this paper, our effective Boltzmann/Keldysh equation 
legally derived via the projection process claims that 
the Berry's curvatures thus introduced indeed govern  
the effective EOM for q.p. in interacting Fermi systems.  

On the other hand, the commutator between 
the anti-hermitian part of the 
self-energy $\Gamma$ and the hermitian part of the 
Green function ${\sf b}$ introduces 
a finite {\it life time} of q.p., i.e. 
the broadening of the spectral functions. 
When it comes to the q.p. closed to a Fermi surface, however, 
the Fourier-transformed $\Gamma$ as a function of $\omega$ 
becomes as small as $O((\omega-\mu)^2,T^2)$. Namely, 
the 2nd expression of eq.~(\ref{1-6b}) dictates that, 
when analytically continued from a Matsubara Green function, 
$\Gamma$ at equilibrium is composed of a delta function, 
which imposes the energy conservation on its internal lines. 
As a result, the momentum integral regions associated with 
the internal lines are restricted to be only near a Fermi surface 
at low temperature~\cite{Mahan}.    
 
$\Gamma$ in those electron-boson coupled systems with 
bosons having a finite excitation energy gets even smaller than this power-low 
decay. When $|\omega-\mu|$ is much smaller than this excitation 
energy $\omega_0$, (the lowest order) perturbative calculations readily 
show that $\hat{\Gamma}$ vanishes exponentially, such as 
$e^{-|\omega-\mu|/\omega_0}$ or $e^{-T/\omega_0}$ (see Appendix D).
Anyway, in both of these two cases, the broadening of the q.p. 
spectral function at sufficiently low $T$ is  
by far smaller than the thermal line-broadening of the 
spectral function $(\sim T)$, which validates the so-called 
adiabatic assumption of Fermi liquid theory.

Meanwhile, the hermitian part of the collisional 
self-energy $\hat\sigma$ remains finite even {\it on} a 
Fermi surface ($\omega = \mu$) at {\it zero} temperature 
($T=0$).  Namely, eq.~(\ref{1-6a}) indicates that,  
when analytically continued from a Matsubara 
Green function, $\hat{\sigma}$ at 
equilibrium is composed of the principal integral   
rather than the delta function and thus free from 
the energy conservation imposed on its internal lines   
(see the appendix D for some example). 
As such, apart from the life time part $\Gamma$, 
its momentum integrals region for the internal lines 
are {\it not} restricted near the Fermi surface, 
which even causes the ultra-violet cut-off 
dependence of $\hat{\sigma}$ at $T=0$.  

Because of these two different features generic 
in $\hat{\Gamma}$ and $\hat{\sigma}$, we ignore in this 
paper the (intrinsic) lifetime effect $\hat{\Gamma}$, 
while fully take into account the renormalization 
effects due to $\hat{\sigma}$.  
Namely, 
instead of eq.~(\ref{1-6c}), we begin with 
the following {\it dissipationless} Keldysh 
equation for spectral functions:
\begin{eqnarray}
[\hat{\sf L},\hat{\sf A}]_{\otimes,-} = \hat{0}. \label{1-7}
\end{eqnarray} 

Even this $SU(N_b)$ dissipationless Keldysh equation is still non-trivial, 
due to the presence of the {\it band index}, whose effect is the central 
issue of this paper. 
In the following, we will present a general method of 
projecting out irrelevant band degrees of freedom 
associated with fully occupied bands and fully empty 
bands, so as to derive perturbatively the 
reduced Keldysh (kinetic) equation only for the relevant 
bands constituting a Fermi surface. 

\section{Reduced Keldysh Equation}

\subsection{gradient expansion}

We will derive these reduced Keldysh equations perturbatively 
{\it with respect to the gradient expansion.}  
The coupling constant of this expansion is a dimensionless 
quantity which measures how much a system is disequilibrated.  
To define this expansion accurately, notice first that 
the lesser and greater Green function {\it at equilibrium}  
acquire the translational invariance in space and time coordinates; 
${\sf g}^{<(>)}(1|1')={\sf g}^{<(>)}_{\alpha_1\alpha_{1'}}(r_1-r_{1'},t_1-t_{1'})$.  
Being given in terms of these Green functions, the 
lesser/greater self-energy and Lagrangian also 
become translationally invariant at equilibrium; 
${\sf L}(1|1')={\sf L}_{\alpha_1\alpha_{1'}}(r_1-r_{1'},t_1-t_{1'})$. 
As such, when Fourier-transformed, the convolution 
encoded in the dissipationless Keldysh equation reduces 
a simple product {\it only with respect to band index};
\begin{eqnarray}
&&[\hat{\sf L}(q,\omega),\hat{\sf A}(q,\omega)]_{-} = \hat{0}, \nonumber \\
&&\left[
\begin{array}{c}
{\sf L}_{\alpha\beta} \\
{\sf A}_{\alpha\beta} \\
\end{array} \right] (q,\omega) \equiv \int dr
dt\ e^{-iqr + \omega t} \left[
\begin{array}{c}
{\sf L}_{\alpha\beta} \\
{\sf A}_{\alpha\beta} \\
\end{array} \right] (r,t).  \nonumber
\end{eqnarray}
Accordingly, we have only to diagonalize the Lagrangian so 
that an arbitrary diagonal $\hat{\sf A}$ in this eigenbasis 
satisfies eq.(\ref{1-7}) at equilibrium. We will regard 
this trivial limit as the non-perturbed case and take into account 
disequilibrations as perturbations.   

When a system is disequilibrated, the lesser/greater 
Green functions generally depend on the 
{\it center of mass coordinate}  
in space and time, i.e. $R\equiv \frac{r_1+r_{1'}}{2}$ and 
$T\equiv \frac{t_1+t_{1'}}{2}$;  
\begin{eqnarray}
&&{\sf g}^{<(>)}(1|1') = {\sf g}^{<(>)}_{\alpha_1\alpha_{1'}}(r,t;R,T). \nonumber 
\end{eqnarray}  
However, as long as a system is not so far from its equilibrium case, 
their dependences on $R$ and 
$T$ are 
slowly varying in comparison with the lattice spacing  
and inverse of band width respectively.  
As such, we could quantify  
the ``distance'' from equilibrium 
by a dimensionless ratio between this slowly 
varying length (time) scale 
and a lattice spacing (inverse of the band width). 
To extract the latter length/time scale, we have only 
to Fourier-transform the relative coordinate, i.e. $r$ and $t$,  
so that the crystal momentum $q$ and energy (frequency) $\omega$ are 
introduced~\cite{note4}; 
\begin{eqnarray}
\hat{\sf g}^{<}(\omega,q;T,R) &=& 
-i \int drdt\ e^{-iqr+i\omega t}\hat{\sf g}^{<}(r,t;R,T),  \nonumber \\
\hat{\sf g}^{>}(\omega,q;T,R) &=& 
i \int drdt\ e^{-iqr+i\omega t}\hat{\sf g}^{>}(r,t;R,T).  \nonumber 
\end{eqnarray}
Then, the derivatives of these Green functions with respect to 
$q$ and $\omega$ are quantities of the 
order of the Fermi length and inverse of a 
band width respectively.  
Thus, the (inner) products between the derivatives of these 
Green functions with respect to $Q\equiv (\omega,q)$  
and those with respect to $X\equiv (T,R)$ can be regarded   
as a small dimensionless quantity,   
as far as a system is only weakly disequilibrated;
\begin{eqnarray}
\hat{\sf g}^{a}\cdot \hat{\sf g}^{b} \gg 
\partial_{X}\hat{\sf g}^{a}\cdot \partial_{Q}\hat{\sf g}^{b}   
\gg \partial_{X}\partial_{X'}\hat{\sf g}^{a}\cdot \partial_{Q}\partial_{Q'}
\hat{\sf g}^b, \cdots . \nonumber    
\end{eqnarray}
The superscripts ``$a$'' and ``$b$'' specify 
the lesser or greater green functions. 
Since Lagrangian and spectral function are given in terms 
of these functions self-consistently,  
we can readily adopt the following relations also:   
\begin{eqnarray}
\hat{\sf L}\cdot\hat{\sf A} &\gg& \partial_{X}\hat{\sf L}
\cdot \partial_{Q}\hat{\sf A}\ ,\   
\partial_{Q}\hat{\sf L}\cdot \partial_{X}\hat{\sf A}\nonumber \\
&\gg& \partial_{X}\partial_{X'}\hat{\sf L}\cdot  
\partial_{Q}\partial_{Q'}\hat{A}\ , \cdots.  \nonumber 
\end{eqnarray} 
 
Observing this, we expand the convolution in the (dissipationless) 
Keldysh equation, in powers of $\partial_{Q}\partial_{X} \equiv - 
\partial_{\omega}\partial_{T} + \partial_{q_j}\partial_{R_j}$;    

\begin{widetext}
\begin{eqnarray}
-\big[\hat{\sf L},\hat{\sf A}\big]_{-} = 
\frac{i}{2}\ \big[\partial_{X_j}\hat{\sf L},\partial_{Q_j}\hat{\sf A}\big]_{+}
-\ \frac{1}{8}\ \Big(\big[\partial_{X_j}\partial_{X_k}\hat{\sf L}
,\partial_{Q_j}\partial_{Q_k}\hat{\sf A}\big]_{-}
-\ \big\{ X_k \leftrightarrow Q_k\big\} \Big)
\ -\ \big\{ X_j \leftrightarrow Q_j\big\} + \cdots.   \label{2-1}
\end{eqnarray}
\end{widetext}
While we kept up to the 2nd order, one could explicitly write down the 
higher than this, by using the following formula for the Moyal product 
\cite{rammer};
\begin{eqnarray}
(A\otimes B)(Q;X)&\equiv& \int d(1-1')\ e^{iQ\cdot \dot(1-1')} \nonumber \\
&&\hspace{1cm} \times \int d\bar{1} A(1|\bar{1})\ B(\bar{1}|1'), \nonumber \\ 
&=& e^{i\frac{1}{2}\big(\partial^{A}_{X}\partial^{B}_{Q}-
\partial^{A}_{Q}\partial^{B}_{X}\big)} A(Q;X)\ B(Q;X), \nonumber 
\end{eqnarray}  
where $\partial^{A}_{X_i}\partial^{B}_{Q_j}\partial^{A}_{Q_m}
\partial^{B}_{X_l}(A B) \equiv (\partial_{X_i}\partial_{Q_m}A)  
(\partial_{Q_j}\partial_{X_l}B)$. 
Note that the $j$ and $k$-summation in eq.(\ref{2-1}) run  
from $0$ to $d$, which will be made 
implicit from now on.  
The (anti-)commutators in eq.(\ref{2-1}) are 
taken {\it only with respect to band indices}, 
while the non-local correlation effect 
encoded into the space-time convolution of eq.(\ref{1-7}) 
is now perturbatively 
taken into account via the gradient expansion. 

\subsection{projection process}
Our projection process is nothing but to solve 
eq.~(\ref{2-1}), 
perturbatively in the gradient expansion, using 
the {\it substitution} method.  
To be more specific, 
we are looking for $\hat{\sf A}$
which satisfies this equation 
at a given order accuracy in the gradient expansion. 
To do this in a well-controlled fashion, we will first 
solve the off-diagonal elements of the spectral function 
in favor of its diagonal elements, by looking into 
the off-diagonal components of the matrix-formed KE  
given in eq.~(\ref{2-1}).  {\it Substituting}  
these solutions back into the diagonal components of 
the same KE, we therefore obtain sort of differential 
equations given {\it only} for the diagonal elements of the 
spectral function (sec.III B,C,D). Then, we will 
further determine appropriate form of these 
diagonal elements, such that  
these differential equations are satisfied (sec. IV). 
   
As in standard perturbation theories, 
we begin with diagonalizing the zero-th order part. 
Introduce the unitary matrix which diagonalizes 
the Lagrangian $\hat{\sf L}$ in the left hand side of 
eq.~(\ref{2-1});  
\begin{eqnarray}
\hat{L}_d \equiv \hat{U}^{\dagger}\hat{\sf L}\ \hat{U}.  
\label{2-1a}
\end{eqnarray} 
Then, a spectral function in this basis has only to 
be diagonal so as to satisfy eq.(\ref{2-1}) at the 
zero-th order; 
\begin{eqnarray}
\hat{A} \equiv \hat{U}^{\dagger}\hat{\sf A}\ \hat{U} 
= \left[ \begin{array}{ccc}
A_1&  & {\bf 0}\\\
& \ddots& \\
{\bf 0}&  & A_{N_b}\\
\end{array} 
\right]. \label{2-1b}
\end{eqnarray} 
We used ``{\sf SansSerif}'' for the spectral/green function 
represented in the old basis, while ``Roman'' for those in 
the new basis.

While each diagonal elements 
can be arbitrary at this level, from the physical point of view, 
they should be delta functions; 
\begin{eqnarray}
A_{\alpha} \equiv \delta (L^{(0)}_{d,\alpha}) 
= Z^{(0)}_{\alpha}\delta(\omega-{\epsilon}^{(0)}_{\alpha,q}). 
\label{2-2}
\end{eqnarray} 
Namely, $\epsilon_{\alpha,q}$ above denotes 
the (renormalized) energy dispersion for the 
$\alpha$-th band q.p., while $Z_{\alpha}$ 
stands for this q.p. spectral weight. 
The superscript $(0)$ simply represents that they 
are quantities of the zero-th order in gradient  
expansion. 
When employing this form as the 0-th order solution, 
we will actually be able to satisfy the Keldysh equation 
up to the 1st order in the gradient expansion (see eq.~(\ref{4-1a-0})). 
Thus, we can justify {\it posteriori} 
that eq.(\ref{2-2}) is the appropriate  
0-th order solution, based on which 
its higher order correction can be built up 
(see also the arguments in section IV).

Under this unitary transformation, the usual derivative 
encoded in Eq.~(\ref{2-1}) is replaced by the ``covariant'' 
derivative; 
\begin{eqnarray}
\hat{U}^{\dagger}\big(\partial_{X} \hat{\sf B}\big)\hat{U}  
=[\hat{D}_X,\hat{B}] &\equiv& \partial_X \hat{B} + 
\hat{\cal A}_X \hat{B} - \hat{B}\hat{\cal A}_X. \label{2-3a} 
\end{eqnarray} 
with $\hat{B} = U^{\dagger}\cdot \hat{\sf B}\cdot U$ and 
$\hat{\cal A}_X=\hat{U}^{\dagger}\partial_X \hat{U}$. 
Namely, in terms of this derivative, 
our matrix-formed differential equation reads  
\begin{eqnarray}
-[\hat{L}_d,\hat{A}] 
= \frac{i}{2}\ \big[[\hat{D}_{X_j},\hat{L}_d],[\hat{D}_{Q_j}\hat{A}]\big]_{+} 
+ \cdots \equiv \hat{F}(\hat{A}).  \label{2-3}   
\end{eqnarray} 

To find the spectral function satisfying Eq.~(\ref{2-3}) 
up to higher order in the gradient expansion, let us next  
look into the off-diagonal components of this 
covariant differential equation; 
\begin{eqnarray}
-L_{d,\alpha}A_{\alpha\beta} + A_{\alpha\beta}L_{d,\beta} 
= F_{\alpha\beta}(\{A_{\gamma}\},\{A_{\gamma\eta}\}). \label{2-4} 
\end{eqnarray}
$F_{\alpha\beta}$ is a functional of a set of diagonal elements 
of the spectral function, i.e. $\{A_{\gamma}\equiv A_{\gamma\gamma}\}$, 
and a set of its off-diagonal elements 
$\{A_{\gamma\eta}\}$ ($\gamma \ne \eta$). 
Notice first that the right hand 
side of eq.~(\ref{2-4}) is at least   
1st order in gradient expansion.   
On the one hand, its left hand side is 
basically proportional to 
a direct band gap between $\alpha$-th 
band and $\beta$-th band, which is finite 
even at equilibrium, i.e. $L_{d,\alpha}\ne L_{d,\beta}$. 
Thus the off-diagonal elements of the spectral function are  
of the order of ${\cal O}(|\partial_X\partial_Q|)$, 
while its diagonal elements remain 
finite even at equilibrium. 

As such, we first solve these off-diagonal elements in 
favor of the diagonal elements of the spectral function, 
{\it iteratively in gradient expansion.}  
Specifically, to the 1st order's accuracy, 
we have only to replace the off-diagonal elements 
in the {\it right hand side}  
of eq.~(\ref{2-4}) by zero;   
\begin{eqnarray}
A_{\alpha\beta}=
F_{\alpha\beta}(\{A_{\gamma}\},\{0\})\cdot  
(L_{d,\beta}-L_{d,\alpha})^{-1} \equiv A^{(1)}_{\alpha\beta}. \label{2-5}
\end{eqnarray}
Using this, one could further obtain the solution of 
$A_{\alpha\beta}$ to the 2nd order accuracy; 
\begin{eqnarray}
A^{(2)}_{\alpha\beta}(\{A_{\gamma}\}) = 
F_{\alpha\beta}(\{A_{\gamma}\},\{A^{(1)}_{\gamma\eta}\}),  \label{2-6} 
\end{eqnarray}
or higher than that, 
\begin{eqnarray}
A^{(n+1)}_{\alpha\beta}(\{A_{\gamma}\}) = 
F_{\alpha\beta}(\{A_{\gamma}\},\{A^{(n)}_{\gamma\eta}\}).  
\label{2-6a} 
\end{eqnarray}
%
When $n$ being infinity, this clearly becomes an exact 
relation between diagonal elements and off-diagonal 
elements.

Substituting these solutions  
into the diagonal components of the covariant differential 
equation, we then have   
$N_b$ {\it decoupled} equations which are given in terms 
{\it only} of a set of diagonal elements 
of the spectral function;
\begin{eqnarray}
0 &=& F_{\alpha}(\{A_{\gamma}\},\{A_{\gamma\eta}\}) \nonumber \\
&=&F_{\alpha}(\{A_{\gamma}\},
\{A_{\gamma\eta}\equiv  A^{(\infty)}_{\gamma\eta}(\{A_{\gamma}\})\}),  
\label{2-7} 
\end{eqnarray}
with $F_{\alpha}\equiv F_{\alpha\alpha}$. 
Then, a remaining task is to determine  
a set of $N_b$ diagonal elements of $\hat{A}$,  
such that they observe these equations. 
$\{A_{\gamma}\}$ thus obtained in combination 
with $A_{\gamma\eta}\equiv  A^{(\infty)}_{\gamma\eta}(\{A_{\gamma}\})$ 
are in principle equivalent to the exact solution of the 
an original dissipationless Keldysh equation, i.e. eq.~(\ref{2-1}). 

\begin{figure}
\begin{center}
\includegraphics[width=0.40\textwidth]{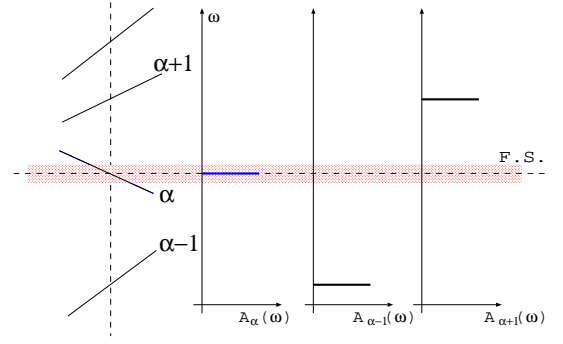}
\end{center}
\caption{A schematic picture of the spectral function 
for the $\alpha$-th band and its neighboring bands.}
\label{fig2}
\end{figure}

However, the exact solution is apparently impossible, since 
it would require us, for example, to perform the 
iteration of eq.~(\ref{2-6a}) at an infinite time. 
Thereby, we are going to indulge ourselves in executing 
this sequence of the process, 
{\it up to a given order in gradient expansion}. 
The highest order up to which we have succeeded in 
obtaining eq.~(\ref{2-7}) is currently the 2nd order.  
Up to this order, we can readily ignore the 3rd 
order gradient expansion term (and higher than that)  
denoted by ``$\cdots$'' in eq.~(\ref{2-1}). 
Furthermore, $F_{\alpha\alpha}$ already containing at least 
one pair of $\partial_X \partial_Q$, the difference between 
eq.~(\ref{2-5}) and eq.~(\ref{2-6}) ends up with 
the 3rd order contributions, when substituted into 
$F_{\alpha\alpha}$ as in eq.~(\ref{2-7}). 
Thus, we are ready to use eq.~(\ref{2-5}). 

The final simplification we will employ 
is that, for given $k$, $R$ and $T$, each quasi-particle 
bands have spectral weights at energetically 
well separated regions from one another. 
To be concrete, let us refer to a band which contains 
a Fermi surface as the $\alpha$-th band. Then,    
the diagonal elements of the spectral function corresponding 
to {\it the other bands}, i.e. $A_{\gamma\ne \alpha}$, 
have negligible weights at the low energy region, i.e. 
$|\omega-{\epsilon}_{\alpha}|\simeq
|\omega-\mu|\ll \min_{\beta}|\Delta_{\alpha\beta}|$. 
This can be seen precisely at 
equilibrium, where $A_{\gamma\ne \alpha}$ is sharply peaked at 
an energy region separated from $\mu$ by the 
direct band gap $\Delta_{\gamma\alpha}$ 
(see Fig.~\ref{fig2}). 
Even off equilibrium, 
higher order gradient expansion corrections 
to $A_{\gamma}$ turn out {\it not} to make additional 
{\it incoherent} weights other than the delta function we originally 
have at equilibrium. Namely, the corrections to $A_{\gamma}$
appear in eq.~(\ref{2-2}) 
{\it only} as the energy dispersion   
{\it shift} and the {\it additive} spectral weight 
(see for example eq.~(\ref{4-7a}));   
\begin{eqnarray}
\epsilon^{(0)}_{\alpha} &\rightarrow& \epsilon^{(0)}_{\alpha} + 
\epsilon^{(1)}_{\alpha},  \nonumber \\
Z^{(0)}_{\alpha} &\rightarrow& Z^{(0)}_{\alpha} + Z^{(1)}_{\alpha}. 
\nonumber   
\end{eqnarray}  
Accordingly, as far as this energy shift, 
i.e. $\epsilon^{(1)}_{\gamma}$, does not change the relative 
position of each quasi-particle energy dispersions, we are {\it posteriori} 
allowed to replace $A_{\gamma\ne \alpha}$ by zero, 
for $|\omega-{\epsilon}_{\alpha}|\ll {\rm min}_{\beta}\Delta_{\alpha\beta}$.  

To summarize the simplifications possible at the 2nd order 
analysis, we have only to derive the following equations;
\begin{eqnarray}
0 &=& F_{\alpha}(\{A_{\alpha},A_{\gamma\ne \alpha}\equiv 0\}, 
\{A^{(1)}_{\gamma\eta}\}),  \label{3-1} \\
A^{(1)}_{\gamma\eta}
&\equiv& F_{\gamma\eta}(\{A_{\alpha}, 
A_{\gamma\ne \alpha}\equiv 0\},\{0\})\cdot(L_{d,\eta} - L_{d,\gamma})^{-1}.  
\nonumber  
\end{eqnarray}
In the next next subsection, 
we will 
substitute the latter 
into the former, so as to obtain the (differential) equation 
only for $A_{\alpha}$. We will 
dub the equation thus obtained 
as a {\it reduced Keldysh equation (RKE)}.    
In the section IV, we will further find a  
$A_{\alpha}$ satisfying this reduced Keldysh equation.  
Thereby, we actually observe that $A_{\alpha}$ 
thus obtained is sharply peaked at 
$\omega=\epsilon_{\alpha}$, while having no 
incoherent weights at high energy sides.   
This observation will support {\it posteriori} 
the logical consistency built in our 
prescription described above.  
 

\subsection{$SU(2)$ FLs and additional coupling constant} 
In the argument of the previous subsection, 
we have implicitly assumed that 
a Fermi surface in question is composed 
only by a single band (especially  
eqs.~(\ref{2-1a},\ref{2-1b})). 
In general, this is true either in those metals 
without any centrosymmetric lattice point or in 
ferromagnetic metals. 
In such FLs, there remains only a charge degree of freedom, 
while the (pseudo-)spin degree of freedom at each $k$-point 
is usually {\it quenched}. This charge degree of freedom 
is then described by the spectral 
function $A_{\alpha}$, which is a {\it scalar} quantity.   
Accordingly, the RKE, i.e. eq.~(\ref{3-1}), 
becomes just a differential equation for this scalar function,   
which we can name as a $U(1)$ RKE.

On the one hand, in usual paramagnetic metals having a centrosymmetric 
lattice point, each $k$-point is (at least) doubly degenerate, describing 
spin-degree of freedom for quasi-particles. Namely, eigenvalues of  
our Lagrangian are always two-folded at equilibrium. 
Thus, a Fermi surface is composed by two degenerate 
bands, which we could name as $SU(2)$ FL; 
\begin{eqnarray}
\hat{L}_d = \hat{U}^{\dagger}\hat{\sf L}\ \hat{U}=\left[ \begin{array}{ccc}
\hat{L}_{d,1}& & {\bf 0}\\ 
&\ddots& \ \\ 
{\bf 0}&  & \hat{L}_{d,N_b}\\ 
\end{array} 
\right]. \nonumber 
\end{eqnarray} 
Here 2 by 2 matrices $\hat{L}_{d,\gamma}$ should be proportional to 
a unit matrix at equilibrium.

As such, the spectral function which satisfies  
eq.~(\ref{2-1}) 
at the zero-th order in gradient expansion 
has only to be block-diagonalized,    
\begin{eqnarray}
\hat{A} = \hat{U}^{\dagger}\hat{\sf A}\ \hat{U}=\left[ \begin{array}{ccc}
\hat{A}_1& & {\bf 0}\\ 
&\ddots& \ \\ 
{\bf 0}&  & \hat{A}_{N_b}\\ 
\end{array} 
\right], \label{3-2}
\end{eqnarray}
with arbitrary $2\times 2$ matrices  
$\hat{A}_{\gamma}$ $(\gamma = 1,\cdots,N_b)$. 

Corresponding to this generic degeneracy 
at the zero-th order, we will derive the RKEs in SU(2) FLs {\it in favor 
of these $2$ by $2$ spectral functions}, i.e. $\hat{A}_{\gamma}$ 
$(\gamma = 1,\cdots,N_b)$, so that 
spin and charge degrees of freedom 
are treated on an equal footing. As will be  
shown later,  
the RKE thus derived becomes a $2$ by $2$ matrix-formed 
differential equation, which we will 
dub as $SU(2)$ RKE.

Apart from relatively minor modifications, 
the derivation of the $SU(2)$ RKE 
also goes along with the same procedure as in $U(1)$ 
case. Specifically, it also 
begins with the inter-band components of KE, which now 
take a $2$ by $2$ matrix-form; 
\begin{eqnarray}
- \hat{L}_{d,\gamma} \hat{A}_{\gamma\eta} 
+ \hat{A}_{\gamma\eta} \hat{L}_{d,\eta} =
\hat{F}_{\gamma\eta}(\{\hat{A}_{\gamma}\},\{\hat{A}_{\gamma\eta}\}),  
\label{3-3} \\
\nonumber 
\end{eqnarray}
where $\hat{A}_{\gamma\eta}$ is a $2$ by $2$ matrix, 
connecting the $\gamma$-th band and $\eta$-th band. 

Being proportional to a unit matrix at equilibrium, $2$ by $2$ matrices  
$\hat{L}_{d,\gamma}$ may be decoupled into its zero-th order part and 
a small degeneracy lifting part;  
\begin{eqnarray}
\hat{L}_{d,\gamma} = L^{(0)}_{d,\gamma} \hat{1} - 
\epsilon_{\gamma} \hat{\sigma}_z.   \label{3-4a} \\
\nonumber 
\end{eqnarray}
A finite $\hat{\epsilon}_{\gamma}\equiv 
\epsilon_{\gamma}\hat{\sigma_z}$ 
is generally originated from    
weak $R$- and $T$-dependences of the Green functions. 
Thus $\epsilon_{\gamma}$ (divided by a 
characteristic band width) 
should be treated as a {\it same order} of quantity 
as $|\partial_X \partial_Q|$.  To convince ourselves of this 
more directly,  consider a specific situation in the presence of 
a {\it small} magnetic field ${\bf b}$, or equivalently 
a {\it slowly varying} magnetic gauge potential.  
In such a case, the Zeeman coupling energy between (bare) spin and 
${\bf b}$ clearly should be included 
in $\hat{\epsilon}_{\gamma}$. 
On the one hand, 
$|\partial_X \partial_Q|$ contribution turns out to be   
proportional to a spatial derivative of the external gauge field, 
which is therefore proportional to ${\bf b}$ also 
(for example compare eq.~(\ref{3-17a}) 
with eqs.~(\ref{3-29c}-\ref{3-29d})).   
As is obvious from this example, 
$|\partial_X \partial_Q|\equiv \lambda_1$ and  
$\epsilon/\Delta\equiv \lambda_2$ should be treated 
as same order of quantities; 
\begin{eqnarray}
\lambda_1 \sim \lambda_2 \sim \lambda. \nonumber 
\end{eqnarray} 
 
Observing the inter-band components of the covariant 
differential equations, i.e. eq.~(\ref{3-3}), we first 
relate the inter-band elements of $\hat{A}$ 
with a set of $N_b$ intra-band elements of $\hat{A}$.  
To the 1st order in $\lambda_1$ or $\lambda_2$, i.e. 
${\cal O}(\lambda_1,\lambda_2)$, we have 
\begin{eqnarray}
\hat{A}_{\gamma\eta}= \hat{F}_{\gamma\eta}(\{\hat{A}_{\gamma}\},
\{\hat{0}\})\cdot (\hat{L}_{d,\eta}-L^{(0)}_{d,\gamma}\hat{1})^{-1} \equiv 
\hat{A}^{(1)}_{\gamma\eta}. \nonumber
\end{eqnarray}  
Substituting these 1st order solutions into the 
{\it intra-band} components of KE, we then  
obtain the following $SU(2)$ RKE to the accuracy of 
${\cal O}(\lambda^2_1,\lambda^2_2,\lambda_1\lambda_2)$; 
\begin{widetext}
\begin{eqnarray}
-\big[\hat{L}_{d,\alpha},\hat{A}_{\alpha}\big] &=&  
\big[\hat{\epsilon}_{\alpha},\hat{A}_{\alpha}\big] = 
\hat{F}_{\alpha}(\{\hat{A}_{\alpha},
\hat{A}_{\gamma\ne \alpha}\equiv 0\},\{\hat{A}^{(1)}_{\gamma\eta}\}),  
\label{3-4} \\
\hat{A}^{(1)}_{\gamma\eta}(\hat{A}_{\alpha})&\equiv& \hat{F}_{\gamma\eta}
(\{\hat{A}_{\alpha},
\hat{A}_{\gamma\ne \alpha}\equiv 0\},\{\hat{A}_{\gamma
\eta}\equiv \hat{0}\}) 
\cdot(\hat{L}_{d,\eta} -L^{(0)}_{d,\gamma}\hat{1})^{-1}. \label{3-5} 
\end{eqnarray} 
\end{widetext}
The intra-band elements of $\hat{A}$ for the bands other than 
the $\alpha$-th band, 
i.e. $\hat{A}_{\gamma \ne \alpha}$, 
were already replaced by zero, 
because of the same reason as we argued in the $U(1)$ case.

In the next subsection, we will 
calculate eq.~(\ref{3-4}) in combination 
with eq.~(\ref{3-5}) more explicitly, so as to 
obtain an actual form of the $SU(2)$ RKE up to 
the order of 
${\cal O}(\lambda^2)$.  
Out of the $SU(2)$ RKE thus derived, 
one can immediately obtain the RKE in $U(1)$ FLs, 
by regarding $\hat{A}_{\alpha}$ as a {\it scalar} function 
and putting $\hat{\epsilon}_{\alpha}$ to be zero.

%
\begin{figure}
\begin{center}
\includegraphics[width=0.45\textwidth]{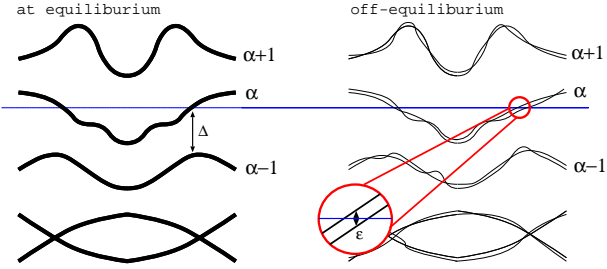}
\end{center}
\caption{A schematic picture of the energy dispersion 
in $SU(2)$ FLs. (Right): Each dispersion is doubly 
degenerate at equilibrium. (Left): When
a system is weakly disequilibrated, the doubly degeneracy 
at each $k$-point is lifted. The associated splitting 
energy, i.e. $\epsilon$, is however much smaller than 
the typical band gap/width at equilibrium, i.e. $\Delta$.}
\label{fig3}
\end{figure}

\subsection{Actual Derivations of $SU(2)$ RKEs } 

In this subsection, we will perform the actual calculation 
of eqs.~(\ref{3-4}) and (\ref{3-5}) in a {\it covariant} way, 
only to arrive at the $SU(2)$ RKE associated 
with the $\alpha$-th band in question.  
By referring ``in a covariant way'', we mean that 
{\it every} step in the following manipulation does {\it not} 
change its {\it explicit} form either under an $SU(2)$ rotation 
$\hat{v}_{\alpha}$ within the $\alpha$-th band or, under an  
$SU(2N_b-2)$ rotation $\hat{V}_{\bar{\alpha}}$ within its 
complementary space; 
\begin{eqnarray}
\hat{U}&\rightarrow& 
\hat{U}\cdot 
\left[ 
\begin{array}{cc}
\hat{v}_{\alpha} & \hat{0} \\
\hat{0} &  \hat{V}_{\bar{\alpha}} \\
\end{array} 
\right]. \label{3-7a}
\end{eqnarray} 
This is because 
any steps of the actual calculation are 
composed either by the {\it inter-band covariant derivatives}  
or by the {\it intra-band covariant derivatives}, 
both of which will be accurately 
defined in the next subsubsection.    
In addition to their definitions, 
some formula frequently used in the next next subsubsection   
will be also summarized in advance for clarity.  

\subsubsection{arithmetic preliminaries}
We have already defined the $SU(2N_b)$ gauge field and 
associated ``covariant'' derivative in eq.~(\ref{2-3a}). 
In an analogous way,  
the $SU(2)$ gauge fields and covariant derivatives 
for the two-fold degenerate $\alpha$-th band are 
defined as follows;
\begin{eqnarray}
&&\big[\hat{D}^{\alpha}_{X},\hat{B}_{\alpha}\big] 
\equiv \partial_{X}\hat{B}_{\alpha} + 
\big[\hat{\cal A}^{\alpha}_{X},\hat{B}_{\alpha}\big], \nonumber  \\ 
&&\ \big[\hat{\cal A}^{\alpha}_{X}\big]_{(\sigma|\sigma')} 
\equiv  \big[\hat{U}^{\dagger}\partial_{X}
\hat{U}\big]_{(\alpha\sigma|\alpha\sigma')}, \label{3-5a} 
\end{eqnarray} 
where $X$ could be any coordinates in the phase space. 
We sub(super)scribe $\alpha$ or $\bar{\alpha}$ such as 
$\hat{B}_{\alpha}$ or $\hat{B}_{\bar{\alpha}}$, 
only to suggest that this matrix  
is a $2$ by $2$ matrix associated with the $\alpha$-th 
band or a $(2N_b-2) \times (2N_b-2)$ matrix in its 
complementary space respectively. 
When sub(super)scribed ``$\alpha\bar{\alpha}$'' 
such as $\hat{B}^{(\alpha\bar{\alpha})}_{\alpha\bar{\alpha}}$, a matrix 
ought to be regarded as a $2\times (2N_b-2)$ matrix. 
   
Apart from the derivative in the {\it full} Hilbert 
space, the covariant derivatives in its {\it subspace} 
do not commute with one another in general, 
\begin{eqnarray}
&&\big[\hat{D}^{\alpha}_{X},\big[\hat{D}^{\alpha}_{X'},\hat{B}_{\alpha}\big]\big]
- \big[\hat{D}^{\alpha}_{X'},\big[\hat{D}^{\alpha}_{X},\hat{B}_{\alpha}\big]\big]
= -i\big[\hat{\Omega}^{\alpha}_{XX'},\hat{B}_{\alpha}\big], \nonumber \\ 
&&\ \ \ \hat{\Omega}^{\alpha}_{XX'}\equiv i\big[\hat{D}^{\alpha}_{X},\hat{D}^{\alpha}_{X'}\big] \nonumber \\
&&\hspace{1.1cm}\ \ =\ i\partial_X \hat{\cal A}^{\alpha}_{X'} 
- i\partial_{X'} \hat{\cal A}^{\alpha}_{X} + i
\big[\hat{\cal A}^{\alpha}_{X},\hat{\cal A}^{\alpha}_{X'}\big] \label{3-6} .
\end{eqnarray} 
As is clear from the latter expression,  
$\hat{\Omega}^{\alpha}_{XX'}$ would be identical to 
zero, if this doubly degenerate $\alpha$-th band were to   
subtend a complete set, i.e. 
$\sum_{\sigma=\pm}|u^{\alpha\sigma}\rangle\langle u^{\alpha\sigma}|\equiv \hat{1}$.  

Note that this derivative and 
the associated curvature $\hat{\Omega}^{\alpha}_{XX'}$ 
transform in a covariant way under the unitary transformation 
in eq.~(\ref{3-7a});  
\begin{eqnarray}
\hat{B}_{\alpha} &\rightarrow& \hat{v}^{\dagger}_{\alpha}\hat{B}_{\alpha} 
\hat{v}_{\alpha}, \nonumber \\
\big[\hat{D}^{\alpha}_{X},\hat{B}_{\alpha}\big]& \rightarrow &
\hat{v}^{\dagger}_{\alpha}\big[\hat{D}^{\alpha}_{X},\hat{B}_{\alpha}\big] 
\hat{v}_{\alpha}, \label{3-6a} \\
\hat{\Omega}^{\alpha}_{XX'}&\rightarrow &
\hat{v}^{\dagger}_{\alpha}\hat{\Omega}^{\alpha}_{XX'}
\hat{v}_{\alpha}, \nonumber 
\end{eqnarray}
while the gauge field $\hat{\cal A}^{\alpha}_{X}$ 
is {\it not}; 
\begin{eqnarray}
\hat{\cal A}^{\alpha}_{X} \rightarrow 
\hat{v}^{\dagger}_{\alpha}\hat{\cal A}^{\alpha}_{X}\hat{v}_{\alpha} 
+ \hat{v}^{\dagger}_{\alpha}\partial_{X}\hat{v}_{\alpha}.  \nonumber 
\end{eqnarray}

In addition to this gauge field in the $\alpha$-th band,  
we can also define the {\it inter-band} 
gauge fields as the $2 \times (2N_b-2)$  
off-diagonal blocks of $\hat{\cal A}_{X}$;
\begin{eqnarray}
\hat{\cal A}_{X}\equiv \hat{U}^{\dagger}\partial_{X}\hat{U}\equiv 
\left[ 
\begin{array}{cc}
\hat{\cal A}^{\alpha}_{X} & \hat{\cal A}^{\alpha\bar{\alpha}}_{X} \\
\hat{\cal A}^{\bar{\alpha}\alpha}_{X} &  \hat{\cal A}^{\bar{\alpha}}_{X} \\
\end{array} 
\right]. \nonumber 
\end{eqnarray}   
Under the unitary transformation defined in eq.~(\ref{3-7a}), 
the inter-band gauge field clearly transforms 
in a  ``covariant'' way; 
\begin{eqnarray}
\hat{\cal A}^{\alpha\bar{\alpha}}_{X} &\rightarrow& 
\hat{v}^{\dagger}_{\alpha}  
\hat{\cal A}^{\alpha\bar{\alpha}}_{X} 
\hat{V}_{\bar{\alpha}}, \nonumber 
\end{eqnarray}
while the {\it intra-band} gauge fields such as $\hat{\cal A}^{\alpha}_{X}$ 
and $\hat{\cal A}^{\bar{\alpha}}_{X}$ are not. 

Using the latter ones, we can further define a  
$2 \times (2N_b-2)$ matrix-form derivative, which 
plays role of a sort of {\it inter-band covariant derivative}; 
\begin{eqnarray} 
\big[\hat{D}^{\alpha\bar{\alpha}}_{X},\hat{B}_{\alpha\bar{\alpha}}\big] 
\equiv \partial_{X}\hat{B}_{\alpha\bar{\alpha}} 
+ \hat{\cal A}^{\alpha}_{X} 
\hat{B}_{\alpha\bar{\alpha}} 
- \hat{B}_{\alpha\bar{\alpha}}
\hat{\cal A}^{\bar{\alpha}}_{X}, \nonumber 
\end{eqnarray} 
where $\hat{B}_{\alpha\bar{\alpha}}$ stands for  
an arbitrary $2\times(2N_b-2)$ matrix. As long as this 
matrix is a covariant quantity,   
i.e. $\hat{B}_{\alpha\bar{\alpha}}
\rightarrow \hat{v}^{\dagger}_{\alpha}\hat{B}_{\alpha\bar{\alpha}} 
\hat{V}_{\bar{\alpha}}$, 
the inter-band derivative above clearly  
transforms in a covariant way under eq.~(\ref{3-7a});
\begin{eqnarray}
\big[\hat{D}^{\alpha\bar{\alpha}}_{X}
,\hat{B}_{\alpha\bar{\alpha}}\big] \rightarrow 
\hat{v}^{\dagger}_{\alpha} \big[\hat{D}^{\alpha\bar{\alpha}}_{X}
,\hat{B}_{\alpha\bar{\alpha}}\big] \hat{V}_{\bar{\alpha}}. \label{3-8}
\end{eqnarray} 

In terms of these quantities and derivatives, let us 
summarize henceforth several formula which become useful in 
the next subsubsection. 
Consider first a inter-band covariant derivative 
of a inter-band gauge field;  
\begin{eqnarray}
&&\big[\hat{D}^{\alpha\bar{\alpha}}_{X}
,\hat{\cal A}^{\alpha\bar{\alpha}}_{X'}\big]_{(\sigma|\beta\sigma')}
\nonumber \\
&&= \big[\big(\partial_{X}\hat{U}^{\dagger}\big)\ \partial_{X'}\hat{U}
\big]_{(\alpha\sigma|\beta\sigma')} 
+ \big[\hat{U}^{\dagger}\partial^2_{XX'}\hat{U}
\big]_{(\alpha\sigma|\beta\sigma')} \nonumber \\
&&\ \ \ \ -  \sum_{\sigma''=\pm}
\big[\big(\partial_{X}\hat{U}^{\dagger}\big)\ 
\hat{U}\big]_{(\alpha\sigma|\alpha\sigma'')} 
\big[\hat{U}^{\dagger}
\partial_{X'}\hat{U}\big]_{(\alpha\sigma''|\beta\sigma')}  \nonumber \\
&&\ \ \ \ + \sum_{\gamma\ne \alpha}\sum_{\sigma''=\pm} 
\big[\big(\partial_{X'}\hat{U}^{\dagger}\big)\ 
\hat{U}\big]_{(\alpha\sigma|\gamma\sigma'')} 
\big[\hat{U}^{\dagger}
\partial_{X}\hat{U}\big]_{(\gamma\sigma''|\beta\sigma')}, \nonumber 
\end{eqnarray}
where $\beta \ne \alpha$ and $\sigma,\sigma'=\pm$.
Then, applying into the last two terms the following identity,  
\begin{eqnarray}
\sum_{\sigma=\pm}|\alpha\sigma\rangle \langle \alpha\sigma|
= \hat{1} - \sum_{\gamma\ne\alpha}\sum_{\sigma=\pm}|\gamma\sigma
\rangle \langle \gamma\sigma|, \nonumber
\end{eqnarray} 
we can exchange the subscripts of the covariant 
derivative and the gauge field with each other;    
\begin{eqnarray}
\big[\hat{D}^{\alpha\bar{\alpha}}_{X}
,\hat{\cal A}^{\alpha\bar{\alpha}}_{X'}\big]_{(\sigma|\beta\sigma')} 
\equiv \big[\hat{D}^{\alpha\bar{\alpha}}_{X'}
,\hat{\cal A}^{\alpha\bar{\alpha}}_{X}\big]_{(\sigma|\beta\sigma')}.   
\label{3-9}
\end{eqnarray}
In the actual calculations, this equality becomes very powerful, 
when combined with the ``decomposition rule'' 
of covariant derivatives such as; 
\begin{eqnarray}
&&\hspace{-0.5cm}\big[\hat{D}^{\alpha}_{X},\hat{B}_{\alpha\bar{\alpha}}\hat{B}'_{\bar{\alpha}}
\hat{B}''_{\bar{\alpha}\alpha}\big] =  
\big[\hat{D}^{\alpha\bar{\alpha}}_{X},\hat{B}_{\alpha\bar{\alpha}}\big]
\hat{B}'_{\bar{\alpha}}\hat{B}''_{\bar{\alpha}\alpha}  \nonumber \\
&&\ \ \ \ +\hat{B}_{\alpha\bar{\alpha}}
\big[\hat{D}^{\bar{\alpha}}_{X},\hat{B}'_{\bar{\alpha}}\big]
\hat{B}''_{\bar{\alpha}\alpha}  
+ \hat{B}_{\alpha\bar{\alpha}}\hat{B}'_{\bar{\alpha}}
\big[\hat{D}^{\bar{\alpha}\alpha}_{X},\hat{B}''_{\bar{\alpha}\alpha}\big]. 
\label{3-10} 
\end{eqnarray} 
One should also note that   
curvatures either in the $\alpha$-th band space or 
in its complementary space can be expressed also in 
terms of {\it inter-band} gauge fields; 
\begin{eqnarray}
\hat{\Omega}^{\alpha}_{XX'}&=& -i \big(
\hat{\cal A}^{\alpha\bar{\alpha}}_{X}
\hat{\cal A}^{\bar{\alpha}\alpha}_{X'} 
-  \hat{\cal A}^{\alpha\bar{\alpha}}_{X'}
\hat{\cal A}^{\bar{\alpha}\alpha}_{X}\big), \label{3-11} \\
\hat{\Omega}^{\bar{\alpha}}_{XX'}&=& -i \big(
\hat{\cal A}^{\bar{\alpha}\alpha}_{X}
\hat{\cal A}^{\alpha\bar{\alpha}}_{X'} 
-  \hat{\cal A}^{\bar{\alpha}\alpha}_{X'}
\hat{\cal A}^{\alpha\bar{\alpha}}_{X}\big). \label{3-12} 
\end{eqnarray}  

\subsubsection{ $SU(2)\times SU(2N_b-2)$ covariant manipulations}

Using the arithmetics described so far, we will study 
eqs.~(\ref{3-4},\ref{3-5}) within the 2nd order accuracy in $\lambda$. 
As is the case for a standard (such as Rayleigh-Schrodinger) 
perturbation theory, our 2nd order 
expression for eqs.~(\ref{3-4},\ref{3-5}) is given both 
by eigen-energies at the zero-th order and 
its eigen-wavefunctions (see eq.~(\ref{3-24}) for example). 
Namely, differences between eigenvalues of $\hat{\sf L}$ enter into 
a sort of ``energy-denominator'', while a ``numerator'' in a usual 
perturbation theory is now transcribed into 
a gauge field (connection),  
which is nothing but the {\it matrix element} of  
our perturbation part (i.e. $\partial_{X}\partial_{Q}$) among 
different eigenbases of $\hat{\sf L}$.    
Accordingly, just 
as in a usual perturbation theory, 
our 2nd order expression for 
eqs.~(\ref{3-4},\ref{3-5}) also 
depends on wavefunctions and eigenvalues 
not only for the $\alpha$-th band in question, 
but also for {\it the bands other than 
the $\alpha$-th band}.     

On the other hand, 
extensive semi-classical analyses 
in a non-interacting system \cite{sn,omn, mnz, cn, bb, si} 
suggest that the low-energy effective theory for (quasi-)particles 
should be constituted {\it only} by 
Bloch wavefunctions and energy dispersions for the $\alpha$-th 
band in question, while   
free from details of the other bands.  

We shall show in this subsubsection that this is indeed the case for  
the $SU(2)$ RKE given in eqs.~(\ref{3-4},\ref{3-5}) at least 
up to 2nd order in $\lambda$. To be more specific, we will 
transform 
eq.~(\ref{3-4}) in combination with eq.~(\ref{3-5})  
into a more {\sl compact} form  
{\sl rigorously up to ${\cal O}(\lambda^2)$}, only to 
find that {\it they actually are given solely in terms of the $SU(2)$  
gauge covariant quantities such as  $\hat{\Omega}^{\alpha}_{XX'}$ and 
$\big[\hat{D}^{\alpha}_{X},\cdots\big]$.} During this transformation, 
several formula described in the previous subsubsection, such as 
eqs.~(\ref{3-9}-
\ref{3-12}), become very useful.  


To see this, let us begin with eq.~(\ref{3-4}), i.e. the $(\alpha,\alpha)$-th  
component (diagonal component) of the original dissipationless 
Keldysh equation;  
\begin{eqnarray}
-\big[\hat{L}_{d},\hat{A}_{\alpha}\big]
= \hat{F}^{(1)}_{\alpha}(\hat{A}_{\alpha}) 
+ F^{(2)}_{\alpha}(\{\hat{A}_{\alpha\eta}\}) 
+ F^{(3)}_{\alpha}(\{\hat{A}_{\eta\delta}\}). \label{3-14b} 
\end{eqnarray} 
For the later clarity, the right hand side was decoupled 
with respect to different elements of $\hat{A}$. 
This becomes possible clearly because our differential equation 
is at most {\it linear} in $\hat{A}$. 
Up to 1st order in $\lambda_1$ 
the first term in the right hand side reads; ~\cite{note1} 
\begin{eqnarray}
&&\hat{F}^{(1)}_{\alpha}(\hat{A}_{\alpha}) = \frac{i}{2}
\Big[\big[\hat{D}^{\alpha}_{X_j},\hat{L}_{d,\alpha}\big],  \big[\hat{D}^{\alpha}_{Q_j},\hat{A}_{\alpha}\big]\Big]_{+} 
- \big\{X_j \leftrightarrow Q_j\big\}  \nonumber \\
&&\hspace{0.2cm} - \big[\hat{\cal M}_{\alpha},\hat{A}_{\alpha}\big]_{-}  
- \frac{1}{4} \big[\hat{\cal N}_{\alpha},
\hat{A}_{\alpha}\big]_{+} + {\cal O}(\lambda^2_1), \label{3-19a}  
\end{eqnarray} 
where we have introduced following $2$ by $2$ hermite 
and anti-hermite matrices respectively; 
\begin{eqnarray}
\hat{\cal M}_{\alpha} &\equiv& \frac{i}{2}
\big\{
\hat{\cal A}^{\alpha\bar{\alpha}}_{Q_j}\big(
\hat{L}_{d,\bar{\alpha}} 
-L^{(0)}_{d,\alpha}\hat{1}\big)
\hat{\cal A}^{\bar{\alpha}\alpha}_{X_j} 
-\{X_j\leftrightarrow Q_j\} \big\}. \label{3-17a} \\ 
\hat{\cal N}_{\alpha} &\equiv& 
\big[\hat{\epsilon}_{\alpha},
\hat{\Omega}^{\alpha}_{X_jQ_j}\big]_{-}. \label{3-18}
\end{eqnarray}
$\hat{L}_{d,\bar{\alpha}}$ denotes a $(2N_b-2)\times (2N_b-2)$ 
diagonal block of $\hat{L}_d$;
\begin{eqnarray}
\hat{L}_d \equiv \left[ \begin{array}{cc}
\hat{L}_{d,\alpha}& {\bf 0}\\ 
{\bf 0}  & \hat{L}_{d,\bar{\alpha}}\\ 
\end{array} 
\right]. \nonumber 
\end{eqnarray}
Observing eq.~(\ref{3-19a}), notice first the commutator 
between $\hat{\cal M}_{\alpha}$ and the $\alpha$-th band spectral 
function, i.e $\hat{A}_{\alpha}$. 
This commutator implies that the former hermitian matrix is 
the 1st order correction to the $\alpha$-th band dispersion.   
Notice also that $\hat{\cal N}_{\alpha}$ enters 
into the {\it anti-commutator} with $\hat{A}_{\alpha}$. 
As will be shown later, this anti-commutator 
lets $\hat{\cal N}_{\alpha}$ play a 
relevant role 
in determining the 1st order gradient expansion 
correction to the spectral {\it weight}  
(see the section.~IV for details).  
\begin{widetext}

The 2nd order contributions in $\hat{F}^{(1)}_{\alpha}(\hat{A}_{\alpha})$  
and $\hat{F}^{(2)}_{\alpha}(\hat{A}_{\alpha\bar{\alpha}})$ and 
$\hat{F}^{(3)}_{\alpha}(\hat{A}_{\bar{\alpha}})$ are given 
as follows;   
\begin{eqnarray}
\hat{F}^{(1)}(\hat{A}_{\alpha})&=& \cdots -\ \frac{1}{8}\ \Big\{
\partial_{X_j}\partial_{X_k}  L^{(0)}_{d,\alpha} \hat{1}  
+ \hat{\cal A}^{\alpha\bar{\alpha}}_{X_j}\big(L^{(0)}_{d,\alpha}\hat{1} - 
\hat{L}_{d,\bar{\alpha}}\big)\hat{\cal A}^{\bar{\alpha}\alpha}_{X_k} 
+ \hat{\cal A}^{\alpha\bar{\alpha}}_{X_k}\big(L^{(0)}_{d,\alpha}\hat{1} - 
\hat{L}_{d,\bar{\alpha}}\big) 
\hat{\cal A}^{\bar{\alpha}\alpha}_{X_j} \Big\} \nonumber \\
&&\hspace{0.7cm} \cdot \Big\{
\big[D^{\alpha}_{Q_j},\big[D^{\alpha}_{Q_k},
\hat{A}_{\alpha}\big]\big] 
+ \hat{\cal A}^{\alpha\bar{\alpha}}_{Q_j}
\hat{\cal A}^{\bar{\alpha}\alpha}_{Q_k}\hat{A}_{\alpha} 
+ \hat{A}_{\alpha}\hat{\cal A}^{\alpha\bar{\alpha}}_{Q_k}
\hat{\cal A}^{\bar{\alpha}\alpha}_{Q_j} \Big\} \nonumber \\ 
&& -\ \frac{1}{8}\ \Big\{- 
{\cal A}^{\alpha\bar{\alpha}}_{X_k}
\big[D^{\bar{\alpha}}_{X_j},
L^{(0)}_{d,\alpha}\hat{1} - \hat{L}_{d,\bar{\alpha}}\big] 
 - {\cal A}^{\alpha\bar{\alpha}}_{X_j}
\big[D^{\bar{\alpha}}_{X_k},L^{(0)}_{d,\alpha}\hat{1} - 
\hat{L}_{d,\bar{\alpha}}\big]   
- \big[D^{\alpha\bar{\alpha}}_{X_j},
\hat{\cal A}^{\alpha\bar{\alpha}}_{X_k}\big]
\big(L^{(0)}_{d,\alpha}\hat{1} - \hat{L}_{d,\bar{\alpha}}\big)\Big\} \nonumber \\
&&\hspace{0.7cm} \cdot\Big\{\big[D^{\bar{\alpha}\alpha}_{Q_j},
\hat{\cal A}^{\bar{\alpha}\alpha}_{Q_k}\big]\hat{A}_{\alpha} 
+  \hat{\cal A}^{\bar{\alpha}\alpha}_{Q_k}  
\big[D^{\alpha}_{Q_j}, \hat{A}_{\alpha}\big] 
 + \hat{\cal A}^{\bar{\alpha}\alpha}_{Q_j}  
\big[D^{\alpha}_{Q_k},\hat{A}_{\alpha}\big]\Big\}  \nonumber \\
&& -\ \big\{X_j\leftrightarrow Q_j|X_k,Q_k\big\}  
-\ \big\{X_j, Q_j|X_k \leftrightarrow Q_k\big\} 
\ +\ \big\{X_j \leftrightarrow Q_j|X_k \leftrightarrow Q_k\big\} \ -\ {\rm h.c.}\ \  
+\  {\cal O}(\lambda^3).  \label{3-20} \\
\hat{F}^{(2)}_{\alpha}(\hat{A}_{\alpha\bar{\alpha}}) &=& 
\frac{i}{2}\big\{\partial_{X_j} L^{(0)}_{d,\alpha} 
\big(\hat{\cal A}^{\alpha\bar{\alpha}}_{Q_j}\hat{A}_{\bar{\alpha}\alpha} 
- \hat{A}_{\alpha\bar{\alpha}}
\hat{\cal A}^{\bar{\alpha}\alpha}_{Q_j}\big)  - \hat{\cal A}^{\alpha\bar{\alpha}}_{X_j}
\big(L^{(0)}_{d,\alpha}\hat{1} - \hat{L}_{d,\bar{\alpha}}\big)  
\big[D^{\bar{\alpha}\alpha}_{Q_j},\hat{A}_{\bar{\alpha}\alpha}\big]
\big\}  \nonumber \\
&& \ - \ \big\{X_j\leftrightarrow Q_j\big\} - {\rm h.c.} 
\ + \ {\cal O}(\lambda^2),   \label{3-22} \\
\hat{F}^{(3)}_{\alpha}(\hat{A}_{\bar{\alpha}}) &=& \frac{i}{2}
\hat{\cal A}^{\alpha\bar{\alpha}}_{Q_j}
\hat{A}_{\bar{\alpha}}\big(L^{(0)}_{d,\alpha}\hat{1} 
-\hat{L}_{d,\bar{\alpha}}\big)
\hat{\cal A}^{\bar{\alpha}\alpha}_{X_j}  
 - \big\{X_j\leftrightarrow Q_j\big\} - {\rm h.c.} \ + \ {\cal O}(\lambda^2). \label{3-23} 
\end{eqnarray}
where ``$\cdots$'' in $\hat{F}^{(1)}$ stands for those 
terms explicit in the right hand side of eq.~(\ref{3-19a}). 
To rewrite the latter two, i.e. $\hat{F}^{(2)}_{\alpha}$ and $\hat{F}^{(3)}_{\alpha}$, 
only in terms of $\hat{A}_{\alpha}$, we first solve $\hat{A}_{\alpha\bar{\alpha}}$ and 
$\hat{A}_{\bar{\alpha}}$ 
in favor of $\hat{A}_{\alpha}$;   
\begin{eqnarray}
\hat{A}_{\alpha\bar{\alpha}}(\hat{A}_{\alpha}) &=&
\frac{i}{2}\big\{ - \hat{A}_{\alpha}
\hat{\cal A}^{\alpha\bar{\alpha}}_{Q_k}
\big(\partial_{X_k}\ L^{(0)}_{d,\alpha}\hat{1} 
+ \big[D^{\bar{\alpha}}_{X_k},\hat{L}_{d,\bar{\alpha}}\big]\big) 
  \ +\ \big[D^{\alpha}_{Q_k},\hat{A}_{\alpha}\big]
\hat{\cal A}^{\alpha\bar{\alpha}}_{X_k}
\big(\hat{L}_{d,\bar{\alpha}} - L^{(0)}_{d,\alpha}\hat{1}
\big) \big\}\cdot(\hat{L}_{d,\bar{\alpha}}- L^{(0)}_{d,\alpha}\hat{1})^{-1} 
\nonumber \\
&&\  -\ \big\{X_k \leftrightarrow Q_k \big\} \ + \ {\cal O}(\lambda^2), 
\label{3-20a} \\
\hat{A}_{\eta\delta}(\hat{A}_{\alpha}) &=& 
\frac{i}{2}\big(\hat{L}_{d,\eta}\hat{\cal A}^{\eta\alpha}_{X_k}
\hat{A}_{\alpha}\hat{\cal A}^{\alpha\delta}_{Q_k} 
- \hat{\cal A}^{\eta\alpha}_{X_k}
\hat{A}_{\alpha}\hat{\cal A}^{\alpha\delta}_{Q_k} \hat{L}_{d,\delta}
\big)\cdot(\hat{L}_{d,\delta}-\hat{L}_{d,\eta})^{-1} 
\ - \ \big\{X_k \leftrightarrow Q_k \big\} \ +
\ {\cal O}(\lambda^2), \nonumber  \\
&=& -\frac{i}{2} \hat{\cal A}^{\eta\alpha}_{X_k}
\hat{A}_{\alpha}\hat{\cal A}^{\alpha\delta}_{Q_k} \ 
- \ \big\{X_k \leftrightarrow Q_k \big\} \ +
\ {\cal O}(\lambda^2), \label{3-21}
\end{eqnarray} 
with $\eta,\delta \ne \alpha$.
One can obtain these relations, by looking into the inter-band components of 
the dissipationless Keldysh equation, precisely as in eq.(\ref{3-5}).
We then substitute these two back into eqs.~(\ref{3-22},\ref{3-23}), only to obtain 
$\hat{F}^{(2)}_{\alpha}$ and $\hat{F}^{(3)}_{\alpha}$ {\it as a functional of 
$\hat{A}_{\alpha}$};  
\begin{eqnarray}
\hat{F}^{(2)}_{\alpha}(\hat{A}_{\alpha}) 
&=& \frac{1}{2}\big(\partial_{X_j} L^{(0)}_{d,\alpha}\big)
\hat{A}_{\alpha} \hat{\cal A}^{\alpha\bar{\alpha}}_{Q_k}
\big[\hat{D}^{\bar{\alpha}}_{X_k}, L^{(0)}_{d,\alpha}\hat{1} + 
\hat{L}_{d,\bar{\alpha}}\big]\big[L^{(0)}_{d,\alpha}\hat{1} - 
\hat{L}_{d,\bar{\alpha}}\big]^{-1}
\hat{\cal A}^{\bar{\alpha}\alpha}_{Q_j} 
+ \frac{1}{2}\big(\partial_{X_j} L^{(0)}_{d,\alpha}\big) 
\big[\hat{D}^{\alpha}_{Q_k},\hat{A}_{\alpha}\big]
\hat{\cal A}^{\alpha\bar{\alpha}}_{X_k}
\hat{\cal A}^{\bar{\alpha}\alpha}_{Q_j}  \nonumber \\
&& - \frac{1}{4}\big[\hat{D}_{Q_j}^{\alpha},\hat{A}_{\alpha}\big]
\hat{\cal A}^{\alpha\bar{\alpha}}_{Q_k}
\big[\hat{D}^{\bar{\alpha}}_{X_k}, L^{(0)}_{d,\alpha}\hat{1} 
+ \hat{L}_{d,\bar{\alpha}}\big]
\hat{\cal A}_{X_j}^{\bar{\alpha}\alpha} 
- \frac{1}{4}\hat{A}_{\alpha}
\big[\hat{D}^{\alpha\bar{\alpha}}_{Q_j},
\hat{\cal A}^{\alpha\bar{\alpha}}_{Q_k}\big]
\big[\hat{D}^{\bar{\alpha}}_{X_k}, L^{(0)}_{d,\alpha}\hat{1} + 
\hat{L}_{d,\bar{\alpha}}\big] 
\hat{\cal A}^{\bar{\alpha}\alpha}_{X_j} \nonumber \\
&& -  \frac{1}{4}\hat{A}_{\alpha}
\hat{\cal A}^{\alpha\bar{\alpha}}_{Q_k}
\Big[\hat{D}^{\bar{\alpha}}_{Q_j},
\big[\hat{D}^{\bar{\alpha}}_{X_k},L^{(0)}_{d,\alpha}\hat{1} 
+ \hat{L}_{d,\bar{\alpha}}\big]\Big]
 \hat{\cal A}^{\bar{\alpha}\alpha}_{X_j} 
- \frac{1}{4} 
\Big[\hat{D}^{\alpha}_{Q_j},
\big[\hat{D}^{\alpha}_{Q_k},\hat{A}_{\alpha}\big]\Big]
\hat{\cal A}^{\alpha\bar{\alpha}}_{X_k}
\big(L^{(0)}_{d,\alpha}\hat{1} - \hat{L}_{d,\bar{\alpha}}\big)
\hat{\cal A}^{\bar{\alpha}\alpha}_{X_j} \nonumber \\
&& - \frac{1}{4} 
\big[\hat{D}^{\alpha}_{Q_k},\hat{A}_{\alpha}
\big]\big[\hat{D}^{\alpha\bar{\alpha}}_{Q_j},
\hat{\cal A}^{\alpha\bar{\alpha}}_{X_k}\big]
\big(L^{(0)}_{d,\alpha}\hat{1} - \hat{L}_{d,\bar{\alpha}}\big)
\hat{\cal A}^{\bar{\alpha}\alpha}_{X_j} 
- \frac{1}{4}
\big[\hat{D}^{\alpha}_{Q_k},\hat{A}_{\alpha}\big]
\hat{\cal A}^{\alpha\bar{\alpha}}_{X_k}
\big[\hat{D}^{\bar{\alpha}}_{Q_j},L^{(0)}_{d,\alpha}\hat{1}
- \hat{L}_{d,\bar{\alpha}}\big]
\hat{\cal A}^{\bar{\alpha}\alpha}_{X_j}  \nonumber \\
&&- \frac{1}{4} \hat{A}_{\alpha}
\hat{\cal A}^{\alpha\bar{\alpha}}_{Q_k}
\big[\hat{D}^{\bar{\alpha}}_{X_k},\ L^{(0)}_{d,\alpha}\hat{1} 
+ \hat{L}_{d,\bar{\alpha}}\big]
\big[\hat{D}^{\bar{\alpha}}_{Q_j},\big(L^{(0)}_{d,\alpha}\hat{1} - \hat{L}_{d,\bar{\alpha}}\big)^{-1}\big]
\big(L^{(0)}_{d,\alpha}\hat{1}- \hat{L}_{d,\bar{\alpha}}\big)
\hat{\cal A}^{\bar{\alpha}\alpha}_{X_j} \nonumber \\
&& + \ \frac{1}{4}\big[\hat{D}^{\alpha}_{Q_k},
\hat{A}_{\alpha}\big]
\hat{\cal A}^{\alpha\bar{\alpha}}_{X_k}
\big(\hat{L}_{d,\bar{\alpha}} - L^{(0)}_{d,\alpha}\hat{1}\big)
\big[\hat{D}^{\bar{\alpha}}_{Q_j},\big(L^{(0)}_{d,\alpha}\hat{1} - \hat{L}_{d,\bar{\alpha}}\big)^{-1}\big]
\big(L^{(0)}_{d,\alpha}\hat{1}- \hat{L}_{d,\bar{\alpha}}\big)
\hat{\cal A}^{\bar{\alpha}\alpha}_{X_j}. \nonumber  \\
&& - \big\{X_{j}\leftrightarrow Q_j|X_k, Q_k\big\} 
- \big\{X_{j}, Q_j|X_k \leftrightarrow Q_k\big\} 
+ \big\{X_{j}\leftrightarrow Q_j|X_k \leftrightarrow Q_k\big\} 
- {\rm h.c.}\ +\ {\cal O}(\lambda^3), \label{3-24} 
\end{eqnarray}
\end{widetext}
and 
\begin{eqnarray} 
\hat{F}^{(3)}_{\alpha}(\hat{A}_{\alpha}) &=& \frac{1}{4}
\hat{\cal A}^{\alpha\bar{\alpha}}_{Q_j}
\hat{\cal A}^{\bar{\alpha}\alpha}_{X_k}\hat{A}_{\alpha}
\hat{\cal A}^{\alpha\bar{\alpha}}_{Q_k} 
\big(L^{(0)}_{d,\alpha}\hat{1} 
-\hat{L}_{d,\bar{\alpha}}\big)
\hat{\cal A}^{\bar{\alpha}\alpha}_{X_j}  \nonumber \\
&& \hspace{-1.4cm} -\ \big\{X_j\leftrightarrow Q_j|X_k,Q_k\big\} 
- \big\{X_j, Q_j|X_k \leftrightarrow Q_k\big\} \nonumber \\
&& \hspace{-1.4cm} +\  
\big\{X_j\leftrightarrow Q_j|X_k \leftrightarrow Q_k\big\} 
- {\rm h.c.} \ +\ {\cal O}(\lambda^3). \label{3-25}
\end{eqnarray}
Eqs.~(\ref{3-20}), (\ref{3-24}) and (\ref{3-25}) in combination with 
eq.~(\ref{3-19a}) are all the terms that enter 
into the right hand side of eq.~(\ref{3-14b}) up to 
${\cal O}(\lambda^2_1,\lambda_1\lambda_2,\lambda^2_2)$.  
. 

Now that we have obtained an explicit form of eq.~(\ref{3-14b}), 
we shall next find out a $2 \times 2$ matrix $\hat{A}_{\alpha}$ 
that satisfies this equation. 
Before doing this, however, 
it would be clearly helpful 
to simplify 
these more than 100 terms.   
In fact, we can further transform all these terms 
precisely into eqs.~(\ref{3-27}-\ref{3-29}), 
whose transparent expression helps   
us to find $\hat{A}_{\alpha}$ 
in sec. IV.    
To be more specific, we can find, for  
any single term in 
eqs.~(\ref{3-20},\ref{3-24},\ref{3-25}), 
several counterpart terms with which it constitutes 
a certain $SU(2)$ gauge covariant quantity  
enumerated in eq.~(\ref{3-27}-\ref{3-29}). 

\paragraph*{\bf Terms proportional to $[\hat{D}^{\alpha}_{Q},\hat{A}_{\alpha}]$}
To see this explicitly, focus first on those terms 
in eqs.~(\ref{3-20},\ref{3-24},\ref{3-25})  
that are linear in  
$[\hat{D}^{\alpha},\hat{A}_{\alpha}]$; 
\begin{eqnarray}
&&\hspace{-0.5cm}\frac{1}{8}\Big\{ 
{\cal A}^{\alpha\bar{\alpha}}_{X_k}
\big[D^{\bar{\alpha}}_{X_j},
L^{(0)}_{d,\alpha}\hat{1} - \hat{L}_{d,\bar{\alpha}}\big] + 
{\cal A}^{\alpha\bar{\alpha}}_{X_j}
\big[D^{\bar{\alpha}}_{X_k},
L^{(0)}_{d,\alpha}\hat{1} - \hat{L}_{d,\bar{\alpha}}\big]
\nonumber \\
&&\hspace{0.5cm} 
+\ \big[D^{\alpha\bar{\alpha}}_{X_j},
\hat{\cal A}^{\alpha\bar{\alpha}}_{X_k}\big]
\big(L^{(0)}_{d,\alpha}\hat{1} - \hat{L}_{d,\bar{\alpha}}\big)\Big\} \nonumber \\
&&\hspace{1.0cm} \times\Big\{
\hat{\cal A}^{\bar{\alpha}\alpha}_{Q_k}  
\big[D^{\alpha}_{Q_j}, \hat{A}_{\alpha}\big] + 
\hat{\cal A}^{\bar{\alpha}\alpha}_{Q_j}  
\big[D^{\alpha}_{Q_k}, \hat{A}_{\alpha}\big] 
\Big\} \nonumber \\
&&+ \frac{1}{2}\big(\partial_{X_j} L^{(0)}_{d,\alpha}\big) 
\big[\hat{D}^{\alpha}_{Q_k},\hat{A}_{\alpha}\big]
\hat{\cal A}^{\alpha\bar{\alpha}}_{X_k}
\hat{\cal A}^{\bar{\alpha}\alpha}_{Q_j}  \nonumber \\
&& - \frac{1}{4}\big[\hat{D}_{Q_j}^{\alpha},\hat{A}_{\alpha}\big]
\hat{\cal A}^{\alpha\bar{\alpha}}_{Q_k}
\big[\hat{D}^{\bar{\alpha}}_{X_k}, L^{(0)}_{d,\alpha}\hat{1} 
+ \hat{L}_{d,\bar{\alpha}}\big]
\hat{\cal A}_{X_j}^{\bar{\alpha}\alpha}  
\nonumber \\ 
&& - \frac{1}{4} 
\big[\hat{D}^{\alpha}_{Q_k},\hat{A}_{\alpha}
\big]\big[\hat{D}^{\alpha\bar{\alpha}}_{Q_j},
\hat{\cal A}^{\alpha\bar{\alpha}}_{X_k}\big]
\big(L^{(0)}_{d,\alpha}\hat{1} - \hat{L}_{d,\bar{\alpha}}\big)
\hat{\cal A}^{\bar{\alpha}\alpha}_{X_j}  \nonumber \\
&&\ + \ \cdots. \label{3-25-1}
\end{eqnarray}
where we have already used 
$[\hat{D}^{\bar{\alpha}}_{Q},(L^{(0)}_{d,\alpha}\hat{1}-\hat{L}_{d,\bar{\alpha}})^{-1}]
\cdot (L^{(0)}_{d,\alpha}\hat{1}-\hat{L}_{d,\bar{\alpha}}) 
= - (L^{(0)}_{d,\alpha}\hat{1}-\hat{L}_{d,\bar{\alpha}})^{-1}
\cdot[\hat{D}^{\bar{\alpha}}_{Q},(L^{(0)}_{d,\alpha}\hat{1}-\hat{L}_{d,\bar{\alpha}})]$.
Note also that 
``$\cdots$'' above indicates those 3 kinds of 
{\it counterpart terms with $X$ and $Q$ exchanged} 
and their {\it hermite conjugate terms};
\begin{eqnarray}
{\rm ``}\cdots {\rm "} 
&\equiv& -\{X_j\leftrightarrow Q_j|X_k,Q_k\} -\{X_j,Q_j|X_k\leftrightarrow Q_k\} 
\nonumber \\ 
&& +  \{X_j\leftrightarrow Q_j|X_k\leftrightarrow Q_k\} - {\rm h.c.} \label{3-25-1-a}
\end{eqnarray}
When taking these terms implicit into account, 
we may rewrite the 1st term in Eq.(\ref{3-25-1}) 
into the following;
\begin{eqnarray} 
&&\hspace{-0.5cm}\frac{1}{8}\Big\{ 
{\cal A}^{\alpha\bar{\alpha}}_{X_k}
\big[\hat{D}^{\bar{\alpha}}_{X_j},
L^{(0)}_{d,\alpha}\hat{1} - \hat{L}_{d,\bar{\alpha}}\big] + 
{\cal A}^{\alpha\bar{\alpha}}_{X_j}
\big[\hat{D}^{\bar{\alpha}}_{X_k},
L^{(0)}_{d,\alpha}\hat{1} - \hat{L}_{d,\bar{\alpha}}\big]
\nonumber \\
&&\hspace{0.25cm} 
+\ \big[\hat{D}^{\alpha\bar{\alpha}}_{X_j},
\hat{\cal A}^{\alpha\bar{\alpha}}_{X_k}\big]
\big(L^{(0)}_{d,\alpha}\hat{1} - \hat{L}_{d,\bar{\alpha}}\big)\Big\} \nonumber \\
&&\hspace{0.5cm} \times\Big\{
\hat{\cal A}^{\bar{\alpha}\alpha}_{Q_k}  
\big[D^{\alpha}_{Q_j}, \hat{A}_{\alpha}\big] + 
\hat{\cal A}^{\bar{\alpha}\alpha}_{Q_j}  
\big[D^{\alpha}_{Q_k}, \hat{A}_{\alpha}\big] \Big\} + \cdots \nonumber \\ 
&&\hspace{-0.25cm} = -\frac{1}{4} \big[\hat{D}^{\alpha}_{Q_j},\hat{A}_{\alpha}\big]
\hat{\cal A}^{\alpha\bar{\alpha}}_{Q_k} 
\big[\hat{D}^{\bar{\alpha}}_{X_k},L^{(0)}_{d,\alpha}\hat{1}- 
\hat{L}_{d,\bar{\alpha}}\big]{\cal A}^{\bar{\alpha}\alpha}_{X_j}
\nonumber \\ 
&& \hspace{-0.25cm} + \frac{1}{4} {\cal A}^{\alpha\bar{\alpha}}_{X_k} 
\big[\hat{D}^{\bar{\alpha}}_{X_j},L^{(0)}_{d,\alpha}\hat{1}- 
\hat{L}_{d,\bar{\alpha}}\big]\hat{\cal A}^{\bar{\alpha}\alpha}_{Q_k} 
\big[\hat{D}^{\alpha}_{Q_j},\hat{A}_{\alpha}\big] \nonumber \\ 
&& \hspace{-0.25cm} + \frac{1}{4}  
\big[\hat{D}^{\alpha\bar{\alpha}}_{X_j},\hat{\cal A}^{\alpha\bar{\alpha}}_{X_k}\big]
\big(L^{(0)}_{d,\alpha}\hat{1}-\hat{L}_{d,\bar{\alpha}}\big)
\hat{\cal A}^{\bar{\alpha}\alpha}_{Q_k}
\big[\hat{D}^{\alpha}_{Q_j},\hat{A}_{\alpha}\big] + \cdots. \nonumber
\end{eqnarray} 
Namely, we replaced  
several terms in the left hand side by either 
their counterparts or 
their hermitian conjugates implicit in ``$\cdots$''. 
We also used eq.~(\ref{3-9}), only to obtain the 
3rd term in the right hand side.  
Since we can regard that {\it these terms were just 
swapped among their 3 other copies and their hermtian 
conjugates}, we can begin with the following, instead of 
eq.~(\ref{3-25-1});  
\begin{eqnarray}
&& {\rm Eq.}\ (\ref{3-25-1}) = \nonumber \\
&&\frac{1}{2}\big(\partial_{X_j} L^{(0)}_{d,\alpha}\big) 
\big[D^{\alpha}_{Q_k}, \hat{A}_{\alpha}\big]
\big(\hat{\cal A}^{\alpha\bar{\alpha}}_{X_k}
\hat{\cal A}^{\bar{\alpha}\alpha}_{Q_j} 
- \hat{\cal A}^{\alpha\bar{\alpha}}_{Q_j}
\hat{\cal A}^{\bar{\alpha}\alpha}_{X_k}\big) \nonumber \\
&&\  +\ \frac{1}{4}\hat{\cal A}^{\alpha\bar{\alpha}}_{X_j}
\big(L^{(0)}_{d,\alpha}\hat{1} - \hat{L}_{d,\bar{\alpha}}\big)
\big[D^{\bar{\alpha}\alpha}_{Q_j},
\hat{\cal A}^{\bar{\alpha}\alpha}_{X_k}\big] 
\big[D^{\alpha}_{Q_k},\hat{A}_{\alpha}\big] \nonumber \\ 
&&\ +\ \frac{1}{4}\hat{\cal A}^{\alpha\bar{\alpha}}_{X_k}
\big[D^{\bar{\alpha}}_{X_j},L_{d,\alpha}\hat{1}
-\hat{L}_{d,\bar{\alpha}}\big]  
\hat{\cal A}^{\bar{\alpha}\alpha}_{Q_k}
\big[D^{\alpha}_{Q_j},\hat{A}_{\alpha}\big] \nonumber \\
&&\ +\ \frac{1}{4}
\big[D^{\alpha\bar{\alpha}}_{X_j},
\hat{\cal A}^{\alpha\bar{\alpha}}_{X_k}\big]
\big(L^{(0)}_{d,\alpha}\hat{1} - \hat{L}_{d,\bar{\alpha}}\big)
\hat{\cal A}^{\bar{\alpha}\alpha}_{Q_k}
\big[D^{\alpha}_{Q_j},\hat{A}_{\alpha}\big]   \nonumber \\
&&\ + \ \ \cdots, \nonumber 
\end{eqnarray}
where ``$\cdots$'' defined in eq.~(\ref{3-25-1-a}).

Compare the 2nd, 3rd and 4-th terms above with 
the decomposition rule for derivatives, i.e. eq.~(\ref{3-10}).  
Namely, we can unify these 3 into a single $\alpha$-th band 
covariant derivative term;  
\begin{eqnarray}
&&\hspace{-1cm}{\rm Eq.}\ (\ref{3-25-1})  
= - \frac{i}{2}\big(\partial_{X_j} L^{(0)}_{d,\alpha}\big)
\big[D^{\alpha}_{Q_k}, \hat{A}_{\alpha}\big]
\hat{\Omega}^{\alpha}_{Q_jX_k} \nonumber \\
&&\hspace{-1.0cm} + \frac{1}{4}\big[\hat{D}^{\alpha}_{X_j},
\hat{\cal A}^{\alpha\bar{\alpha}}_{X_k} 
\big(L^{(0)}_{d,\alpha}\hat{1} - 
\hat{L}_{d,\bar{\alpha}}\big)\hat{\cal A}^{\bar{\alpha}\alpha}_{Q_k}
\big]\big[D^{\alpha}_{Q_j}, \hat{A}_{\alpha}\big] + \cdots. \label{3-26} 
\end{eqnarray}
When combined explicitly with its counterpart with $X_k$ and $Q_k$ exchanged, 
the 2nd term above is expressed by $\hat{\cal M}_{\alpha}$ 
(see eq.~(\ref{3-17a})).  
Accordingly, 
eq.~(\ref{3-25-1}) can be rigorously transformed into the following 
$SU(2)$ covariant quantities,   
\begin{eqnarray}
{\rm Eq.}\ (\ref{3-25-1}) &=& -\frac{i}{2}\big(\partial_{X_{j}}L^{(0)}_{d,\alpha}\big)
\Big[\hat{\Omega}^{\alpha}_{Q_jX_k},
\big[\hat{D}^{\alpha}_{Q_k},\hat{A}_{\alpha}\big]\Big]_{+} \nonumber \\
&& \hspace{-2cm} -\  \big\{X_{j} \leftrightarrow Q_j | X_k,Q_k\big\} 
 - \  \big\{X_{j},Q_j | X_k \leftrightarrow Q_k\big\} \nonumber \\
&& \hspace{-2cm} + \ \big\{X_{j} \leftrightarrow Q_j | X_k \leftrightarrow Q_k\big\} 
\nonumber \\
&&\hspace{-2cm} -\ \frac{i}{2}\Big[\big[\hat{D}^{\alpha}_{X_j},\hat{\cal M}_{\alpha}\big],
\big[\hat{D}^{\alpha}_{Q_j},\hat{A}_{\alpha}\big]\Big]_{+}  
- \big\{Q_j \leftrightarrow X_j\big\}. \label{3-26-1}
\end{eqnarray}
\paragraph*{\bf Terms proportional to $\hat{A}_{\alpha}$} 
Employing a similar manipulation, we can further simplify all 
the remaining terms, which are linear in 
either (a) $\hat{A}_{\alpha}$ itself or (b)  
$\big[D^{\alpha}_{Q_j},\big[D^{\alpha}_{Q_k},\hat{A}_{\alpha}\big]\big]$.      
As will be shown next, 
the latter one can be easily proved to be zero in total up to the order 
of ${\cal O}(\lambda^2)$. 
Thus, we will henceforth look into those terms in 
eqs.~(\ref{3-20},\ref{3-24},\ref{3-25}), which 
are proportional to $\hat{A}_{\alpha}$;   
\begin{eqnarray}
&&\hspace{-0.5cm} -\ \frac{1}{8}\Big\{\hat{\cal A}^{\alpha\bar{\alpha}}_{X_j}
\big(L^{(0)}_{d,\alpha}\hat{1} - \hat{L}_{d,\bar{\alpha}}\big)
\hat{\cal A}^{\bar{\alpha}\alpha}_{X_k} +   
\big(X_j \leftrightarrow X_k\big)\Big\} \nonumber \\
&&\hspace{0.5cm} \times \Big\{ \hat{\cal A}^{\alpha\bar{\alpha}}_{Q_j}
\hat{\cal A}^{\bar{\alpha}\alpha}_{Q_k}\hat{A}_{\alpha} 
+ \hat{A}_{\alpha}\hat{\cal A}^{\alpha\bar{\alpha}}_{Q_k}
\hat{\cal A}^{\bar{\alpha}\alpha}_{Q_j}\Big\} \nonumber \\
&&\hspace{-0.5cm} +\ \frac{1}{8}\Big\{ 
{\cal A}^{\alpha\bar{\alpha}}_{X_k}
\big[D^{\bar{\alpha}}_{X_j},
L^{(0)}_{d,\alpha}\hat{1} - \hat{L}_{d,\bar{\alpha}}\big] + 
\big(X_j\leftrightarrow X_k\big)\Big\} \nonumber \\
&&\hspace{0.5cm} \times \big[D^{\bar{\alpha}\alpha}_{Q_j},
\hat{\cal A}^{\bar{\alpha}\alpha}_{Q_k}\big]\hat{A}_{\alpha} \nonumber \\
&&\hspace{-0.5cm} -\ \frac{1}{4}\hat{A}_{\alpha}
\big[\hat{D}^{\alpha\bar{\alpha}}_{Q_j},
\hat{\cal A}^{\alpha\bar{\alpha}}_{Q_k}\big]
\big[\hat{D}^{\bar{\alpha}}_{X_k}, L^{(0)}_{d,\alpha}\hat{1} + 
\hat{L}_{d,\bar{\alpha}}\big] 
\hat{\cal A}^{\bar{\alpha}\alpha}_{X_j} \nonumber \\
&&\hspace{-0.5cm} -\ \frac{1}{4}\hat{A}_{\alpha}
\hat{\cal A}^{\alpha\bar{\alpha}}_{Q_k}
\Big[\hat{D}^{\bar{\alpha}}_{Q_j},
\big[\hat{D}^{\bar{\alpha}}_{X_k},L^{(0)}_{d,\alpha}\hat{1} 
+ \hat{L}_{d,\bar{\alpha}}\big]\Big]
 \hat{\cal A}^{\bar{\alpha}\alpha}_{X_j}  \nonumber \\ 
&&\hspace{-0.5cm} +\ \frac{1}{4} 
\hat{\cal A}^{\alpha\bar{\alpha}}_{Q_j}
\hat{\cal A}^{\bar{\alpha}\alpha}_{X_k}\hat{A}_{\alpha}
\hat{\cal A}^{\alpha\bar{\alpha}}_{Q_k} 
\big(L^{(0)}_{d,\alpha}\hat{1} 
-\hat{L}_{d,\bar{\alpha}}\big)
\hat{\cal A}^{\bar{\alpha}\alpha}_{X_j} + \cdots, \label{3-27-1}
\end{eqnarray}
where ``$\cdots$'' defined in eq.~(\ref{3-25-1-a}). 
We have already neglected those 
which are canceled either by their counterparts 
or by their hermite conjugates.   

Notice first that, to 2nd order in $\lambda$, we can 
regard that {\it $\hat{A}_{\alpha}$ in eq.~(\ref{3-27-1}) 
commutes with other $2 \times 2$ matrices}. 
This is because $\hat{A}_{\alpha}$ reduces to a unit matrix 
at equilibrium, while eq.~(\ref{3-27-1}) is already {\it at 
least 2nd order} in $\lambda_1$. Thus, commutators between 
$\hat{A}_{\alpha}$ and other $2$ by $2$ matrices 
inevitably end up with the 3rd order contributions 
in $\lambda$, e.g.  
\begin{eqnarray}
&&\hspace{-0.4cm}\big(\partial_{X_k}L^{(0)}_{d,\alpha}\big)
\hat{A}_{\alpha}\big[\hat{D}^{\alpha\bar{\alpha}}_{Q_k},
\hat{\cal A}^{\alpha\bar{\alpha}}_{X_j}\big]
\hat{\cal A}^{\bar{\alpha}\alpha}_{Q_j} \nonumber \\
&&\hspace{-0.2cm}= \big(\partial_{X_k}L^{(0)}_{d,\alpha}\big)
\big[\hat{D}^{\alpha\bar{\alpha}}_{Q_k},
\hat{\cal A}^{\alpha\bar{\alpha}}_{X_j}\big]
\hat{\cal A}^{\bar{\alpha}\alpha}_{Q_j}\hat{A}_{\alpha} 
+ {\cal O}(\lambda^2_1\lambda_2,\lambda^3_1). \label{3-27-1-a}
\end{eqnarray}  

Observing this, one can easily unify      
the 2nd term and (the hermitian conjugate of) 3rd term 
in eq.~(\ref{3-27-1}), so that they are given solely in 
terms of the Berry's curvature for the $\alpha$-th band,      
\begin{eqnarray}
&&\hspace{-0.5cm}  \frac{1}{8}\big\{ 
{\cal A}^{\alpha\bar{\alpha}}_{X_k}
\big[D^{\bar{\alpha}}_{X_j},
L^{(0)}_{d,\alpha}\hat{1} - \hat{L}_{d,\bar{\alpha}}\big] + 
\big(X_j\leftrightarrow X_k\big)\big\} 
\big[D^{\bar{\alpha}\alpha}_{Q_j},
\hat{\cal A}^{\bar{\alpha}\alpha}_{Q_k}\big]\hat{A}_{\alpha} \nonumber \\ 
&&\hspace{-0.3cm} +\ \frac{1}{4}\hat{\cal A}^{\alpha\bar{\alpha}}_{X_j}
\big[\hat{D}^{\bar{\alpha}}_{X_k}, L^{(0)}_{d,\alpha}\hat{1} + 
\hat{L}_{d,\bar{\alpha}}\big]\big[\hat{D}^{\bar{\alpha}\alpha}_{Q_j},
\hat{\cal A}^{\bar{\alpha}\alpha}_{Q_k}\big]
\hat{A}_{\alpha} \nonumber \\  
&& \hspace{0.5cm} = \frac{1}{2}\big(\partial_{X_k}L^{(0)}_{d,\alpha}\big)
\hat{\cal A}^{\alpha\bar{\alpha}}_{X_j} 
\big[\hat{D}^{\bar{\alpha}\alpha}_{Q_k},\hat{A}^{\bar{\alpha}\alpha}_{Q_j}\big]
\hat{A}_{\alpha}. \nonumber 
\end{eqnarray}
Namely, when combined with its counterpart with $X_j$ and $Q_j$ exchanged  
and their hermitian conjugate, ${\cal O}(\lambda^2)$-contribution of 
the right hand side above can be expressed only by $\hat{\Omega}^{\alpha}_{X_jQ_j}$; 
\begin{eqnarray}
&&\hspace{-0.5cm}\frac{1}{2}\big(\partial_{X_k}L^{(0)}_{d,\alpha}\big)\big\{
\hat{\cal A}^{\alpha\bar{\alpha}}_{X_j} 
\big[\hat{D}^{\bar{\alpha}\alpha}_{Q_k},\hat{A}^{\bar{\alpha}\alpha}_{Q_j}\big]
\hat{A}_{\alpha} +  \hat{A}_{\alpha}
\big[\hat{D}^{\alpha\bar{\alpha}}_{Q_k},\hat{A}^{\alpha\bar{\alpha}}_{X_j}\big]
\hat{\cal A}^{\bar{\alpha}\alpha}_{Q_j}\big\} \nonumber \\
&&\hspace{-0.3cm} - \ \big\{X_j \leftrightarrow Q_j|X_k,Q_k\big\} \nonumber \\
&&\hspace{-0.3cm} = 
\frac{i}{2}\big(\partial_{X_k}L^{(0)}_{d,\alpha}\big)
\big[\hat{D}^{\alpha}_{Q_k},\hat{\Omega}^{\alpha}_{X_jQ_j}\big]\hat{A}_{\alpha} + 
{\cal O}(\lambda^3), \label{3-27-2}
\end{eqnarray}
where we used eqs.~(\ref{3-27-1-a},\ref{3-10},\ref{3-11}).  

The 1st, 4-th and 5-th terms in eq.~(\ref{3-27-1}) are described   
in terms of $\hat{\cal M}_{\alpha}$ and $\hat{\Omega}_{X_jQ_j}$ 
in total. To see this, let us begin with the 4th term;
\begin{eqnarray}
&&\hspace{-0.5cm} 
-\ \frac{1}{4}\hat{A}_{\alpha}
\hat{\cal A}^{\alpha\bar{\alpha}}_{Q_k}
\Big[\hat{D}^{\bar{\alpha}}_{Q_j},
\big[\hat{D}^{\bar{\alpha}}_{X_k},L^{(0)}_{d,\alpha}\hat{1} 
+ \hat{L}_{d,\bar{\alpha}}\big]\Big]
 \hat{\cal A}^{\bar{\alpha}\alpha}_{X_j} + \cdots \nonumber \\
&&\hspace{-0.1cm} =\ -\frac{1}{8}\hat{A}_{\alpha}
\hat{\cal A}^{\alpha\bar{\alpha}}_{Q_k}
\big[\hat{D}^{\bar{\alpha}}_{Q_j},
\big[\hat{D}^{\bar{\alpha}}_{X_k},L^{(0)}_{d,\alpha}\hat{1} 
+ \hat{L}_{d,\bar{\alpha}}\big]\big]
\hat{\cal A}^{\bar{\alpha}\alpha}_{X_j} \nonumber \\
&&\hspace{0.2cm} +  \frac{1}{8}
\hat{\cal A}^{\alpha\bar{\alpha}}_{Q_j}
\big[\hat{D}^{\bar{\alpha}}_{X_j},
\big[\hat{D}^{\bar{\alpha}}_{Q_k},L^{(0)}_{d,\alpha}\hat{1} 
+ \hat{L}_{d,\bar{\alpha}}\big]\big]
\hat{\cal A}^{\bar{\alpha}\alpha}_{X_k}
\hat{A}_{\alpha}  + \cdots \nonumber \\
&&\hspace{-0.1cm}  =\ -\frac{1}{8}\hat{A}_{\alpha}
 \hat{\cal A}^{\alpha\bar{\alpha}}_{Q_k}
 \Big(\big[\hat{D}^{\bar{\alpha}}_{Q_j},
 \big[\hat{D}^{\bar{\alpha}}_{X_k},L^{(0)}_{d,\alpha}\hat{1}+
 \hat{L}_{d,\bar{\alpha}}\big]\big] \nonumber \\
&& \hspace{0.2cm}  - \big[\hat{D}^{\bar{\alpha}}_{X_k},
 \big[\hat{D}^{\bar{\alpha}}_{Q_j},L^{(0)}_{d,\alpha}\hat{1}+
 \hat{L}_{d,\bar{\alpha}}\big]\big]\Big)\hat{\cal A}^{\bar{\alpha}\alpha}_{X_j}  
 + {\cal O}(\lambda^3) + \cdots  
 \nonumber \\
&&\hspace{-0.1cm} = \frac{i}{8}\hat{A}_{\alpha}\hat{\cal A}^{\alpha\bar{\alpha}}_{Q_k}
\big[\hat{\Omega}^{\bar{\alpha}}_{X_kQ_j},L^{(0)}_{d,\alpha}\hat{1}
-\hat{L}_{d,\bar{\alpha}}\big]_{-}
\hat{\cal A}^{\bar{\alpha}\alpha}_{X_j} + \cdots \nonumber \\
&&\hspace{-0.1cm} = \frac{1}{8}\hat{A}_{\alpha}
\big[\hat{\cal A}^{\alpha\bar{\alpha}}_{Q_k}
\hat{\cal A}^{\bar{\alpha}\alpha}_{X_k},
\hat{\cal A}^{\alpha\bar{\alpha}}_{Q_j}\big(L^{(0)}_{d,\alpha}\hat{1}-
\hat{L}_{d,\bar{\alpha}}\big)\hat{\cal A}^{\bar{\alpha}\alpha}_{X_j}\big]_{-} \nonumber \\
&& \hspace{-0.1cm} - \frac{1}{8}\hat{A}_{\alpha}
\big[\hat{\cal A}^{\alpha\bar{\alpha}}_{Q_k}
\hat{\cal A}^{\bar{\alpha}\alpha}_{Q_j},
\hat{\cal A}^{\alpha\bar{\alpha}}_{X_k}\big(L^{(0)}_{d,\alpha}\hat{1}-
\hat{L}_{d,\bar{\alpha}}\big)\hat{\cal A}^{\bar{\alpha}\alpha}_{X_j}\big]_{-} + \cdots, 
\label{3-27-2-b}
\end{eqnarray}
where ``$\cdots$'' stands for 3 other counterparts and their hermitian conjugates. 
In the 1st equality, we have swapped $1/2$ of the 1st term in the left hand side 
for its counterpart 
in (the hermitian conjugates of) 
$\{X_{j}\leftrightarrow Q_j|X_k\leftrightarrow Q_k\}$.   
In the 2nd equality, we have ignored 
${\cal O}(\lambda^2_1\lambda_2,\lambda^3_1)$-  
contribution associated with the commutators 
between $\hat{A}_{\alpha}$ and other 
$2\times 2$ matrices. We further used 
eqs.~(\ref{3-6},\ref{3-12}),   
only to reach the final expression. 

Within ${\cal O}(\lambda^2)$, the 2nd term in eq.~(\ref{3-27-2-b}) is 
canceled by the 1st and 5th term in eq.~(\ref{3-27-1});  
\begin{eqnarray}
&&\hspace{-0.1cm} ({\rm 1st}) + ({\rm 4th}) + ({\rm 5th})\ {\rm in}\  {\rm eq.~(\ref{3-27-1})} \nonumber \\ 
&&\hspace{0.1cm} \ = \frac{1}{8}\hat{A}_{\alpha}\big[
\hat{\cal A}^{\alpha\bar{\alpha}}_{Q_k}
\hat{\cal A}^{\bar{\alpha}\alpha}_{X_k},
\hat{\cal A}^{\alpha\bar{\alpha}}_{Q_j}
\big(L^{(0)}_{d,\alpha}\hat{1}-
\hat{L}_{d,\bar{\alpha}}\big)
\hat{\cal A}^{\bar{\alpha}\alpha}_{X_j}\big]_{-}  \nonumber \\
&&\hspace{0.7cm} + {\cal O}(\lambda^3) + \cdots. \label{3-27-3}
\end{eqnarray}
Then notice that, when combined with its 3 other counterparts implicit in ``$\cdots$'',    
the coefficient of $\hat{A}_{\alpha}$ above clearly reduces 
to the commutator between $\hat{\Omega}^{\alpha}_{Q_kX_k}$  
and $\hat{\cal M}_{\alpha}$
(see eqs.~(\ref{3-11},\ref{3-17a})). 
In combination with eq.~(\ref{3-27-2}), this
dictates that ${\cal O}(\lambda^2)$-contributions 
in eq.~(\ref{3-27-1}) are indeed given only by 
$\hat{\cal M}_{\alpha}$ and $\hat{\Omega}^{\alpha}_{X_jQ_j}$;  
\begin{eqnarray}
&&\hspace{-0.4cm}{\rm eq.}\ {\rm (\ref{3-27-1})} = \frac{i}{4}
\big(\partial_{X_k}L^{(0)}_{d,\alpha}\big)\big[
\big[\hat{D}^{\alpha}_{Q_k},\hat{\Omega}^{\alpha}_{X_jQ_j}\big]
,\hat{A}_{\alpha}\big]_{+} \nonumber \\
&&\hspace{0.1cm}\ - \ \{X_k\leftrightarrow Q_k\} \ 
- \frac{1}{4}
\big[\big[\hat{\cal M}_{\alpha},\hat{\Omega}^{\alpha}_{X_jQ_j}\big]_{-}
,\hat{A}_{\alpha}\big]_{+} + {\cal O}(\lambda^3). \nonumber \\
\label{3-27-4} 
\end{eqnarray}  
\paragraph*{\bf Terms proportional to 
$[\hat{D}^{\alpha}_{Q},[\hat{D}^{\alpha}_{Q'},\hat{A}_{\alpha}]]$} 
Those terms in eqs.~(\ref{3-20},\ref{3-24},\ref{3-25}) 
which are linear in the 2nd covariant derivative  
vanish up to ${\cal O}(\lambda^2)$.  One can easily see this, 
by noting that any commutator between $\hat{A}_{\alpha}$ and other matrices 
in eqs.~(\ref{3-20},\ref{3-24},\ref{3-25}) ends up with 
${\cal O}(\lambda^2_1\lambda_2,\lambda^3_1)$;
\begin{eqnarray}
&&\hspace{-0.4cm} -\ \frac{1}{8}\ \Big\{
\partial_{X_j}\partial_{X_k}  L^{(0)}_{d,\alpha} \hat{1}  
+ \hat{\cal A}^{\alpha\bar{\alpha}}_{X_j}\big(L^{(0)}_{d,\alpha}\hat{1} - 
\hat{L}_{d,\bar{\alpha}}\big)\hat{\cal A}^{\bar{\alpha}\alpha}_{X_k}  \nonumber \\
&& \hspace{0.2cm} + \ 
\hat{\cal A}^{\alpha\bar{\alpha}}_{X_k}
\big(L^{(0)}_{d,\alpha}\hat{1} - \hat{L}_{d,\bar{\alpha}}\big)  
\hat{\cal A}^{\bar{\alpha}\alpha}_{X_j} \Big\}
\cdot \big[D^{\alpha}_{Q_j},\big[D^{\alpha}_{Q_k},
\hat{A}_{\alpha}\big]\big]  \nonumber \\
&&\hspace{0.3cm} - \frac{1}{4} 
\big[\hat{D}^{\alpha}_{Q_j},
\big[\hat{D}^{\alpha}_{Q_k},\hat{A}_{\alpha}\big]\big]
\hat{\cal A}^{\alpha\bar{\alpha}}_{X_k}
\big(L^{(0)}_{d,\alpha}\hat{1} - \hat{L}_{d,\bar{\alpha}}\big)
\hat{\cal A}^{\bar{\alpha}\alpha}_{X_j} + \cdots \nonumber \\
&&\hspace{-0.1cm}= 
- \frac{1}{8}\big\{\hat{\cal A}^{\alpha\bar{\alpha}}_{X_j} 
\big(L^{(0)}_{d,\alpha}\hat{1} - 
\hat{L}_{d,\bar{\alpha}}\big)
\hat{\cal A}^{\bar{\alpha}\alpha}_{X_k} 
 - \big\{X_k \leftrightarrow X_j\big\} \big\}\nonumber \\
&&\hspace{0.5cm}\times 
\partial_{Q_j}\partial_{Q_k}\big(\hat{A}_{\alpha}\big) +  
{\cal O}(\lambda^3) + \cdots  = {\cal O}(\lambda^3), \nonumber 
\end{eqnarray}
where ``$\cdots$'' already defined in eq.~(\ref{3-25-1-a}). 

\subsubsection{$SU(2)$ RKE up to the 2nd order accuracy and its physical implications} 
According to the analyses up to the previous  subsubsection such as 
eqs.~(\ref{3-19a},\ref{3-26-1},\ref{3-27-4}),  
the ${\cal O}(\lambda^2)$-contributions of 
eqs.~(\ref{3-20},\ref{3-24},\ref{3-25}) can be rigorously transformed 
into the following;
\begin{widetext}
\begin{eqnarray}
\hat{\cal L}_{1}(\hat{A}_{\alpha}) 
+ \hat{\cal L}_{2}(\hat{A}_{\alpha})&\equiv &  
 \big[\hat{L}_{d,\alpha},\hat{A}_{\alpha}\big] +   
\hat{F}^{(1)}_{\alpha}(\hat{A}_{\alpha}) + 
\hat{F}^{(2)}_{\alpha}(\hat{A}_{\alpha}) + 
\hat{F}^{(3)}_{\alpha}(\hat{A}_{\alpha}) + {\cal O}(\lambda^3).  \label{3-27} \\
\hat{\cal L}_{1}(\hat{A}_{\alpha}) &=& - 
\big[\hat{\epsilon}_{\alpha}+\hat{\cal M}_{\alpha}, \hat{A}_{\alpha}\big]_{-} 
+\ i\big(\partial_{X_k}L^{(0)}_{d,\alpha}\big)
\big[\hat{D}^{\alpha}_{Q_k},\hat{A}_{\alpha}\big]  
- i\big(\partial_{Q_k} L^{(0)}_{d,\alpha}\big) 
\big[\hat{D}^{\alpha}_{X_k},\hat{A}_{\alpha}\big] \label{3-28} \\
\hat{\cal L}_{2}(\hat{A}_{\alpha}) &=& 
 - \frac{1}{4}\Big[
\big[\hat{\epsilon}_{\alpha},\hat{\Omega}^{\alpha}_{X_j,Q_j}\big]_{-},
\hat{A}_{\alpha}\Big]_{+}   
\ - \frac{1}{4} \Big[ 
\big[\hat{\cal M}_{\alpha}, \hat{\Omega}^{\alpha}_{X_j,Q_j}\big]_{-},
\hat{A}_{\alpha}\Big]_{+} \nonumber \\
&&\ + \frac{1}{4}\Big[i\big(\partial_{X_k}L^{(0)}_{d,\alpha}\big) 
\big[\hat{D}^{\alpha}_{Q_k}, \hat{\Omega}^{\alpha}_{X_j,Q_j}\big] 
- i\big(\partial_{Q_k}L^{(0)}_{d,\alpha}\big)
\big[\hat{D}^{\alpha}_{X_k}, \hat{\Omega}^{\alpha}_{X_j,Q_j}\big] 
,\hat{A}_{\alpha}\Big]_{+}\nonumber \\
&&\ - \frac{i}{2}\Big[\big[\hat{D}^{\alpha}_{X_j}, 
\hat{\epsilon}_{\alpha} + \hat{\cal M}_{\alpha}\big],
\big[\hat{D}^{\alpha}_{Q_j},\hat{A}_{\alpha}\big]\Big]_{+} 
+ \frac{i}{2}\Big[\big[\hat{D}^{\alpha}_{Q_j},
\hat{\epsilon}_{\alpha} + \hat{\cal M}_{\alpha}\big]
,\big[\hat{D}^{\alpha}_{X_j},\hat{A}_{\alpha}\big]\Big]_{+} \nonumber \\ 
&&\ -\ \frac{i}{2}\big(\partial_{X_k} L^{(0)}_{d,\alpha}\big)
\Big[\hat{\Omega}^{\alpha}_{Q_k,X_j},
\big[\hat{D}^{\alpha}_{Q_j},\hat{A}_{\alpha}\big]\Big]_{+}   
+ \frac{i}{2} \big(\partial_{Q_k} L^{(0)}_{d,\alpha}\big)
\Big[\hat{\Omega}^{\alpha}_{X_k,X_j},
\big[\hat{D}^{\alpha}_{Q_j},\hat{A}_{\alpha}\big]\Big]_{+}  \nonumber \\ 
&&\ \ +\ \frac{i}{2}\big(\partial_{X_k} L^{(0)}_{d,\alpha}\big)  
\Big[\hat{\Omega}^{\alpha}_{Q_k,Q_j},
\big[\hat{D}^{\alpha}_{X_j},\hat{A}_{\alpha}\big]\Big]_{+}   
- \frac{i}{2}\big(\partial_{Q_k} L^{(0)}_{d,\alpha}\big)
\Big[\hat{\Omega}^{\alpha}_{X_k,Q_j},
\big[\hat{D}^{\alpha}_{X_j},\hat{A}_{\alpha}\big]\Big]_{+}. \label{3-29}
\end{eqnarray} 
\end{widetext}
${\cal L}_{i}$ above stands for the 
${\cal O}(\lambda^i)$-contribution.
Observing these results, 
notice that the first term in $\hat{\cal L}_{2}(\hat{\cal A}_{\alpha})$ 
is nothing but the anti-commutator 
between $\hat{\cal N}_{\alpha}$ and $\hat{A}_{\alpha}$, encoded 
in ${\cal O}(\lambda_1)$-contribution of $F^{(1)}(\hat{A}_{\alpha})$ 
(see eqs.~(\ref{3-19a},\ref{3-18})). On the one hand, 
the 2nd term in $\hat{\cal L}_2(\hat{\cal A}_{\alpha})$ is  
obtained from the ${\cal O}(\lambda^2_1)$-contribution  
in $\hat{F}^{(1)}$, $\hat{F}^{(2)}$ and $\hat{F}^{(3)}$, 
i.e. 
eq.~(\ref{3-27-4}). 
In spite of these apparently different origins, 
$\hat{\epsilon}_{\alpha}$ and $\hat{\cal M}_{\alpha}$ 
in these first two terms appear 
in a totally {\it parallel} fashion. We can see 
a similar feature also in the 5-th and 6-th term in 
${\cal L}_{2}(\hat{A}_{\alpha})$. These observations   
not only imply the consistency of our 
derived $SU(2)$ RKE but also dictate that 
$\hat{\cal M}_{\alpha}$ indeed plays role of 
the ${\cal O}(\lambda_1)$-correction to 
the quasi-particle energy dispersion. 
Therefore, its non-trivial matrix 
structure as well as that  
of $\hat{\epsilon}_{\alpha}$ 
induces the spin-precession of quasi-particles.        

When a disequilibration is created only by 
the external electromagnetic fields,  
$\hat{\cal M}_{\alpha}$ defined in eq.~(\ref{3-17a}) 
is composed by the {\it spatial (magnetic)} component 
and {\it temporal (electric)} one.  
The former is the Zeeman coupling 
between an external magnetic field and internal magnetic   
moment associated quasi-particle wavepacket. 
The later one is that between an applied 
electric field and internal electric dipole moment.   
To see these internal dipole moments 
explicitly, consider a situation  
in the presence of physical (i.e. external) 
electromagnetic gauge fields $({\rm a}_0,{\bf a})$, 
with corresponding physical electromagnetic fields 
${\bf b}=\nabla \times {\bf a}$, ${\bf e}=
\nabla_{R}{\rm a}_{0} - \partial_T {\bf a}$.  
Then, following the standard recipe, let us  
introduce the canonical momentum $k\equiv q+{\bf a}(T,R)$ 
and canonical frequency 
$\omega' \equiv\omega - {\rm a}_{0}(R)$ ~\cite{LL};
\begin{eqnarray}
|u^{\alpha\sigma}_{X;Q}\rangle \equiv 
|u^{(0),\alpha\sigma}_{k\equiv q+{\rm a},\omega'\equiv \omega - {\rm a}_{0}}\rangle.
\nonumber 
\end{eqnarray}  
Substituting this into eq.~(\ref{3-17a}), one can then 
re-express $\hat{\cal M}_{\alpha}$, such that 
it is given solely by the partial derivative 
with respect to these canonical quantities. Namely,  
this ${\cal O}(\lambda_1)$-correction  
to a q.p. energy dispersion precisely reduces into 
an inner product between external electromagnetic fields 
and sort of internal dipole moments; 
\begin{eqnarray}
&&\hspace{-0.3cm}\big[\hat{\cal M}_{\alpha}\big]
={\bf b}\cdot \big[\hat{\bf M}_{\alpha}\big] 
+ {\bf e}\cdot \big[\hat{\bf P}_{\alpha}\big] \nonumber \\ 
&&\hspace{-0.3cm}\big[\hat{\bf M}_{\alpha,m}\big]_{\sigma\sigma'} 
= \frac{i\epsilon_{mnl}}{2} \nonumber \\
&&\hspace{1.5cm}\ \times\  
\big\langle 
\partial_{k_n} u^{(0),\alpha\sigma}_{k,\omega'}
\big| L^{(0)}_{d,\alpha} -\hat{\sf L} 
\big| 
\partial_{k_l} u^{(0),\alpha\sigma'}_{k,\omega'} 
\big\rangle, \label{3-29c} \\
&&\hspace{-0.3cm}
\big[\hat{\bf P}_{\alpha,m}\big]_{\sigma\sigma'}  
= \frac{i}{2}
\big(
\big\langle 
\partial_{\omega'} u^{(0),\alpha\sigma}_{k,\omega'}
\big|L^{(0)}_{d,\alpha} -\hat{\sf L} 
\big| 
\partial_{k_m} u^{(0),\alpha\sigma'}_{k,\omega'} 
\big\rangle  \nonumber \\
&&\hspace{1.2cm}\ -\ 
\big\langle 
\partial_{k_m} u^{(0),\alpha\sigma}_{k,\omega'}
\big|L^{(0)}_{d,\alpha} -\hat{\sf L} 
\big| 
\partial_{\omega'} u^{(0),\alpha\sigma'}_{k,\omega'} 
\big\rangle \big).\label{3-29d}
\end{eqnarray}

Note that the magnetic dipole moment 
${\bf M}_{\alpha}$ above reproduces eq.~(\ref{1-2m}) 
in a non-interacting limit. 
On the other hand, the electric dipole moment 
${\bf P}_{\alpha}$ has 
no non-interacting counterpart, since it 
is purely 
associated with the energy-derivative of the quasi-particle 
Bloch wavefunction. 

As is trivial from its coupling with 
${\bf e}$, when considered in $U(1)$ FLs, 
this electric dipole moment ${\bf P}_{\alpha}$ is 
literally time-reversally even while parity odd;  
\begin{eqnarray}
{\bf P}_{\alpha}(k) &=& {\bf P}_{\alpha}(-k) \ \ \ T{\rm -reversal},  
\nonumber \\ 
{\bf P}_{\alpha}(k)&=&-{\bf P}_{\alpha}(-k) \ \ I{\rm -invserse}.
\nonumber 
\end{eqnarray} 
Therefore, in $SU(2)$ FLs where 
both of these symmetries are guaranteed, ${\bf P}_{\alpha}$ 
becomes traceless just in a same way as ${\bf M}_{\alpha}$ does;    
\begin{eqnarray}
{\rm Tr}\big[{\bf P}_{\alpha}(k)\big]\equiv 
{\rm Tr}\big[{\bf M}_{\alpha}(k)\big]\equiv 0.  \label{3-29e}
\end{eqnarray} 

When translated into the   
the $SU(2)$ effective EOM, 
these two dipole moments, first of all, enter 
into that for the $CP^1$ vector;  
\begin{eqnarray}
i\frac{d{\bf z}}{dt} = \Big\{ \hat{\bf M}_{\alpha}\cdot {\bf b} + 
\hat{\bf P}_{\alpha}\cdot {\bf e} + \cdots\Big\} {\bf z}, \nonumber 
\end{eqnarray}
which describes the spin-precession due to the 
external electromagnetic fields. 
%
Observing eqs.~(\ref{3-27}-\ref{3-29}), 
note also that every derivative term found 
in a conventional Keldysh equation is now replaced by  
the corresponding $SU(2)$ covariant derivative in our $SU(2)$ RKE; 
\begin{eqnarray} 
\partial_{X} (\cdots) \rightarrow 
\big[\hat{D}^{\alpha}_{X}, \cdots\big] \equiv \partial_{X}(\cdots) 
+ \big[\hat{\cal A}^{\alpha}_{X},\cdots\big]. \nonumber  
\end{eqnarray} 
Therein, its usual derivative part 
describes the charge degree of freedom,  
while the commutator with $\hat{\cal A}^{\alpha}_{X}$ 
stands for the precession of quasi-particle spin due to 
this gauge field. 
In this sense, our derived RKE treats the charge and spin 
degrees of freedom {\it on a equal footing way}, 
by using the $SU(2)$ covariant derivative.

To uncover the physical significance of remaining terms in  
$\hat{\cal L}_{2}$ such as the 1st 4 terms and 
the final 4 terms, we further need to solve this 
reduced Keldysh equation {\it in favor of} $\hat{A}_{\alpha}$ 
(sec. IV) and to derive the effective Boltzmann equation  
(sec. V) respectively.

\section{Perturbative solution for RKEs}
\subsection{prescription} 
When obtaining the reduced Keldysh equation described in 
the previous section, we have chosen 
the spectral function {\sl at equilibrium} as  
a unit matrix with its coefficient to be 
a delta function of $L^{(0)}_{d,\alpha}$;
\begin{eqnarray}
\hat{A}_{\alpha} = \delta(L^{(0)}_{d,\alpha})\hat{1} 
= Z^{(0)}_{\alpha}\delta(\omega-\epsilon^{(0)}_{\alpha})
\hat{1} \equiv 
f_{0}(L^{(0)}_{d,\alpha})\hat{1}. 
\label{4-1} 
\end{eqnarray}  
This is not only because such solutions are physically sensible,  
but also because, starting from this zero-th order spectral functions, 
we can indeed keep on satisfying the RKE up  
to higher order in $\lambda$. 
We will prove this point, by 
solving actually the derived RKE   
in favor for $\hat{A}_{\alpha}$ perturbatively 
in $\lambda$  
(see eq.~(\ref{4-4a-1})). 
   
Our choice of the spectral function at equilibrium  
clearly satisfies the derived $SU(2)$ RKE 
{\sl up to the 1st order}; 
\begin{eqnarray}
&&\hspace{-0.7cm}\hat{\cal L}_{1}(f_0\hat{1}) 
 = i \Big\{\big(\partial_{X_k}L^{(0)}_{d,\alpha}\big)
\big(\partial_{Q_k}L^{(0)}_{d,\alpha}\big)- 
\{X_k \leftrightarrow Q_k\}\Big\}f'_{0}\hat{1} \nonumber \\
&&\hspace{0.6cm} = \hat{0}.   \label{4-1a-0} 
\end{eqnarray}
However, it does not up to the 2nd order, 
i.e. $\hat{\cal L}_2(f_0\hat{1}) \ne \hat{0}$. 
To resolve this, we will introduce the 
1st order correction to the spectral functions;
\begin{eqnarray}
\hat{A}_{\alpha} = f_{0}\hat{1} + \hat{f}_{1} + \cdots, \label{4-1a}
\end{eqnarray} 
such that $\hat{A}_{\alpha}$ satisfies the $SU(2)$ RKEs even 
{\sl up to the 2nd order 
accuracy}; 
\begin{eqnarray}
\hat{0} = \hat{\cal L}_{1}(\hat{f}_1) + \hat{\cal L}_{2}(f_0\hat{1}). \label{4-2}
\end{eqnarray}
In general, we could further obtain  
the higher order correction of the spectral function  
$\hat{A}_{\alpha}$, provided that the RKE  
is given up to the 3rd order's accuracy or higher than 
that, i.e. 
$\hat{\cal L} = \hat{\cal L}_1 + \hat{\cal L}_2 
+ \hat{\cal L}_{3} + \hat{\cal L}_4 + \cdots$.  
Namely, its 3rd order part $\hat{\cal L}_3$ determines 
the the 2nd order correction to $\hat{A}_{\alpha}$ 
in terms of $\hat{f}_1$ and $f_0$;  
\begin{eqnarray}
\hat{0} &=& \hat{\cal L}_{1}(\hat{f}_2) + \hat{\cal L}_2(\hat{f}_1) 
+ \hat{\cal L}_3(f_0\hat{1}), \label{4-3} 
\end{eqnarray}  
The following equations further specify the higher order 
corrections such as $\hat{f}_3, \cdots$, in 
terms of $\hat{f}_2$,$\hat{f}_1$ and $f_0$ iteratively;
\begin{eqnarray}
\hat{0} &=& \hat{\cal L}_{1}(\hat{f}_3) + \hat{\cal L}_2(\hat{f}_2) 
+ \hat{\cal L}_3(\hat{f}_1) + \hat{\cal L}_4(f_0\hat{1}), \label{4-3a} \\   
\hat{0} &=& \cdots. \nonumber 
\end{eqnarray} 
However, our derived RKE being exact up to 
${\cal O}(\lambda^2_1,\lambda_1\lambda_2,\lambda^2_2)$,  
the highest order to which accuracy 
we could determine $\hat{A}_{\alpha}$ is $\hat{f}_1$.
\subsection{first order correction to $\hat{A}_{\alpha}$}
As we will show below, even up to this lowest order analysis,  
we obtain a non-trivial correction  to the spectral function, 
having an interesting physical implication (next subsection).  
Notice first the last {\bf 4} terms in eq.~(\ref{3-29}) 
do not contribute at all to ${\cal L}_{2}(f_0\hat{1})$, 
since they are antisymmetrized with respect to the exchange 
between $X$ and $Q$;   
\begin{eqnarray}
{\cal L}_{2}(f_0\hat{1}) &=& -\frac{f_0}{2} 
\Big\{\big[\hat{\epsilon}_{\alpha} +\hat{\cal M}_{\alpha},
\hat{\Omega}^{\alpha}_{X_jQ_j}\big]_{-} + 
\nonumber \\
&&\hspace{-1.8cm} - i\big(\partial_{X_k}L^{(0)}_{d,\alpha}\big)
\big[\hat{D}^{\alpha}_{Q_k},\hat{\Omega}^{\alpha}_{X_jQ_j}\big] 
+ i\big(\partial_{Q_k}L^{(0)}_{d,\alpha}\big) 
\big[\hat{D}^{\alpha}_{X_k},\hat{\Omega}^{\alpha}_{X_jQ_j}\big] 
\Big\} \nonumber \\   
&& \hspace{-1.8cm} - f'_{0}  
\Big\{ i\big(\partial_{Q_j}L^{(0)}_{d,\alpha}\big)
\big[\hat{D}^{\alpha}_{X_j},\hat{\epsilon}_{t,\alpha}\big]  
-  i\big(\partial_{X_j}L^{(0)}_{d,\alpha}\big)
\big[\hat{D}^{\alpha}_{Q_j},\hat{\epsilon}_{t,\alpha}\big] 
\Big\}.  
\nonumber 
\end{eqnarray}   
where $\hat{\epsilon}_{t,\alpha} \equiv \hat{\epsilon}_{\alpha} + \hat{\cal M}_{\alpha}$. 
Then, the last two terms above are readily canceled in eq.~(\ref{4-2}), 
when $\hat{f}_{1} = -\hat{\epsilon}_{t,\alpha}f'_{0}$ is substituted into 
$\hat{\cal L}_1$;
\begin{eqnarray}
&&\hspace{-0.3cm}\hat{\cal L}_{1}(-\hat{\epsilon}_{t,\alpha}f'_{0}) 
+ \hat{\cal L}_{2}(f_0\hat{1}) = \nonumber \\ 
&&\hspace{0.4cm}-\ \frac{f_0}{2}\Big\{\big[\hat{\epsilon}_{t,\alpha},
\hat{\Omega}^{\alpha}_{X_jQ_j}\big]_{-} - \ i\big(\partial_{X_k}L^{(0)}_{d,\alpha}\big)
\big[\hat{D}^{\alpha}_{Q_k},\hat{\Omega}^{\alpha}_{X_jQ_j}\big] \nonumber \\
&&\hspace{1.5cm}+ \ i\big(\partial_{Q_k}L^{(0)}_{d,\alpha}\big)
\big[\hat{D}^{\alpha}_{X_k},\hat{\Omega}^{\alpha}_{X_jQ_j}\big]\Big\}. \label{4-3a-1} 
\end{eqnarray} 
Since $\hat{\epsilon}_{t,\alpha}$ is regarded as a small quantity, 
$-\hat{\epsilon}_{t,\alpha}f'_0$ is nothing but the 
${\cal O}(\lambda)$-correction  to 
the {\it argument} of $f_{0}$,  
i.e. $\hat{L}^{(0)}_{d,\alpha}$.  

To set off also those terms in 
$\hat{\cal L}_2(f_0\hat{1})$ which are linear in $f_0$ {\it itself}, 
we need to introduce a correction to an {\it overall coefficient} 
of $f_0$, in addition to that to its argument. 
Namely, consider the following $\hat{f}_1$;
\begin{eqnarray}
\hat{f}_{1} \equiv - 
\hat{\epsilon}_{t,\alpha} f'_{0} 
- \frac{1}{2}\hat{\Omega}^{\alpha}_{X_jQ_j} f_0. \label{4-4}
\end{eqnarray} 
Then, when entering into $\hat{\cal L}_{1}$, 
the 2nd term above totally cancel all the terms in the 
right hand side of eq.~(\ref{4-3a-1}). 
 
Notice that, this cancellation becomes possible, only because 
the {\it relative ratio} among the coefficients of first 3 terms 
in $\hat{\cal L}_{2}(\hat{A}_{\alpha})$ perfectly match with 
the ratio among the corresponding three terms in 
$\hat{\cal L}_{1}(\hat{A}_{\alpha})$; 
\begin{eqnarray} 
&&\hspace{-0.4cm}\hat{\cal L}_{2}(\hat{A}_{\alpha}) =  - \frac{1}{4}\Big[
\big[\hat{\epsilon}_{\alpha},\hat{\Omega}^{\alpha}_{X_j,Q_j}\big]_{-},
\hat{A}_{\alpha}\Big]_{+}   \nonumber \\ 
&&\hspace{1.2cm} - \frac{1}{4} \Big[ 
\big[\hat{\cal M}_{\alpha}, \hat{\Omega}^{\alpha}_{X_j,Q_j}\big]_{-},
\hat{A}_{\alpha}\Big]_{+} \nonumber \\ 
&&\hspace{1.2cm} + \frac{i}{4}\Big[\big(\partial_{X_k}L^{(0)}_{d,\alpha}\big) 
\big[\hat{D}^{\alpha}_{Q_k}, \hat{\Omega}^{\alpha}_{X_j,Q_j}\big],
\hat{A}_{\alpha}\Big]_{+} \nonumber \\
&&\hspace{1.2cm} - \big\{X_k \leftrightarrow Q_k\big\} + \cdots,\nonumber  \\
&&\hspace{-0.6cm}\leftrightarrow \ 
\ \hat{\cal L}_{1}(\hat{A}_{\alpha}) = -   
\big[\hat{\epsilon}_{\alpha}, \hat{A}_{\alpha}\big]_{-} 
- \big[\hat{\cal M}_{\alpha}, \hat{A}_{\alpha}\big]_{-}  
\nonumber \\
&&\hspace{1.6cm} +  i 
\big(\partial_{X_k}L^{(0)}_{d,\alpha}\big)
\big[\hat{D}^{\alpha}_{Q_k},\hat{A}_{\alpha}\big]  
- \big\{X_k \leftrightarrow Q_k\big\}. \nonumber  
\end{eqnarray} 
If either {\sl signs} or {\sl coefficients}
of any one of the 3 terms in $\hat{\cal L}_{2}$ 
were not to meet with those of corresponding 3 terms 
in $\hat{\cal L}_1$,  we could never have any 
$\hat{f}_1$ satisfying eq.~(\ref{4-2}).  
In other words, there is no a priori guarantee 
for the existence of  
$\hat{f}_1$ which satisfies eq.~(\ref{4-2}).    
In spite of this, both {\sl signs} and 
{\sl coefficients} in $\hat{\cal L}_{2}$ 
completely met with those in $\hat{\cal L}_{1}$,  
we safely have eq.~(\ref{4-4}) as an appropriate 
${\cal O}(\lambda)$-correction to $\hat{A}_{\alpha}$. 
Reversely speaking, this perfect coincidence between 
${\cal L}_1$ and ${\cal L}_2$ indicates the validity 
and the consistency of our derived $SU(2)$ RKE.

\subsection{Physical Implications}  
\begin{figure}
\begin{center}
\includegraphics[width=0.4\textwidth]{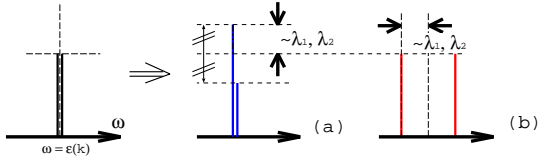}
\end{center}
\caption{A schematic picture of spectral functions in 
$SU(2)$ case. (a): The correction due to the 
Berry's curvature changes its weight. Furthermore, 
the change for the up-spin weight and that for the 
down-spin are in general different. (b): Usual 1st order 
correction change an energy spectrum.}
\label{fig4}
\end{figure}

Let us next argue the physical consequence of our solution 
$\hat{A}_{\alpha}$;
\begin{eqnarray}
\hat{A}_{\alpha} = f_0(L^{(0)}_{d,\alpha}) \hat{1} 
- \hat{\epsilon}_{t,\alpha}f'_{0}(L^{(0)}_{d,\alpha}) 
-\frac{1}{2}\hat{\Omega}^{\alpha}_{X_jQ_j} 
f_0(L^{(0)}_{d,\alpha}). \label{4-4a-1}
\end{eqnarray}
As was already mentioned, 
the 2nd term above clearly stands for 
the renormalization of the energy dispersion. Namely, 
diagonalizing 
$\hat{\epsilon}_{\alpha} + \hat{\cal M}_{\alpha} \equiv  
\hat{\epsilon}_{t,\alpha}$,  
we can transcribe this correction into that of 
the arguments of $f_0$,
\begin{eqnarray}
f_{0}\hat{1} - \hat{\epsilon}_{t,\alpha}f'_{0}
\simeq \left[\begin{array}{cc}
	f_0(L^{(0)}_{d,\alpha} - \epsilon_{t,\alpha,+}) &  \\
		 & f_0(L^{(0)}_{d,\alpha} - \epsilon_{t,\alpha,-}) \\   
\end{array} \right]. \nonumber  
\end{eqnarray}
A finite $\epsilon_{t,\alpha,+}-\epsilon_{t,\alpha,-}$ thereby 
lifts the doubly degeneracy at equilibrium (Fig.~\ref{fig4}(b)). 

On the other hand, the last term in eq.~(\ref{4-4a-1})
changes the {\it weight} of the spectral function.    
To see this, let us integrate eq.~(\ref{4-4a-1}) with 
respect to frequency;
\begin{eqnarray}
\int_{-\Lambda}^{\Lambda}\ \hat{A}_{\alpha} d\omega 
= Z^{(0)}_{\alpha}\times\big\{1-\frac{1}{2}
(\hat{\Omega}_{X_jQ_j})_{|\omega=\epsilon^{(0)}_{\alpha}}
\big\},  \label{4-4a}
\end{eqnarray}
where $Z^{(0)}_{\alpha}$ stands for the 
residue at $L^{(0)}_{d,\alpha}=0$. This clearly 
dictates that, apart from the conventional 
renormalization factor $Z^{(0)}_{\alpha}$, 
the integrated spectral weight acquires an 
additional ${\cal O}(\lambda)$-correction 
which is given by the $SU(2)$ Berry's 
curvature. 

This curvature correction   
$\hat{\Omega}^{\alpha}_{X_jQ_j}$  
has a {\it spin-selective} effect. Namely, 
choosing the basis diagonalizing this $SU(2)$ 
Berry's curvature, we have, 
\begin{eqnarray}
&&\hspace{-0.5cm}
\int_{-\Lambda}^{\Lambda} \ \hat{A}_{\alpha} 
d\omega \sim  
Z^{(0)}_{\alpha}\times\left[\begin{array}{cc}
	1-\frac{1}{2}
	\Omega^{\alpha,+}_{X_jQ_j} & 0 \\ 
		0 & 1-\frac{1}{2}
		\Omega^{\alpha,-}_{X_jQ_j}  \\   
\end{array} \right]_{|\omega=\epsilon^{(0)}_{\alpha}}, \label{4-5} 
\end{eqnarray}
where the up-chirality electron and down one 
generally acquire the different 
correction to its spectral weight with each other.     

Observing eq.~(\ref{4-5}), we can propose 
several photoemission experiments as a candidate 
experimental tool to visualize the 
{\it dual $SU(2)$ electromagnetic fields} in 
a momentum resolved way. To see this, 
consider the situation 
in the presence of applied electromagnetic 
fields, i.e. ${\bf b} = \nabla \times {\bf a}$ and 
${\bf e} = \nabla {\rm a}_{0} - \partial_{T}{\bf a}$. 
Then, using the canonical momentum and frequency 
introduced previously, eq.(\ref{4-4a}) reduces into 
the {\it vector product} between the real   
electromagnetic fields and 
the a sort of dual version of $SU(2)$ 
electromagnetic fields; 
\begin{eqnarray}
Z_{\alpha} &\equiv& 
Z^{(0)}_{\alpha}\times
\big\{1-\frac{1}{2}(\hat{\Omega}^{\alpha}_{X_jQ_j})_{|\omega=
\epsilon^{(0)}_{\alpha}}\big\} 
\nonumber \\
&=& Z^{(0)}_{\alpha}\times 
\big\{1+\frac{1}{2}\bar{\cal B}^{\alpha}\cdot {\bf b}
+\frac{1}{2}\bar{\cal E}^{\alpha}\cdot{\bf e}\big\}. 
\label{4-6} 
\end{eqnarray}
Here the dual quantities in 
the right hand side are estimated on shell 
at equilibrium;  
\begin{eqnarray}
&&\hspace{-0.1cm}
\bar{\cal B}^{\alpha}_{m} = i\epsilon_{mnl}\big\{\partial_{k_n}{\cal A}^{\alpha}_l
+ {\cal A}^{\alpha}_n
{\cal A}^{\alpha}_l\big\}_{|\omega=\epsilon^{(0)}_{\alpha}},  
\label{4-6a} \\
&&\hspace{-0.1cm}
\bar{\cal E}^{\alpha}_{m} = i\big\{\partial_{0}{\cal A}^{\alpha}_m 
- \partial_m {\cal A}^{\alpha}_{0} 
+ \big[{\cal A}^{\alpha}_0, {\cal A}^{\alpha}_{m}\big]
\big\}_{|\omega=\epsilon^{(0)}_{\alpha}},  \label{4-6b} \\ 
&&\hspace{0.9cm} 
\big[{\cal A}^{\alpha}_{m}\big]_{\sigma\sigma'}
\equiv \langle u^{(0),\alpha\sigma}_{k,\omega}|
\partial_{k_m} u^{(0),\alpha\sigma'}_{k,\omega}\rangle,  
\nonumber \\
&&\hspace{0.9cm}  
\big[{\cal A}^{\alpha}_{0}\big]_{\sigma\sigma'}
\equiv \langle u^{(0),\alpha\sigma}_{k,\omega}| 
\partial_{\omega} u^{(0),\alpha\sigma'}_{k,\omega}\rangle, \nonumber 
\end{eqnarray}
with $m=1,\cdots d$.

As is clear from their coupling with real electromagnetic 
fields, $\bar{\cal B}^{\alpha}_m(k)$ and 
$\bar{\cal E}^{\alpha}_m(k)$ as functions of $k$ 
obey precisely same symmetries as ${\bf M}_{\alpha,m}(k)$ 
and ${\bf P}_{\alpha,m}(k)$ do respectively. 
Thus, in a $SU(2)$ FL having time-reversal and  
spatial inversion symmetry, these $SU(2)$ matrices 
become both traceless; 
\begin{eqnarray}
{\rm Tr}\big[\bar{\cal E}^{\alpha}_m(k)\big] 
\equiv {\rm Tr}\big[\bar{\cal B}^{\alpha}_m(k)\big] \equiv 0 
\label{4-7}
\end{eqnarray}
just as in eq.~(\ref{3-29e}). 
This suggests that the (integrated) spectral weights for  
the doubly degenerate quasi-particles, 
subjected under applied electromagnetic fields,  
acquire the 1st order corrections having 
{\it a reverse sign} with 
each other (see Fig.~\ref{fig4}(a)). 

The spectral weight for 
quasi-particle is in principle detectable in a momentum 
resolved way in terms of photoemission experiments  
such as angle resolved photoemission spectroscopy (ARPES) 
and resonating inelastic X-ray scattering (RIXS).  
The above theoretical observation therefore leads us to 
raise the {\sl spin-resolved} ARPES as the potential 
candidate tool to 
measure the dual $SU(2)$ electromagnetic fields.  
Namely, observing the change of the 
spin-resolved spectral weight at  
each $k$-point under small ${\bf e}$ or ${\bf b}$, 
we can identify the dual electric or magnetic 
field as its linear response. 
Furthermore, eq.~(\ref{4-7})  
claims that, {\it maximizing the measured 
linear responses with respect 
to the resolved-spin's direction}, one can 
even determine the spin-quantization axis for the  
eigenbasis diagonalizing $\bar{\cal B}^{\alpha}$ 
or $\bar{\cal E}^{\alpha}$. Thus, 
provided that these processes are performed for all the 
components of ${\bf e}$ and ${\bf  b}$ separately, 
we could determine, {\it as $2$ by $2$ matrices},  
all the ${\bf 2}\times {\bf  d}$ 
components of these $SU(2)$ Berry's curvatures at  
an arbitrary $k$ point on a Fermi surface.   

The (spin-resolved) ARPES experiment  
under external electromagnetic 
fields is, however, a difficult experiment.  
A more relatively less unrealistic proposal 
might be detecting {\sl abelian} Berry's 
curvatures in $U(1)$ FLs, such as ferromagnetic metals 
or paramagnetic metals without any centrosymmetric lattice 
point.  In such $U(1)$ FLs, we have only to regard all the 
$SU(2)$ hermitian matrices in 
eqs.~(\ref{3-27}-\ref{3-29}) as a scalar quantity, 
to obtain the $U(1)$ RKEs and its 
solutions up to the 1st order accuracy \cite{rapicom}; 
\begin{eqnarray}
A_{\alpha}&=&(1-\frac{1}{2}\Omega^{\alpha}_{X_jQ_j}) 
\delta(L_{d,\alpha}-{\cal M}_{\alpha}) \nonumber \\
&=& \big(1+\frac{1}{2}\bar{\cal B}^{\alpha}\cdot {\bf b} 
+ \bar{\cal E}^{\alpha}\cdot {\bf e}\big)Z_{\alpha}
\delta(\omega - \epsilon_{\alpha}).  \label{4-7a}  
\end{eqnarray} 
Note that the dual abelian electromagnetic fields 
are again defined to be on-shell as in $SU(2)$ case;
\begin{eqnarray}
&&\hspace{-0.3cm}\bar{\cal B}^{\alpha}_{j} \equiv 
i\epsilon_{jml}
\big(\partial_{k_m}{\cal A}^{\alpha}_{l}
\big)_{|\omega=\epsilon_{\alpha}}, \   
\bar{\cal E}^{\alpha}_{j}  \equiv i 
\big(\partial_{\omega}{\cal A}^{\alpha}_j - 
\partial_{j} {\cal A}^{\alpha}_{0}
 \big)_{|\omega =\epsilon_{\alpha}}, \nonumber  \\
&& \hspace{-0.3cm}{\cal A}^{\alpha}_{\mu} 
\equiv \langle u^{(0),\alpha}|
\partial_{\mu}u^{(0),\alpha}\rangle,  
\label{4-8} 
\end{eqnarray}
with $\mu = \omega,k$. 

In such a $U(1)$ FL, the time-reversal (spatial inversion) 
counterpart being already lifted at equilibrium,  
these linear responses against ${\bf b}$ and ${\bf e}$ 
can be measured only by the spectral weight 
{\it itself} for a given Fermi surface.  
Namely, the spin-filter are already implemented 
in matters themselves.  
Thereby, in stead of ARPES experiments, 
the resonating inelastic X-ray scattering (RIXS) 
experiment, which is by far compatible with 
applied electromagnetic fields, is a more 
promising photoemission experiment. 




\section{$U(1)$ effective Boltzmann equation and EOM for quasi-particles}


In the previous section, we clarified the physical implications 
of the first 4 terms of $\hat{\cal L}_2$ given in eq.~(\ref{3-29}). 
We will now study on quasi-particle dynamics, only to interpret 
remaining 4 terms in $\hat{\cal L}_2$ such as  
\begin{eqnarray}
\hat{\cal L}_2 &=& \cdots  -\frac{i}{2}\big(\partial_{X_k}L^{(0)}_{d,\alpha}\big)
\Big[\hat{\Omega}^{\alpha}_{Q_kX_j},
\big[\hat{D}^{\alpha}_{Q_j},\hat{A}_{\alpha}\big]\Big] \nonumber \\ 
&&\ \ -\ \{X_j \leftrightarrow Q_j|X_k,Q_k\} -\{X_j,Q_j|X_k\leftrightarrow Q_k\} 
\nonumber \\ 
&&\ \ +\ \{X_j\leftrightarrow Q_j|X_k\leftrightarrow Q_k\}. \label{5-0} 
\end{eqnarray}
 
Strictly speaking, when it comes to an EOM for quasi-particles,    
we in principle have to begin with the Keldysh equation for the 
lesser (or greater) Green functions $\hat{\sf g}^{<(>)}$, i.e. 
\begin{eqnarray}
&&\big[{G_0}^{-1} - \Sigma^{\rm HF} - \sigma , {\sf g}^{<(>)} \big]_{\otimes,-} 
-\big[\Sigma_{\rm c}^{<(>)} , {\sf b} \big]_{\otimes,-} \nonumber \\
&&\ \ = \ \frac{1}{2} \big[\ {\Sigma}_{\rm c}^{>} , {\sf g}^{<} \big]_{\otimes,+} 
-\frac{1}{2} \big[\ {\Sigma}_{\rm c}^{<} , {\sf g}^{>} \big]_{\otimes,+}. \label{5-0-a}
\end{eqnarray}
which is different from that for the spectral function $\hat{\sf A}$.  

As far as a system is not so far from its equilibration, however, 
this difference could be regarded as small     
at $T\simeq 0$ and $\omega \simeq \mu$.  
To be more specific, thanks to the  boundary condition 
imposed on the Matsubara self-energy and its continuity,  
both lesser and greater collisional self-energies 
$\hat{\Sigma}^{<,>}_{\rm c}(\omega)$ above   
become small around $\omega \simeq \mu$ (see the arguments in 
Appendix A2d and A3). 
As a result, we could begin with the following dissipationless 
equation, instead of eq.~(\ref{5-0-a});  
\begin{eqnarray}
\big[\hat{G_0}^{-1} - \hat{\Sigma}^{\rm HF} - \hat{\sigma} , \hat{\sf g}^{<(>)} \big]_{\otimes,-} 
= \hat{0}. \label{5-1-a} 
\end{eqnarray}  
Note that a small 
$\hat{\Sigma}^{<,>}_{\rm c}(\omega)_{|\omega\simeq \mu}$  
does {\it not} necessarily lead to a small 
$\hat{\sigma}(\omega)_{|\omega\simeq \mu}$. Namely, 
the latter one is given by the {\it energy- (principal) integral} of the 
former two; 
\begin{eqnarray}
\hat{\sigma}(\omega) \equiv \int_{-\infty}^{+\infty} 
\frac{d\omega'}{2\pi}
\frac{\cal P}{\omega-\omega'}\big(
\hat{\Sigma}^{<}_{\rm c}(\omega')  
+ \hat{\Sigma}^{>}_{\rm c}(\omega')
\big). \nonumber 
\end{eqnarray}  

With a help of this biased treatment of  
the collisional self-energies for different frequency regions, 
however, 
we can obtain clear physical interpretations for all the 
remaining terms in $\hat{\cal L}_2$ given in eq.~(\ref{5-0}).  
Namely, we can apply precisely a same projection process 
onto eq.~(\ref{5-1-a}) as we did for $\hat{\sf A}$. Thus, 
we begin with the exactly same reduced  
Keldysh equation for $g^{<}_{\alpha}$ as in  
eqs.~(\ref{3-27}-\ref{3-29}), with 
$g^{<}_{\alpha}$ being the $(\alpha,\alpha)$-th element 
of $N_b$ times $N_b$ matrix 
$\hat{g}^{<} \equiv \hat{U}^{\dagger} \hat{\sf g}^{<}\ \hat{U}$.
By constructions, this diagonal element $g^{<}_{\alpha}$ 
can be decomposed into the product between the 
{\it generalized Fermi distribution function} $f_{\alpha}$ 
(scalar quantity) and the spectral function for the 
$\alpha$-th band, later of which was already derived in the 
previous section~\cite{note4};   
\begin{eqnarray}
g^{<}_{\alpha}(Q;X)= A_{\alpha}(Q;X) 
f_{\alpha}(Q;X). \label{5-2}
\end{eqnarray}    

Generally speaking, based on this  
decomposition, one must derive a coupled effective Boltzmann 
equation for spin and charge, out of the $SU(2)$ RKEs 
for $g^{<}_{\alpha}$.  
However, since this analysis is somehow  
involved, we will henceforth restrict ourselves to 
$U(1)$ FLs, only to derive effective Boltzmann 
equation for charge degree of freedom, i.e. EOM 
only for $f_{\alpha}$.     
 
In $U(1)$ case, substituting eq.~(\ref{5-2}) 
into the abelian RKE, 
we first   
obtain; 
\begin{eqnarray}
&&\hspace{-0.2cm} 0\  = \ \big\{{\cal L}_{1}(A_{\alpha}) 
 + {\cal L}_{2}(A_{\alpha})\big\}\times f_{\alpha} \nonumber \\  
&& \hspace{-0.4cm}  + \  A_{\alpha}\times \big\{\partial_{X_j}(L_{d,\alpha}-
{\cal M}_{\alpha})\partial_{Q_j}f_{\alpha}  \nonumber \\ 
&&\hspace{-0.4cm} - \  
\partial_{X_k}\Omega^{{\alpha}}_{Q_kX_j}\partial_{Q_j}f_{\alpha}\  
+\ \partial_{Q_k}\Omega^{\alpha}_{X_kX_j}\partial_{Q_j}f_{\alpha} 
- (X_j \leftrightarrow Q_j)\big\}. \nonumber 
\end{eqnarray}
Notice that the spectral function $A_{\alpha}$ 
was determined  such that 
${\cal L}_1(A_{\alpha}) + {\cal L}_2(A_{\alpha}) \equiv 0$ 
(up to the 2nd order in gradient expansion). 
Thus, in the right hand side, we can safely drop 
those terms which are proportional to $f_{\alpha}$ itself.    
Notice also that the spectral function $A_{\alpha}$  
thus determined is sharply peaked at $\omega = \epsilon_{\alpha}$.  
Thereby, we have only to integrate this equation 
over frequency, 
such that $f_{\alpha}$ is replaced by the physical 
{\it quasi-particle occupation number in the $q$-$R$ space};
\begin{eqnarray}
n_{\alpha}(q;R,T) \equiv f_{\alpha}
(q,\omega\equiv\epsilon_\alpha;R,T).  \nonumber 
\end{eqnarray}

The EOM 
thus obtained reads as follows;
\begin{eqnarray}
&&\big(1-\bar{\Omega}^{\alpha}_{T\epsilon_{\alpha}} 
+ (\partial_{R_j}\epsilon_{\alpha})\ 
\bar{\Omega}^{\alpha}_{q_j\epsilon_{\alpha}} 
- (\partial_{q_j}\epsilon_{\alpha})\ 
\bar{\Omega}^{\alpha}_{R_j\epsilon_{\alpha}}\big)\ 
\partial_T n_\alpha \nonumber \\
&&=\ \big(\partial_{R_j}\epsilon_{\alpha} 
+ (\partial_{T}\epsilon_{\alpha})\  
\bar{\Omega}^{\alpha}_{\epsilon_{\alpha} R_j} 
- (\partial_{R_k}\epsilon_{\alpha})\ 
\bar{\Omega}^{\alpha}_{q_kR_j} 
+ \ \bar{\Omega}^{\alpha}_{TR_j}  
 \nonumber \\
&&\hspace{0.5cm}\ +\  
(\partial_{q_k}\epsilon_{\alpha})\ 
\bar{\Omega}^{\alpha}_{R_kR_j}\big)
\ \partial_{q_j}n_\alpha - \{ q_j \leftrightarrow R_j\}, \nonumber  
\end{eqnarray}      
which we can regard as the Boltzmann equation for 
the $\alpha$-th band quasi-particles. 
Note that the partial $q$, $R$ and $T$-derivatives 
encoded into the curvatures therein     
apply only onto their {\it explicit} dependences  
and do {\it not} apply to their arguments of 
$\epsilon_{\alpha}$, e.g. 
\begin{eqnarray}
\bar{\Omega}^{\alpha}_{q_{i}R_{j}} \equiv 
\big(\Omega^{\alpha}_{q_iR_j}\big)_{|\omega=\epsilon_{\alpha}}\ , \  
\bar{\Omega}^{\alpha}_{\epsilon_{\alpha} R_j} \equiv 
\big(\Omega^{\alpha}_{\epsilon_{\alpha} R_j}\big)_{|\omega=\epsilon_{\alpha}}.  
\nonumber 
\end{eqnarray} 

As was explained in the introduction, however, 
one can also introduce the curvature  
defined in the {\it co-dimensional space} associated with 
$\omega\equiv \epsilon_{\alpha}(q,R,T)$, e.g.   
\begin{eqnarray}
\tilde{\Omega}^{\alpha}_{q_iR_j} &\equiv & 
i \big(\partial_{q_i}\tilde{\cal A}^{\alpha}_{R_j} 
- \partial_{R_j}\tilde{\cal A}^{\alpha}_{q_i}\big), \nonumber \\
\tilde{\cal A}^{\alpha}_{R_j} &\equiv& 
\langle \tilde{u}^{\alpha}|\partial_{R_j} \tilde{u}^{\alpha}
\rangle, \ |\tilde{u}^{\alpha}\rangle \equiv 
|u^{\alpha}\rangle_{|\omega=\epsilon_{\alpha}}. \label{5-3} 
\end{eqnarray}
Then the Boltzmann equation above 
can be also expressed solely in terms of these curvatures 
in the $q$-$R$-$T$ space. 
Specifically, normalizing the 
coefficient of $\partial_T n_{\alpha}$, we obtain 
the following up to the 2nd order in $\lambda$; 
\begin{eqnarray}
&& \hspace{-0.8cm} 0 = \partial_{T}n_{\alpha} + \nonumber \\
&& \hspace{-0.9cm} 
 +\ \big\{\partial_{R_j}\epsilon + \tilde{\Omega}^{\alpha}_{TR_j} 
 - (\partial_{R_k}\epsilon)\tilde{\Omega}^{\alpha}_{q_k R_j} 
 + (\partial_{q_k}\epsilon)\tilde{\Omega}^{\alpha}_{R_k R_j}\big\} 
\partial_{q_j}n_{\alpha} \nonumber \\
&& \hspace{-0.9cm} 
 -\ \big\{\partial_{q_j}\epsilon + \tilde{\Omega}^{\alpha}_{Tq_j} 
 - (\partial_{R_k}\epsilon)\tilde{\Omega}^{\alpha}_{q_k q_j} 
 + (\partial_{q_k}\epsilon)\tilde{\Omega}^{\alpha}_{R_k q_j}\big\} 
\partial_{R_j}n_{\alpha}. \label{5-4} 
\end{eqnarray}  
Accordingly, comparing this Boltzmann equation with the  
continuity equation, i.e. 
$0 = \partial_{T}n_{\alpha} + (\partial_{T}q)\partial_{q} n_{\alpha}
+ (\partial_{T}R)\partial_{R} n_{\alpha}$, one can  
readily read {\it the effective EOM for the quasi-particle}; 
\begin{eqnarray}
\frac{dR_j}{dT} = - \partial_{q_j}\epsilon - \tilde{\Omega}^{\alpha}_{Tq_j} 
+ (\partial_{R_k}\epsilon_{\alpha})\ 
\tilde{\Omega}^{\alpha}_{q_k q_j} 
- (\partial_{q_k}\epsilon_{\alpha})\  
\tilde{\Omega}^{\alpha}_{R_k q_j},  \nonumber \\
\frac{dq_j}{dT} = \partial_{R_j}\epsilon + \tilde{\Omega}^{\alpha}_{TR_j} 
- (\partial_{R_k}\epsilon_{\alpha})\ 
\tilde{\Omega}^{\alpha}_{q_k R_j} 
+ (\partial_{q_k}\epsilon_{\alpha})\ 
\tilde{\Omega}^{\alpha}_{R_k R_j}.  \nonumber 
\end{eqnarray}  

This effective EOM for quasi-particles 
proves, at least to the  
accuracy of 2nd order in $\lambda$,   
that the EOM valid in ``non-interacting'' Fermi 
system \cite{sn} can be also generalized into 
``interacting'' Fermi systems with {\it those curvatures 
defined in the co-dimensional space}, such as   
eq.~(\ref{5-3}).  
However, observing that the 1st order correction 
to the renormalization factor is characterized by 
the Berry's curvature {\it in Euclidean $q$-$\omega$-$R$-$T$ 
space} instead of that in co-dimensional space 
(see eq.~(\ref{4-7a})),  
studying the 3rd order correction to this EOM is 
still an interesting open question (see also sec.~IV).       

As an immediate application of this 
effective EOM,  we can again consider 
the case with external electromagnetic fields;   
\begin{eqnarray} 
\frac{dR}{dT} &=& {\bf v}_{\alpha} + 
\tilde{\cal B}^{\alpha} \times \frac{dk}{dT}, \label{5-5} \\
\frac{dk}{dT} &=& - {\bf e} + {\bf b} \times  \frac{dR}{dT}. \label{5-6}
\end{eqnarray}
where ${\bf v}_{\alpha}\equiv \partial_{k}\epsilon_{\alpha}$ 
and $\tilde{\cal B}^{\alpha}$ reads, 
\begin{eqnarray}
\tilde{\cal B}^{\alpha} &\equiv& 
\bar{\cal B}^{\alpha} - \bar{\cal E}^{\alpha} 
\times {\bf v}_{\alpha}. \label{5-6a}
\end{eqnarray}  
Note that $\bar{\cal B}^{\alpha}$ and $\bar{\cal E}^{\alpha}$ 
are curvatures in Euclidean $k$-$\omega$ space 
(eqs.(\ref{1-2},\ref{1-2-a}) and eqs.(\ref{1-4},\ref{1-4-a}) respectively), while 
$\tilde{\cal B}^{\alpha}$ is the curvature  
defined in the co-dimensional space (eqs.~(\ref{codim1},\ref{codim2})).  
Observing eq.~(\ref{5-5}), one then see that 
the {\it intrinsic} AHE in $U(1)$ FLs  
should be defined in terms of both 
electromagnetic fields in the dual space;
\begin{eqnarray}
\sigma_{jm} = \epsilon_{jml}\frac{e^2}{\hbar}
\sum_{\alpha} \int \frac{dk}{(2\pi)^d}
\big(\bar{\cal B}^{\alpha} - 
\bar{\cal E}^{\alpha}\times {\bf v}_{\alpha}\big)_{l} 
n_{\rm f}(\epsilon_{\alpha,k}),  \label{5-7}
\end{eqnarray}
where $n_{\rm f}$ denotes a Fermi distribution function.

Notice that the terminology ``intrinsic'' is now generalized 
into a slightly wider sense. Namely, the above expression 
for the Hall conductivity contains not only  
the ``$k$-space magnetic field'' effect, which is 
already present in a ``non-interacting'' limit,   
but also the ``many-body effect'' via the  
corresponding ``electric field'' component. 
Based on the Ishikawa-Matsuyama formula 
and Fermi liquid assumptions, one 
can also see that this ``many-body'' 
correction indeed takes over (a part of)  
the so-called vertex correction to 
the static transverse conductivity (see appendix B). 
This consistency check at the linear response regime 
strongly supports the validity of our derived effective 
EOM i.e. eqs.(\ref{5-5}) and (\ref{5-6}). 

\section{Summary and open problems}

To extract an information of low-energy effective theory 
in a generic {\it multiple-band} interacting Fermi systems, 
we have derived the Reduced Keldysh Equation (RKE) which
effectively describes the charge and spin degrees of freedom for a
specific {\it single-band} forming a Fermi surface.  Our derivation is
perturbative with respect to the gradient expansion whose coupling
constant measures how a system is disequilibrated.  It is, however,
non-perturbative in the
electron-electron interactions.  Instead, it relies only on the
``adiabatic assumption'' which is also utilized to validate the usual
Fermi liquid framework.  This assumption claims that, when it comes to
the low-frequency region at sufficiently low temperature, the intrinsic
life time of a quasi-particle (due to electron-electron interactions),
which is ${\cal O}(T^{-2},(\omega-\mu)^{-2})$, becomes much longer than
the inverse of thermal broadening of the spectral functions ($\sim {\cal
  O}(T^{-1})$ ).  This assumption usually provides a finite energy
region within which Fermi liquid theory works.  Based on the same
spirit and as a sort of zero-th order approximation, we completely ignored
the life-time part (anti-hermitian part) of the collisional
self-energies, while studying on {\it its hermitian part} on a general
ground.

Out of the RKE thus derived, we have succeeded in extracting several
intriguing physical implications both in $SU(2)$ and $U(1)$ FLs. A first
observation is that the linear response of the spectral weight for a
quasi-particle with respect to an applied electromagnetic field, ${\bf
  e}$/${\bf b}$, is characterized by (what we call) the {\it dual
  electromagnetic fields}, i.e.  $\bar{\cal E}^{\alpha}$/$\bar{\cal
  B}^{\alpha}$.  Since the linear response condition of small ${\bf e}$ and
${\bf b}$ precisely coincides with the condition of validity of the
gradient expansion, these linear response expressions are 
{\it asymptotically exact} for any 
quasi-particle on a Fermi surface at zero temperature.  Based on this
theoretical observation, we also tried to give some rough idea of how to
measure the $SU(2)$ Berry's curvatures in terms of photoemission
experiments, so as to provide some future directions for the (spin)
galvanomagnetic community (see sec.~IV).

Another important achievement is our derivation of the effective
equation of motion (EOM) for a quasi-particle in a $U(1)$ FL {\sl up to
  the 2nd order accuracy in gradient expansion} (see sec.~V).  From this
equation, one sees that {\sl both} the ``$k$-space magnetic field'' and
its corresponding temporal component, dubbed as ``electric field'',
contribute to the Lorentz force which acts on quasi-particles in
$k$-space.  In the linear response regime, this effective EOM is indeed
consistent with the exact many-body formula for the static transverse
conductivity (see appendix B).  

We conclude with a discussion of future problems and issues.  First, one
ambiguity remains unresolved in our derivation of the reduced EOM.
Specifically, working only to 2nd order in the gradient expansion, we
cannot exclude the possibility that ${\bf v}_{\alpha}$ in
eq.~(\ref{5-6a}) may be more generally replaced by $\frac{dR}{dT}$.
These two forms are equivalent up to 2nd order but not more generally. To fix
this ambiguity, a further analysis including 3rd order effects in the
gradient expansion would be necessary, which we leave for future work.
   
A more pressing issue is the well-controlled treatment of the
anti-hermitian part of the collisional self-energies.  Having completely
ignored the corresponding terms {\it by hand}, our results are valid, in
a strict sense, only for those quasi-particles exactly {\it on} a Fermi
surface and at {\it zero} temperature.  Since any physical quantities do
involve non-zero excitation of quasiparticles (even in linear response),
this is clearly not satisfactory.  While one might na\"ively expect that
collisional effects can be included by direct analogy to classical
kinetics (e.g. in the relaxation time or more sophisticated
approximations), the possibility of some more interesting interplay
between collisions and the Berry phase physics captured herein cannot be
excluded.  This may be of particular interest if one considers the {\sl
  instabilities} of conventional metals toward ordered states such as
superconductors, dielectrics, etc.  A treatment of both the
anti-hermitian part (decay of quasi-particles) and hermitian part
(adiabatic transport of quasi-particles) of the RKE {\it on an equal
  footing} is certainly warranted in the future \cite{haldane2}.

\acknowledgments 

The authors are pleased to acknowledge F. Duncan M.  Haldane, Qian Niu,
Congjun Wu, Jiangping Hu, Akio Kimura, Takayuki Kiss, Syunsuke Tsuda,
Allan H. MacDonald and Akira Furusaki for their 
discussions and encouragements.  The 
authors especially would like to thank Q. Niu for drawing our attention 
to the temporal component of $\hat{\cal M}_{\alpha}$,  and to C.j. Wu 
for the discussion about the source of the dual electric fields.  
R. S. appreciated informations from A. Kimura, S. Tsuda and T.
Kiss about photoemission experiments under external electromagnetic fields. 
This work was supported by NSF Grant DMR04-57440, PHY99-07949,
and the Packard Foundation. R.S. was supported by JSPS 
Research Fellowships for Young Scientists.

\appendix 
\section{Brief review of Keldysh equation and collisional self-energies} 

To make this paper self-contained, we will briefly 
review the derivation of the Keldysh equation~\cite{KB}, defining 
several kinds of self-energies introduced in section II, 
such as $\hat{\Sigma}^{\rm HF},$ 
$\hat{\sigma}$ and $\hat{\Sigma}^{<,>}_{\rm c}$.  
In addition to this, we will also look into the  
boundary condition especially imposed on the collisional 
self-energy $\hat{\Sigma}^{<,>}_{\rm c}$. 
This boundary condition with additional arguments validates the 
dissipationless Keldysh equation for green functions, i.e. 
eqs.~(\ref{5-1-a},\ref{6-17a}), starting from which we derived 
the $U(1)$ effective Boltzmann equation in sec.V.   

In the next subsection, we will begin with  
the time-ordered Green function defined 
on the {\it imaginary time domain}  
$G(1,1')$ (see eq.~(\ref{6-1}) for its definition).    
Specifically, the Dyson equation for this 
temperature (Matsubara) Green function is 
derived first. The Hartree-Fock and collisional 
self-energy encoded there are also  
iteratively defined in terms of this Matsubara  
Green function (see eqs.~(\ref{6-5}-\ref{6-6a})).    

The time-ordered Green function 
as a function of imaginary 
time $t$ is {\it analytic}  
separately in two region:   
${\rm Im}\ t\in [-\beta,0]$ and $[0,\beta]$. 
In other words, 
we can introduce two functions, 
usually dubbed as {\it lesser} and {\it greater}  
Green functions, which coincide 
with this time-ordered one  
{\sl and} which are analytic 
in these two regions respectively  
 \begin{eqnarray} 
G(1,1') &=&\left\{ \begin{array}{ll}
G^{>}(1,1')& \hspace{0.5cm}{\rm for} 
\hspace{0.5cm} {\rm Im}\ t_1 < {\rm Im}\ t_{1'}, \\  
- G^{<}(1,1')& \hspace{0.5cm}{\rm for} 
\hspace{0.5cm} {\rm Im}\ t_{1'} < {\rm Im}\ t_1, \\ 
\end{array} \right. \nonumber   
\end{eqnarray}
(compare eq.~(\ref{6-1}) with 
eqs.~(\ref{6-7-a},\ref{6-7})).       
In the 2nd subsection of this appendix,  
keeping the analyticity 
of these two, we will extend 
its time domain from the 
imaginary time into the 
{\it real time domain}. 
Correspondingly, 
the Dyson equation is also analytically continued 
onto the real time domains  
(see eqs.~(\ref{6-14},\ref{6-15a},\ref{6-15b})).   

We will see in the final subsection that periodic 
boundary condition imposed along the 
{\it imaginary time axis} relates the lesser and greater collisional 
self-energies defined {\it on the real time domain} 
with each other. 
(see eq.~(\ref{6-18a})). 
This relation guarantees that, at zero temperature 
and at equilibrium,  
lesser and greater self-energies have no weight 
at $\omega < \mu$ and $\omega > \mu$ respectively, 
i.e. eqs. (\ref{6-16},\ref{6-17}). 
These observations allow us to approximate 
the Keldysh equation derived further into the so-called 
dissipationless Keldysh equation 
(see the arguments from eq.~(\ref{6-15a}) to eq.~(\ref{6-17a})).

\subsection{Dyson equation for temperature Green function} 
Let us begin with the imaginary time Green function defined in the ``interaction'' representations:
\bea
G(1,1';\phi,t_0)&\equiv& \frac{1}{i}\frac{\bla 
{\cal T}\big\{\hat{S}\ \psi(1)\psi^{\dagger}(1')\big\}\bra}
{\bla {\cal T}\big\{\hat{S}\big\}\bra} \nonumber \\ 
 &=&  \frac{1}{i}\frac{{\rm Tr}\big[e^{-\beta({\cal H}-\mu N)}  
{\cal T}\big\{\hat{S}\ \psi(1)\psi^{\dagger}(1')\big\}\big]}
{{\rm Tr}\big[e^{-\beta ({\cal H}-\mu N)} {\cal T}\big\{\hat{S}\big\}\big]},
 \nonumber \\
\hat{S} &\equiv& {\rm exp}\Big[-i\ \int_{t_0}^{t_0-i\beta} d2\ \phi(2)\cdot \hat{n}(2)\Big]. 
\label{6-1} 
\eea
Note that the argument of the fermion operator ``$j$'' 
is an abbreviation of $(r_j,t_j,\alpha_j)$.  
Accordingly, $\int_{t_0}^{t_0-i\beta} dj$ includes not only 
the integral with respect to the imaginary time $t_j$, 
but also the summation over the band indices $\alpha_j$ and 
the integral over the spatial coordinate;
\bea
\int_{t_0}^{t_0-i\beta} dj \equiv \sum_{\alpha_j} 
\int_{t_0}^{t_0-i\beta}dt_j \int d r_j. \nonumber 
\eea
$``{\cal T}$'' is the imaginary-time-ordering 
operator along $\big[t_0,t_0-i\beta\big]$,   
where $t_0$ is always real-valued  
(In the next subsection, we will set  
$t_0$ to $-\infty$, {\sl after} analytically 
continue $t_1$ and $t_{1'}$ onto the real time domain.);   
\bea
{\cal T}\big\{\psi(1)\ \psi(1')\big\} 
\equiv \left\{ \begin{array}{ll}
\psi(1)\ \psi(1')  &\hspace{0.1cm} 
{\rm for}\hspace{0.1cm} 
{\rm Im}\ t_1 < {\rm Im}\ t_{1'}, \\
- \psi(1')\ \psi(1) &\hspace{0.1cm} 
{\rm for}\hspace{0.1cm} 
{\rm Im}\ t_{1'} < {\rm Im}\ t_{1}. \\
\end{array} \right. \nonumber 
\eea 
The time dependence of the fermion 
operator $\psi(1)\equiv \psi_{\alpha_1}(r_1,t_1)$ is specified by 
${\cal H}\equiv {\cal H}_0 + {\cal H}_1$: 
\begin{eqnarray}
-i\partial_{t_1} \psi(1) = \big[{\cal H},\psi(1)\big],  \nonumber 
\end{eqnarray}
with ${\cal H}_0$ and ${\cal H}_1$ given in 
eqs.~(\ref{non-interacting},\ref{interacting}). 
Then taking the $t_1$-derivative of eq.~(\ref{6-1}), 
we have  the following EOM for this 1-point Green function;   
\bea
&&\hspace{-0.6cm}
\int_{t_0}^{t_0-i\beta}d\bar{1}\ 
G^{-1}_{0}(1,\bar{1}) 
G(\bar{1},1')\  =\ \delta(1-1')\nonumber \\ 
&&\hspace{-0.2cm} -i\int d2 V(1,2) 
G_2(1,2,1',2+)_{|t_2=t_1-i|\epsilon|},  
\label{6-2} 
\eea
where the inverse of a bare Green function 
$G^{-1}_{0}(1,1')$ and two-point Green 
function are defined as follows; 
\bea
&&\hspace{-0.1cm} G^{-1}_{0}(1,1') \equiv  
\big[(i\partial_{t_1}   
- \phi(1)) \delta_{\alpha_1\alpha_{1'}}
- \hat{H}_{0}\big] \delta(1-1'), \nonumber \\ 
&&\hspace{-0.6cm} G_2(1,2,1',2') \equiv - 
\frac{\bla{\cal T}\big\{\hat{S} 
\psi(1) \psi(2) \psi^{\dagger}(2') \psi^{\dagger}(1')
\big\}\bra}{\bla{\cal T}\big\{\hat{S}\big\}\bra} 
. \nonumber 
\eea    
``2+'' in eq.~(\ref{6-2}) means 
that its temporal argument is chosen to be 
infinitesimally {\it later} than $t_2$   
along the imaginary time axis; 
$2+\equiv (\alpha_2,r_2,t_2-i|\epsilon|)$.  

The auxiliary scalar potential $\phi(2)$ entering 
into eq.~(\ref{6-1}) 
as an ``interaction''    
is utilized so as to describe  
the above 2-point Green function in terms of 
the {\it self-energy} $\Sigma$ and 
1-point Green's function $G$;   
\bea 
&&\hspace{-0.9cm} 
\int_{t_0}^{t_0-i\beta}d\bar{1}\  
\big[G^{-1}_{0}(1,\bar{1}) 
- \Sigma(1,\bar{1})\big]G(\bar{1},1') 
= \delta(1-1'). \label{6-3} 
\eea
Notice first that the functional derivative 
of the 1-point Green function with respect to 
$\phi(2)$ brings about the 2-point Green 
function;   
\bea
&&\hspace{-0.6cm}
G_2(1,2,1',2+) = \big[G(2,2+) - \frac{\delta}{\delta \phi(2)}
\big] G(1,1').  \nonumber 
\eea    
Thus, substituting this into eq.~(\ref{6-2}), we 
obtained a closed equation for the  
1-point Green function;  
\bea
&&\hspace{-0.9cm}
\int_{t_0}^{t_0-i\beta}d\bar{1}\ G^{-1}_{0}(1,\bar{1})
G(\bar{1},1') = \delta(1-1') \nonumber \\
&&\hspace{-1.2cm} - i \int d2\ V(1,2) \Big[G(2,2+) 
- \frac{\delta}{\delta \phi(2)}\Big]_{|t_2=t_1-i|\epsilon|} 
G(1,1') \label{6-4}.   
\eea

This self-consistent equation can be 
readily transcribed into  
that for the self-energy $\Sigma$.  
Matrix-multiplying eq.~(\ref{6-4}) 
by $G^{-1} \equiv G^{-1}_{0} - \Sigma$, 
we first obtain the following; 
\bea
&&\hspace{-0.2cm}\Sigma(1,1') = 
-i\delta(1-1') \int d2\ V(1,2) 
G(2,2+)_{|t_2=t_1-i|\epsilon|} \nonumber \\
&&\hspace{-0.2cm} - i \int_{t_0}^{t_0-i\beta} d\bar{1} 
\int d2\ V(1,2)  
G(1,\bar{1})
\frac{\delta G^{-1}(\bar{1},1')}
{\delta \phi(2)}_{|t_2=t_1-i|\epsilon|}.   \nonumber 
\eea 
Using $G^{-1} = G^{-1}_{0} - \Sigma$ and $\frac{\delta G^{-1}_0(1,1')}
{\delta \phi(2)} = - \delta(1-1')\delta(1-2)$ 
in the right hand side above, 
the following iterative equation for the 
self-energy $\Sigma$ is derived: 
\bea
\Sigma(1,1') &=& 
-i\delta(1-1') \int d2\ V(1,2) 
G(2,2+)_{|t_2=t_1-i|\epsilon|} \nonumber \\
&&\hspace{-1.5cm} +\ i \delta(t_1-t_{1'}) V(1,1') 
G(1,1')_{|t_{1'}=t_1-i|\epsilon|} \nonumber \\
&&\hspace{-1.5cm} +\ i \int_{t_0}^{t_0-i\beta} 
d\bar{1} \int d2\ 
V(1,2) G(1,\bar{1})
\frac{\delta \Sigma(\bar{1},1')}
{\delta \phi(2)}_{|t_2=t_1-i|\epsilon|}. \label{6-5}
\eea 

Being proportional to $\delta(t_1-t_{1'})$, 
the first 2 terms are temporally instantaneous, which
we usually call the Hartree-Fock term;
\begin{eqnarray}
\Sigma^{\rm HF} &\equiv& -i\delta(1-1') \int d2\ V(1,2) 
G(2,2+)_{|t_2=t_1-i|\epsilon|} \nonumber \\
&& \ \ +\ i \delta(t_1-t_{1'}) V(1,1') 
G(1,1')_{|t_{1'}=t_1-i|\epsilon|}.  \label{6-6}
\end{eqnarray}
A successive iteration through the 
3rd term of eq. (\ref{6-5}) leads
to the self-energy with 
higher order in $V$, which 
is clearly temporally non-instantaneous.  
We dub this non-instantaneous 
self-energy as the collisional 
self-energy henceforth; 
\bea
\Sigma_{\rm c}(1,1') \equiv \Sigma(1,1') - \Sigma^{\rm HF}(1,1'). 
\label{6-6a}
\eea 
Eq.~(\ref{6-3}) in combination 
with eq.~(\ref{6-5}) is the Dyson equation for the 
imaginary-time 1-point Green function.  
 
\subsection{Analytic continuations}
\subsubsection{analytic continuation of $G^{<}$ and $G^{>}$}
Real-time Green functions are analytically continued from the 
following 
lesser and greater Green functions defined on the imaginary time 
domain; 
\begin{widetext}
\bea
G^{>}(1,1';\phi,t_0) &=& - i\frac{\bla {\cal U}(t_0,t_0-i\beta)
\ {\cal U}^{-1}(t_0,t_{1})\psi(1)\ 
{\cal U}(t_0,t_{1})\ 
{\cal U}^{-1}(t_0,t_{1'}) \psi^{\dagger}(1')\ {\cal U}(t_0,t_{1'}) \bra}
{\bla {\cal U}(t_0,t_0-i\beta)\ \bra}, \label{6-7-a} \\ 
G^{<}(1,1';\phi,t_0) &=& i\frac{\bla\ {\cal U}(t_0,t_0-i\beta)\  
{\cal U}^{-1}(t_0,t_{1'})\psi^{\dagger}(1')\ {\cal U}(t_0,t_{1'})
\ {\cal U}^{-1}(t_0,t_{1})  \psi(1)\ 
{\cal U}(t_0,t_{1}) \bra} 
{\bla {\cal U}(t_0,t_0-i\beta) \bra}.  \label{6-7} 
\eea   
\end{widetext}
${\cal U}(t,t')\equiv {\cal T}\big\{\exp\big[-i\int_{t}^{t'}d1\phi(1)
\cdot \hat{n}(1)\big]\big\}$  
characterizes the temporal evolution along the imaginary time axis 
$[t_0,t_0-i\beta]$
and thus is not unitary operator. 

Note that, at equilibrium {\sl and} 
with $\phi\equiv 0$, these two functions 
are analytic in ${\rm Im}\ (t_1-t_{1'})\in [-\beta,0]$ and $[0,\beta]$   
respectively, regardless of ${\rm Re}\ t_1$ and ${\rm Re}\ t_{1'}$.  
Then, assuming that these analyticities also hold true 
in a weakly disequilibrated system with finite $\phi$,  
we will introduce {\it separately} the lesser and 
greater Green functions defined on the {\it real time domain} 
as follows;
\begin{eqnarray}
\hspace{-0.2cm}\left\{\begin{array}{l} 
\hspace{0.0cm}{\sf g}^{<}_{\alpha_1\alpha_{1'}}(r_1,r_{1'}:{\rm Re}\ t_1,{\rm Re}\ t_{1'};\phi) \\
\hspace{0.2cm}\equiv \  \lim_{t_{0}\to -\infty} 
G^{<}(1,1';\phi,t_0)_{|\ {\rm Im}\ t_1 ={\rm Im}\ t_{1'} + |\epsilon| = |\epsilon| },  \\  
\hspace{0.0cm}{\sf g}^{>}_{\alpha_1\alpha_{1'}}(r_1,r_{1'}:{\rm Re}\ t_1,{\rm Re}\ t_{1'};\phi) \\
\hspace{0.2cm} \equiv \  \lim_{t_{0}\to -\infty}
G^{>}(1,1';\phi,t_0)_{|\ {\rm Im}\ t_1+|\epsilon|={\rm Im}\ t_{1'}=0},  \\ 
\end{array}  
\right.
\label{6-7a-}
\end{eqnarray}
by taking $\epsilon$ infinitesimally small. 
Choose the test scalar field $\phi(1)$  
such that it vanishes at ${\rm Re}\ t_1\rightarrow -\infty$. 
Then, ${\cal U}(t_0,t_0-i\beta)$ 
appearing in both the denominator and the numerator of 
eqs.~(\ref{6-7-a},\ref{6-7}) 
reduces to unit, when $t_0 \rightarrow -\infty$. 
Accordingly, the two functions in the right hand 
side are solely defined on the real domain;      
\begin{eqnarray}
{\sf g}^{>}(1,1';\phi)&=& - i\bla \psi_{\phi}(1)\ 
\psi_{\phi}^{\dagger}(1') \bra, \label{6-7a} \\ 
{\sf g}^{<}(1,1';\phi) &=& i\bla\ \psi_{\phi}^{\dagger}(1')\ 
\psi_{\phi}(1) \bra  \label{6-8},    
\end{eqnarray} 
with $\psi_{\phi}(1) \equiv {\cal U}^{\dagger}(-\infty,t_1)
\psi(1) {\cal U}(-\infty,t_1)$. 
Now that ${\cal U}(-\infty,t)$ is a unitary operator,  
we can construct a real-time-ordered Green function 
in terms of eqs.~(\ref{6-7a}-\ref{6-8});   
\begin{eqnarray}
{\sf g}(1,1';\phi)&\equiv& i\ \bla {\cal T}\big\{\psi_{\phi}(1)\ \psi_{\phi}(1')\big\}\bra 
\nonumber \\
&\equiv& \left\{ \begin{array}{ll} 
{\sf g}^{>}(1,1';\phi)& \hspace{0.2cm}{\rm for} \hspace{0.2cm} t_1 < t_{1'}, \\  
- {\sf g}^{<}(1,1';\phi)& \hspace{0.2cm}{\rm for}\hspace{0.2cm} t_{1'} < t_1. \\ 
\end{array} \right. \label{6-9}   
\end{eqnarray}

\subsubsection{analytic continuation of $\hat{\Sigma}^{<}_{\rm c}$ and $\hat{\Sigma}^{>}_{\rm c}$}
The Dyson equation being 
composed also of the self-energy, 
let us next look into the analytic continuation 
of the collisional self-energy.   
The analytic property of the collisional 
self-energy can be obtained from 
that of Green functions at equilibrium. 
At equilibrium, both the 1-point
Green function $G(1,1')$  and  the 2-point 
Green function   
$G_{2}(1,2,1',2+i|\epsilon|)_{|t_2=t_1-i|\epsilon|}$  
are analytic in two regions, i.e.  
${\rm Im}\ (t_1-t_{1'})\in [-\beta,0]$ 
and $[0,\beta]$, separately. 
Compare these analyticities with 
the relation among the self-energy, 
1-point Green function and 2-point Green function;   
\begin{eqnarray}
&&\int_{t_0}^{t_0-i\beta} d2\ V(1,2)\ G_{2}(1,2,1',2+i|\epsilon|)_{|t_2=t_1-i|\epsilon|} \nonumber \\
&&\equiv \int_{t_0}^{t_0-i\beta} 
d2\ \Sigma(1,2)\cdot G(2,1'). \label{6-10} 
\end{eqnarray}  
Then one can readily see that $\Sigma_{\rm c}(1,1')$ is 
also analytic both at these two regions,  
but separately.  Thus, we can introduce the 
{\it lesser} and {\it greater} self-energy, such that 
they are analytic in the following two regions 
respectively;  
\begin{eqnarray}
\Sigma_{\rm c}(1,1') \equiv \left\{ \begin{array}{ll}
 \Sigma_{\rm c}^{>}(1,1') & \hspace{0.1cm}{\rm for} 
 \hspace{0.2cm} {\rm Im}\ t_1 < {\rm Im}\ t_{1'}, \\  
 - \Sigma_{\rm c}^{<}(1,1') & \hspace{0.1cm}{\rm for} 
 \hspace{0.2cm} {\rm Im}\ t_{1'} < {\rm Im}\ t_1. \\ 
\end{array} \right. \nonumber  
\end{eqnarray}  
Exactly in a same way as we did for the 1-point 
Green function, we can then formally extend 
these two self-energies onto the real time domain 
separately ;
\begin{eqnarray}
\hspace{-0.2cm}\left\{ 
\begin{array}{l} 
{{\Sigma}_{\rm c}^{<}}_{\alpha_1,\alpha_{1'}}
(r_1,r_{1'}:{\rm Re}\ t_1,{\rm Re}\ t_{1'};\phi) \\
\hspace{0.2cm} \equiv \lim_{t_{0}\rightarrow -\infty} 
{\Sigma}_{\rm c}^{<}(1,1';\phi,t_0)
_{|\ {\rm Im}\ t_1 ={\rm Im}\ t_{1'} + |\epsilon| = |\epsilon| } \\  
{{\Sigma}_{\rm c}^{>}}_{\alpha_1,\alpha_{1'}}
(r_1,r_{1'}:{\rm Re}\ t_1,{\rm Re}\ t_{1'};\phi) \\
\hspace{0.2cm}  \equiv \lim_{t_{0}\rightarrow -\infty}
{\Sigma}_{\rm c}^{>}
(1,1';\phi,t_0)_{|\ {\rm Im}\ t_1+|\epsilon|={\rm Im}\ t_{1'}=0}. \\ 
\end{array} 
\right.  \label{6-10a-}
\end{eqnarray}

\subsubsection{analytic continuation of Dyson equation}
\begin{figure}
\begin{center}
\includegraphics[width=0.45\textwidth]{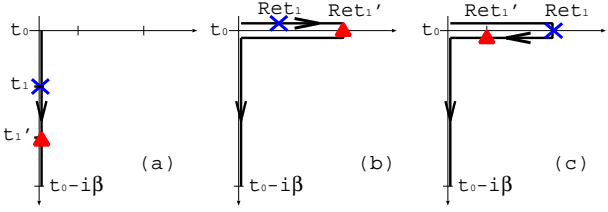}
\end{center}
\caption{(a): The integral path for 
eq.~(\ref{6-10a}). (b): Integral path for eq.~(\ref{6-12}) in 
the case of Re $t_1 <$ Re ${t_1}'$. (c) That for Re $t_1 >$ Re ${t_1}'$.}
\label{fig5}
\end{figure}

Now that both the self-energy and Green functions are 
analytically continued onto the real-time domain, 
we will derive the Dyson equation for these real-time functions.  
We begin with that for the imaginary time domain;
\begin{eqnarray} 
&&\big(\hat{G}_{0}^{-1}-\hat{\Sigma}^{\rm HF}\big)\cdot G^{<}(1,1')
\nonumber \\
&& =\int_{t_0}^{t_0-i\beta} d\bar{1}\ 
\Sigma_{\rm c}(1,\bar{1})\ G(\bar{1},1'). \label{6-10a}
\end{eqnarray}
Decompose first the right hand side into three terms,  
such that each term is expressed solely 
in terms of lesser / greater Green 
function / collisional  self-energy.   
Namely, depending on whether $t_{\bar{1}}$ 
locates within $[t_0,t_{1}]$, 
$[t_{1},t_{1'}]$ or $[t_{1'},t_0-i\beta]$, 
the right hand side can be divided into the following three 
parts; 
\begin{eqnarray}
&&\hspace{-0.5cm}
\big(\hat{G}_{0}^{-1}-\hat{\Sigma}^{\rm HF}\big)\cdot G^{<}(1,1') 
\nonumber \\
&&\hspace{-0.5cm} =\int_{t_0}^{t_{1}}d\bar{1}\ 
\Sigma_{\rm c}^{>}(1,\bar{1}) G^{<}(\bar{1},1') 
\  + \int_{t_{1}}^{t_{1'}}d\bar{1}\ 
\Sigma_{\rm c}^{<}(1,\bar{1}) G^{<}(\bar{1},1') \nonumber \\
&&\hspace{-0.1cm}  + \int_{t_{1'}}^{t_0-i\beta}d\bar{1}\ 
\Sigma_{\rm c}^{<}(1,\bar{1}) G^{>}(\bar{1},1'), 
 \label{6-11}
\end{eqnarray}     
where the integral path is still along $[t_0,t_0-i\beta]$ 
(see Fig.~\ref{fig5}(a)). 
Then, we will deform this integral path 
so that both $t_1$ and $t_{1'}$ reach the real-time 
axis. During this deformation, however,  
to keep the analyticity for all the Green 
functions and self-energies in the both 
right and left hand sides,  
we have to observe the following condition;
\begin{eqnarray}
-\beta<{\rm Im}\ t_{1'}<{\rm Im}\ t_{1} < 0.   \nonumber 
\end{eqnarray}
Due to this condition, for given ``target'' 
Re $t_1$ and Re $t_{1'}$, our integral path 
is uniquely deformed into a ``L'' shaped 
path depicted in Fig.~\ref{fig5}(b-c); 
\begin{eqnarray}
&&\hspace{-0.5cm}\big(\hat{G}_{0}^{-1}-\hat{\Sigma}^{\rm HF}\big)\cdot 
G^{<}(1,1') \nonumber \\
&&\hspace{-0.6cm} =\int_{t_0}^{t_1}d\bar{1}\ 
{\Sigma}_{\rm c}^{>}(1,\bar{1}) G^{<}(\bar{1},1') 
 + \int_{t_1}^{t_{1'}}d\bar{1}\ 
{\Sigma}_{\rm c}^{<}(1,\bar{1}) G^{<}(\bar{1},1') + 
\nonumber \\
&&\hspace{-0.7cm}\int_{t_{1'}}^{t_0-i0} d\bar{1}\ 
{\Sigma}_{\rm c}^{<}(1,\bar{1}) G^{>}(\bar{1},1') 
 + \int_{t_0-i0}^{t_0-i\beta}d\bar{1}\ 
{\Sigma}_{\rm c}^{<}(1,\bar{1})G^{>}(\bar{1},1'). \nonumber \\
\label{6-12} 
\end{eqnarray} 

Note that the 1st three integral paths in the right hand 
side are all along the real time domain, while the last one 
is strictly along $[t_0,t_0-i\beta]$.  
So as to eliminate the final term, we will 
take  $t_0$  to be $-\infty$. In this limit,  
$G^{>}(\bar{1},1';\phi)$ in the last term 
should vanish, since 
$|{\rm Re}\ t_{\bar{1}}- {\rm Re}\ t_{1'}|  
\rightarrow \infty$.   
This procedure simultaneously 
completes the analytic continuations 
of all the Green functions / collisional 
self-energies encoded in this equation 
(see eqs.~(\ref{6-7a-},\ref{6-10a-}));
\begin{eqnarray}
&&\hspace{-0.5cm}
\big(\hat{G}_{0}^{-1}-\hat{\Sigma}^{\rm HF}\big)\cdot {\sf g}^{<}(1,1') 
\nonumber \\
&&\hspace{-0.5cm} =\int_{-\infty}^{t_1}d\bar{1} 
{\Sigma}_{\rm c}^{>}(1,\bar{1})\ 
{\sf g}^{<}(\bar{1},1') \  + \ \int_{t_1}^{t_{1'}}d\bar{1}  
{\Sigma}_{\rm c}^{<}(1,\bar{1})\ {\sf g}^{<}(\bar{1},1')  \nonumber \\
&&\hspace{0.0cm} +\ \int_{t_{1'}}^{-\infty} d\bar{1}\ 
{\Sigma}_{\rm c}^{<}(1,\bar{1})\ {\sf g}^{>}(\bar{1},1')  \nonumber \\
&&\hspace{-0.5cm} 
= \int_{-\infty}^{t_1}d\bar{1}\  
\big[{\Sigma}_{\rm c}^{>}(1,\bar{1}) - 
{\Sigma}_{\rm c}^{<}(1,\bar{1})\big]\ 
{\sf g}^{<}(\bar{1},1) \nonumber \\
&&\hspace{0.0cm} 
- \int_{-\infty}^{t_{1'}}d\bar{1}\ 
{\Sigma}_{\rm c}^{<}(1,\bar{1})\  
\big[{\sf g}^{>}(\bar{1},1')-{\sf g}^{<}(\bar{1},1')\big]. \label{6-13}
\end{eqnarray}   
In a similar way, we can easily obtain 
an equation of motion for the greater Green function; 
\begin{eqnarray}
&&\hspace{-0.5cm}
\big(\hat{G}_{0}^{-1}-\hat{\Sigma}^{\rm HF}\big) 
{\sf g}^{>}(1,1') \nonumber \\ 
&&\hspace{-0.5cm} =\ \int_{-\infty}^{t_1}d\bar{1} 
\big[{\Sigma}_{\rm c}^{>}(1,\bar{1}) - 
{\Sigma}_{\rm c}^{<}(1,\bar{1})\big]\ {\sf g}^{>}(\bar{1},1) 
\nonumber \\
&&\hspace{-0.0cm} 
- \int_{-\infty}^{t_{1'}}d\bar{1} 
{\Sigma}_{\rm c}^{>}(1,\bar{1})\  
\big[{\sf g}^{>}(\bar{1},1')-{\sf g}^{<}(\bar{1},1')\big]. \label{6-14}
\end{eqnarray}   

At equilibrium, both Green 
function and self-energy become translationally 
invariant in space and time, e.g.   
\begin{eqnarray}
{\sf g}^{<(>)}(1,1';\phi\equiv 0) &=& 
{\sf g}^{<(>)}_{\alpha_1,\alpha_{1'}}(r_1-r_{1'},t_1-t_{1'}),  
\nonumber \\
{\Sigma}_{\rm c}^{<(>)}(1,1';\phi\equiv 0) &=& 
{{\Sigma}_{\rm c}^{<(>)}}_{\alpha_1,\alpha_{1'}}(r_1-r_{1'},t_1-t_{1'}). \nonumber 
\end{eqnarray}  
Thus, when Fourier-transformed with respect to  
these relative coordinates,  the convolutions 
appearing in eq.~(\ref{6-14}) would have 
reduced to a {\sl simple product}, 
if the associated integral region 
were to range  $[-\infty,\infty]$.  
In reality, however, their 
temporal integral regions do not range 
from $-\infty$ to $+\infty$.   
As a result, when Fourier transformed, 
they end up with several principal integrals 
with respect to the frequency.     
The {\it hermitian} part of 
the collisional self-energy and 
Green function {\it solely} 
take over these principal integral parts. 
To be specific, introduce the following 
two functions; 
\begin{eqnarray}
{\sf b}(1,1') &\equiv& \frac{1}{2}\frac{t_1-t_{1'}}{|t_{1}-t_{1'}|}
\big[ {\sf g}^{>}(1,1')-{\sf g}^{<}(1,1')\big], \nonumber \\ 
\sigma(1,1') &\equiv& 
\frac{1}{2}\frac{t_1-t_{1'}}{|t_1-t_{1'}|}
\big[ {\Sigma}_{\rm c}^{>}(1,1')-
{\Sigma}_{\rm c}^{<}(1,1') \big].  \label{6-15}
\end{eqnarray}   
Then we can formally rewrite the above equation in such 
a way that their convolutions with respect to time 
are always taken over $[-\infty,\infty]$;
\begin{eqnarray}
&&\hspace{-1.0cm} 
\big(\hat{G}_{0}^{-1}-\hat{\Sigma}^{\rm HF}\big)\otimes {\sf g}^{<} =
\sigma\otimes {\sf g}^{<} + {\Sigma}_{\rm c}^{<}\otimes {\sf b}
\nonumber \\
&&\hspace{2.2cm} 
+\frac{1}{2} \big({\Sigma}_{\rm c}^{>}\otimes {\sf g}^{<}  
- {\Sigma}_{\rm c}^{<}\otimes {\sf g}^{>}\big)\label{6-15a}, \\ 
&&\hspace{-1.0cm} 
\big(\hat{G}_{0}^{-1}-\hat{\Sigma}^{\rm HF}\big)\otimes {\sf g}^{>} =
\sigma\otimes {\sf g}^{>} + {\Sigma}_{\rm c}^{>}\otimes {\sf b}
\nonumber \\
&&\hspace{2.2cm} 
+\frac{1}{2} \big({\Sigma}_{\rm c}^{>}\otimes {\sf g}^{<}  
- {\Sigma}_{\rm c}^{<}\otimes {\sf g}^{>}\big).\label{6-15b}     
\end{eqnarray}    
where we used the following abbreviated notations;
\begin{eqnarray}
(A\otimes B)(1,1')\equiv \int_{-\infty}^{\infty} 
d\bar{1}\ A(1,\bar{1}) 
B(\bar{1},1').  \nonumber 
\end{eqnarray}  


\subsubsection{dissipationless Keldysh equation}
Eqs.~(\ref{6-15a},\ref{6-15b}) are what we call the Keldysh equation, 
from which we readily obtain eq.~(\ref{1-6})\cite{note3}. Notice that 
the re-expression in terms of ${\sigma}$ and $\sf b$ is not 
just for its formality, but is in fact an important  
step to approximate this Keldysh equation, based on 
physical arguments. 
As we see in the next subsection, at zero temperature 
and at equilibrium, the periodic boundary condition 
along the imaginary time domain ensures that, 
when Fourier-transformed with respect to 
their relative coordinates, the lesser/greater 
collisional self-energy vanishes   
for those $\omega$ greater/less than 
chemical potential $\mu$ respectively;   
\begin{eqnarray}
\hat{\Sigma}^{<}_{\rm c}(q,\omega) &=& \hat{0} \hspace{0.3cm} {\rm for} 
\hspace{0.2cm} \omega > \mu, \label{6-16} \\ 
\hat{\Sigma}^{>}_{\rm c}(q,\omega) &=& \hat{0} \hspace{0.3cm} {\rm for} 
\hspace{0.2cm} \omega < \mu. \label{6-17}  
\end{eqnarray}
Thereby, provided that a system is 
weakly disturbed from this equilibrated situation, 
we could still assume that these relations 
hold true for the collisional self-energy, 
with $\mu$ generalized into a function 
depending on $R$ and $T$; $\mu \rightarrow \mu(R,T)$.  
Observing further the continuity of these 
collisional self-energies around 
$\omega \simeq \mu$, we could regard 
both of them to be sufficiently small 
quantities at these low frequency regions, 
i.e. $\omega \simeq \mu$. 

The hermitian part of the 
self-energy $\hat{\sigma}(\omega)$, however, 
is not necessarily a small quantity, 
even for $\omega \simeq \mu$.  
Namely, when Wigner-transformed, 
those lesser/greater self-energies for high 
frequency regions do contribute to  
$\hat{\sigma}(\omega)$ at the low 
frequency region \cite{note4};
\begin{eqnarray}
&&\hspace{-0.7cm}
\hat{\sigma}(q,R;\omega,T) 
= \nonumber \\
&&\hspace{-0.3cm}
\int_{-\infty}^{+\infty}\ \frac{d\omega'}{2\pi} 
\frac{\cal P}{\omega-\omega'}
\big(\hat{\Sigma}^{<}_{\rm c}+ 
\hat{\Sigma}^{>}_{\rm c}\big)(q,R;\omega',T). \nonumber  
\end{eqnarray}

Due to this different behavior between  
$\sigma(\omega)$ and 
$\Sigma^{<(>)}_{\rm c}(\omega)$ 
around $\omega \simeq \mu$,  
we can approximate eqs.~(\ref{6-15a},\ref{6-15b}) into 
the following dissipationless Keldysh equations;
\begin{eqnarray}
&& \big(\hat{G}_{0}^{-1}-\hat{\Sigma}^{\rm HF} - 
\hat{\sigma}\big)\otimes \hat{\sf g}^{<(>)} = 0. 
\label{6-17a} 
\end{eqnarray}
which is supposed to be still valid for 
$\omega \simeq \mu$ at sufficiently 
low temperature.  

\subsection{periodic boundary conditions}
Observing eq.~(\ref{6-10}), we will first 
read the boundary condition (b.c.) for the 
collisional self-energy, out of  
those for the 1-point and 2-point Green 
functions. 
Notice that the 1-point Matsubara 
Green functions defined in eqs.~(\ref{6-7-a}-\ref{6-7}) 
obey the following boundary condition at 
equilibrium;
\begin{eqnarray} 
G^{<}(1,1')=-e^{\beta\mu}G^{>}(1-i\beta,1') \label{6-18} 
\end{eqnarray}   
for $-\beta < {\rm Im}$ $t_{1'} < {\rm Im}$ $t_{1} < 0$ and 
arbitrary ${\rm Re}$ $t_{1}$ and ${\rm Re}$ $t_{1'}$.   
In a similar way, we can also see from its definition 
that the 2-point Matsubara Green function observes, 
\begin{eqnarray}
&& G_{2}(1,2,1',2+i|\epsilon|)_{|t_2=t_1-i|\epsilon|} \nonumber \\
&&\ \ =\ -e^{\beta\mu}
G_{2}(1-i\beta,2,1',2+i|\epsilon|)_{|t_2=t_1-i\beta-i|\epsilon|}  
\nonumber 
\end{eqnarray}  
for $-\beta<{\rm Im}\ t_{1'}<{\rm Im}\ t_1< 0$ and arbitrary ${\rm Re}\ t_1$ 
and ${\rm Re}\ t_{1'}$. 
When compared with eq.~(\ref{6-10}),  
these two then lead to the following 
boundary condition for the collisional self-energy; 
\begin{eqnarray}
\Sigma_{c}(1,1')  =
 -e^{\beta\mu}\Sigma_{c}(1-i\beta,1'). \nonumber
\end{eqnarray}

While being imposed on the imaginary time 
direction, this equation becomes 
useful, when Fourier-transformed with 
respect to ${\rm Re}t_1-{\rm Re}t_{1'}$~\cite{note4};
\begin{eqnarray}
\Sigma^{<}_{\rm c}(r_1,r_{1'};\omega) = e^{-\beta(\omega - \mu)} 
\Sigma^{>}_{\rm c}(r_1,r_{1'};\omega). \label{6-18a} 
\end{eqnarray} 
Namely, eq.~(\ref{6-18a}) requires both of them to strictly 
observe eqs.~(\ref{6-16},\ref{6-17}) 
at zero temperature, since either $\Sigma^{<}_{\rm c}$ or 
$\Sigma^{>}_{\rm c}$ can not be singular in a usual metal.  
 
\section{Consistency with Ishikawa-Matsuyama-Haldane formula}
The conductivities being given by the current-current correlation  
functions, they are usually subject under the vertex corrections 
in interacting Fermi systems. 
However, when it comes to the static and transverse component, 
the Ward identity relates this vertex part with  
the derivative of the 1-point Green function, such that the Hall 
conductivity is expressed solely in terms of the 
single-particle Green function;
\begin{eqnarray}
\sigma_{\lambda} = \frac{e^2}{2\hbar}
\frac{\epsilon_{\lambda\mu\nu}}{(2\pi)^d} 
\int dk \int \frac{d\omega}{2\pi} e^{i\omega 0+} 
{\rm Tr}\big[\frac{\partial \hat{g}}{\partial \omega} 
\frac{\partial \hat{g}^{-1}}{\partial k_{\mu}}\hat{g} 
\frac{\partial \hat{g}^{-1}}{\partial k_{\nu}}\big], \label{8-1}
\end{eqnarray}
which is known as the Ishikawa-Matsuyama formula~\cite{IMH,haldane1}.      
In this appendix, we will show that our derived  
expression for the $U(1)$ Hall conductivity, 
i.e. eq.~(\ref{5-7}), is in indeed consistent 
with this many-body formula, using only Fermi 
liquid assumptions. 
Namely, we will assume that, when diagonalized, 
each eigenvalue for the time-ordered 1-point 
Green function has a pole sufficiently closed to 
the real-axis, such that the corresponding 
quasi-particle life time is infinitely long;
\begin{eqnarray}
&&\hspace{-0.4cm}
\big[\hat{g}_{d}\big]_{\alpha\alpha}\equiv 
\big[\hat{U}^{-1}\hat{g}\ \hat{U}\big]_{\alpha\alpha} \nonumber \\
&&\ \ = \frac{1}{\omega - E_{\alpha,k}(\omega) - i{\rm sign}(\omega -\mu)\cdot 0+}. 
\nonumber 
\end{eqnarray}   
We can identify $\hat{U}$ above as the unitary matrix 
diagonalizing our Lagrangian $\hat{\sf L}$.  

Specifically, let 
us choose the basis in eq.~(\ref{8-1}), such that 
the Green function is diagonalized;  
\begin{widetext}
\begin{eqnarray}
&&\hspace{-1.6cm}\sigma_{\lambda}= \frac{e^2}{2\hbar}
\frac{\epsilon_{\lambda\mu\nu}}{(2\pi)^d} 
\int dk \int \frac{d\omega}{2\pi}  
e^{i\omega 0+}\ {\rm Tr}\Big[
\big\{(\partial_{\omega}\hat{U})\ \hat{g}_{d}\ \hat{U}^{-1} 
+ \hat{U}\ (\partial_{\omega}\hat{g}_{d})\ \hat{U}^{-1} 
+ \hat{U}\ \hat{g}_{d}\ (\partial_{\omega}\hat{U}^{-1})\big\} 
\nonumber \\ 
&&\hspace{3.4cm} 
\big\{(\partial_{k_{\mu}}\hat{U})\ \hat{g}^{-1}_{d}\ \hat{U}^{-1}  
+ \hat{U}\ (\partial_{k_{\mu}}\hat{g}^{-1}_{d})\ \hat{U}^{-1} 
+ \hat{U}\ \hat{g}^{-1}_{d}\ (\partial_{k_{\mu}}\hat{U})^{-1}\big\} 
\nonumber \\
&& \hspace{3.6cm} 
\hat{U}\ \hat{g}_{d}\ \hat{U}^{-1}
\big\{(\partial_{k_{\nu}}\hat{U})\ \hat{g}^{-1}_{d}\ \hat{U}^{-1} 
+ \hat{U}\ (\partial_{k_{\nu}}\hat{g}^{-1}_{d})\ \hat{U}^{-1} 
+  \hat{U}\ \hat{g}^{-1}_{d}\ (\partial_{k_{\nu}}\hat{U}^{-1})\big\} 
\Big]. \label{8-2}
\end{eqnarray}
\end{widetext}

These 27 terms can be classified into 3 types,  
according to the matrix structure within the trace. 
To see this, notice first that, irrespective of 
being differentiated or not, $\hat{g}_d$ and $\hat{g}^{-1}_d$ 
alternate each other within the trace. When we have either 
$\hat{U}^{-1}(\partial_{Q}\hat{U})\equiv {\cal A}_{Q}$ or 
$(\partial_{Q}\hat{U}^{-1})\cdot (\partial_{Q'}\hat{U})$ between 
a pair of neighboring $\hat{g}_d$ and $\hat{g}^{-1}_{d}$, 
this pair of green function and its inverse  
can not be {\it directly-connected}. On the one hand, 
when both $\hat{U}$ and $\hat{U}^{-1}$ between a pair of 
$\hat{g}_d$ and $\hat{g}^{-1}_{d}$ are free from 
the derivative, these two can be clearly 
directly-connected.  We will first 
classify all the terms appearing within  
the above integrand in terms of the  
number of directly-connected pairs of 
$\hat{g}_d$ and $\hat{g}^{-1}_{d}$.

 
One class is those terms having no pair of directly 
connected Green function and its inverse.   
For example, the following term belongs to 
this class;
\begin{eqnarray}
&&\hspace{-0.6cm}{\rm Tr}\big[\hat{U}
(\partial_{\omega}\hat{g}_{d}) 
\hat{U}^{-1} \hat{U}  
\hat{g}^{-1}_{d}(\partial_{k_{\mu}}\hat{U}^{-1}) 
\hat{U}\hat{g}_{d} \hat{U}^{-1}
(\partial_{k_{\nu}}\hat{U})
\hat{g}^{-1}_{d}\hat{U}^{-1} \big] \nonumber \\
&&={\rm Tr}\big[
(\partial_{\omega}\hat{g}^{-1}_{d})\ 
\hat{\cal A}_{k_{\mu}}\ \hat{g}_{d}\  
\hat{\cal A}_{k_{\nu}} \big], \nonumber
\end{eqnarray} 
where one pair of $\hat{g}_{d}$ and $\hat{g}^{-1}_d$
was directly-connected only to reduce into a unit, i.e. 
$\hat{g}_d\cdot\hat{g}^{-1}_{d}=1$. Such an annihilated 
pair is {\it not} regarded as a directly-connected pair.  
Among 27 terms enumerated above, 
there exist 16 terms belonging 
to this class, all of which 
can be summarized into several total 
derivatives;  
\begin{eqnarray}
&&\partial_{k_{\nu}}\Big\{{\rm Tr}
\big[\hat{\cal A}_{\omega}\ \hat{g}_{d}\ 
\hat{\cal A}_{k_{\mu}}\ \hat{g}^{-1}_{d}\big]\Big\} \nonumber \\
&&\ \ + \ \partial_{k_{\mu}}\Big\{{\rm Tr}
\big[\hat{\cal A}_{k_{\nu}}\ \hat{g}_{d}\ 
\hat{\cal A}_{\omega}\ \hat{g}^{-1}_{d}\big]\Big\} \nonumber \\
&& \ \ \ \ + \ \partial_{\omega}\Big\{{\rm Tr}
\big[\hat{\cal A}_{k_{\nu}}\ \hat{g}_{d}\ 
\hat{\cal A}_{k_{\mu}}\ \hat{g}^{-1}_{d}\big]\Big\} \rightarrow 0.  
\nonumber 
\end{eqnarray}
Being contour-integrated, all these terms 
vanish as indicated.    

The 2nd class is those terms having two directly connected 
pairs. Among 27 terms enumerated above, 
there clearly exists only one such a term;
\begin{eqnarray}
&&\hspace{-0.5cm}
{\rm Tr}\big[(\partial_{\omega}\hat{g}_{d})\ 
(\partial_{k_{\mu}}\hat{g}^{-1}_{d})\ 
\hat{g}_{d}\ 
(\partial_{k_{\nu}}\hat{g}^{-1}_{d})\big] \nonumber \\
&& \ \ =\  {\rm Tr}\big[(\partial_{\omega}\hat{g}_{d})\ 
(\partial_{k_{\nu}}\hat{g}^{-1}_{d})\ 
\hat{g}_{d}\ 
(\partial_{k_{\mu}}\hat{g}^{-1}_{d})\big], \nonumber 
\end{eqnarray}
Then, observing the overall factor $\epsilon_{\lambda\mu\nu}$ 
in eq.~(\ref{8-2}), we can readily drop this term. 

The 3rd class consists of 10 terms 
having only one directly connected pair.  
4 terms being canceled by one another, we have 
the following 6 terms remained;
\begin{eqnarray}
&&\hspace{-0.5cm}-2\ {\rm Tr}
\big[\hat{g}^{-1}_{d}(\partial_{\omega} \hat{g}_{d})
\ (\partial_{k_{\mu}}\hat{U}^{-1})
(\partial_{k_{\nu}}\hat{U})\big]  \nonumber \\
&& \hspace{-0.3cm}-2\ {\rm Tr}
\big[\hat{g}^{-1}_{d}(\partial_{k_{\nu}} \hat{g}_{d})
\ (\partial_{\omega}\hat{U}^{-1})
(\partial_{k_{\mu}}\hat{U})\big] \nonumber \\
&& \hspace{-0.1cm}+2\ {\rm Tr}
\big[\hat{g}^{-1}_{d}(\partial_{k_{\nu}} \hat{g}_{d})
\ (\partial_{k_{\mu}}\hat{U}^{-1})
(\partial_{\omega}\hat{U})\big]. \label{8-3}
\end{eqnarray}  

Notice that, when either 
the band index $\alpha$ or its momentum 
$k$ denotes a {\it filled} Bloch state, 
the diagonalized time-ordered  
Green functions has a pole at the {\it upper} half 
plane in the complex $\omega$-plane, say, 
$\omega = \epsilon_{\alpha,k}+i0+$; 
\begin{eqnarray}
\big[\hat{g}^{-1}_{d}(\partial_{k_{\mu}}
\hat{g}_{d})\big]_{\alpha\alpha} 
\equiv  
\frac{(\partial_{k_{\mu}}{\epsilon_{\alpha,k}})}
{\omega - \epsilon_{\alpha,k} - i0+},  \label{8-3a}
\end{eqnarray}   
with $\epsilon_{\alpha,k} - E_{\alpha,k}(\epsilon_{\alpha,k})=0$. 
Note that, in the right hand side, we have assumed the 
infinite life-time for these Bloch states.  
This assumption corresponds to ignoring 
the anti-Hermitian part of the 
collisional self-energy, i.e. $\hat{\Gamma}$. 

Substituting eq.~(\ref{8-3}) into eq.~(\ref{8-2}), 
we then integrate over the frequency, such that the 
summation/integral regions over $\alpha/k$ are 
restricted within the filled Bloch states. 
Using further eq.~(\ref{8-3a}), we then observe that 
the Hall conductivity is indeed characterized by the
dual version of $U(1)$ electromagnetic fields 
introduced in the text; 
\begin{eqnarray}
&&\hspace{0.5cm}\sigma_{\lambda}= 
\frac{e^2}{\hbar}\frac{1}{(2\pi)^d} \sum_{\alpha} \int dk 
\big\{\bar{\cal B}^{\alpha}
- (\bar{\cal E}^{\alpha}\times {\bf v}_{\alpha})
\big\}_{\lambda}, \nonumber \\ 
&&\hspace{0.5cm}\bar{\cal B}^{\alpha}_{\lambda}=i
\epsilon_{\lambda\mu\nu}
\big\{\big[(\partial_{k_{\mu}}\hat{U})^{\dagger} 
(\partial_{k_{\mu}}\hat{U})\big]_{\alpha\alpha}
\big\}_{|\omega = \epsilon_{\alpha,k}}, \nonumber \\
&&\hspace{0.5cm}\bar{\cal E}^{\alpha}_{\lambda} = 
i \big\{\big[(\partial_{\omega}\hat{U})^{\dagger} 
(\partial_{k_{\lambda}}\hat{U}) - 
{\rm c.c.} \big]_{\alpha\alpha}
\big\}_{|\omega=\epsilon_{\alpha,k}}, \nonumber 
\end{eqnarray}    
with 
${\bf v}_{\alpha,\lambda}=
(\partial_{k_{\lambda}}\epsilon_{\alpha,k})$. 

\section{Ampere's law}
In a non-interacting Fermi system, 
the Gauss law solely determines the 
distribution of the $U(1)$ magnetic field. 
When a doubly degeneracy point  
is formed by two neighboring energy dispersions, 
say $\alpha$-th and $(\alpha+1)$-th band 
at $k=k^{\rm mm}$,  
\begin{eqnarray}
L_{d,\alpha}(k^{\rm mm}) = L_{d,\alpha+1}(k^{\rm mm}),  \nonumber 
\end{eqnarray} 
this degeneracy point in a 3-dimensional $k$-space 
becomes a source of the spatial divergence of 
the $U(1)$ magnetic field associated with 
these two bands; 
\begin{eqnarray}
{\nabla}_{k}\cdot \bar{\cal B}^{\alpha} = {\rm sign}
\{\det V^{\alpha}\}\delta^{3}(k-k^{\rm mm}) \equiv 
\rho_{\alpha}(k).   \label{9-0} 
\end{eqnarray}
The sign of the magnetic charge is 
given by the asymptotic form of the 
effective 2 by 2 Hamiltonian around this 
degeneracy point; 
\begin{eqnarray}
\hat{\sf L}^{(\alpha,\alpha+1)}(k) \simeq 
\sum_{\mu,\nu=x,y,z} 
(k_{\mu}-k^{\rm mm}_{\mu})V^{\alpha}_{\mu\nu}
\hat{\sigma}_{\nu}. \label{9-0-a}
\end{eqnarray} 
where $\hat{\sigma}_{\mu}$ stands for the 
Pauli matrices~\cite{Berry}.

\begin{figure}
\begin{center}
\includegraphics[width=0.18\textwidth,angle=-90]{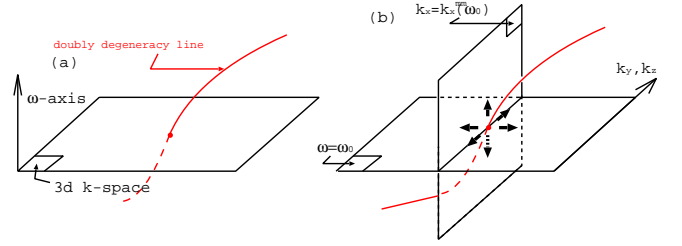}
\caption{(a) $L_{d,\alpha}$ and $L_{d,\alpha+1}$ 
are degenerated at a line in the dual 
$(3+1)$-dimensional space, which we dub as 
doubly degeneracy line (red curved line).  
The plane  in this figure corresponds to 
a 3-dimensional space, while its complementary axis 
is specified by the four-th direction. 
(b) $k_x = k^{\rm mm}_{x}(\omega_0)$ plane 
and $\omega = \omega_0$ plane share a ``magnetic charge'' at 
$(\omega,k)=(\omega_0,k^{\rm mm}(\omega_0))$. The relative sign 
between the charge viewed from the former and that from the latter 
is determined by the sign of the $x$-component of 
magnetic charge current (compare eq.~(\ref{9-0}) and eq.~(\ref{9-5a})).} 
\label{fig6}
\end{center}
\end{figure}

As was discussed in this paper, 
eigenvalues of the Lagrangian $\hat{\sf L}$ in 
Fermi liquid acquire 
another argument $\omega$ in addition 
to the crystal momentum $k$.    
The doubly degeneracy point 
in a 3-dimensional space, however, is 
by construction stable against any four-th axis. 
Thus, along this $\omega$-direction, 
this degeneracy point forms a 
degeneracy {\it line} (see Fig.~\ref{fig6}(a)); 
\begin{eqnarray}
L_{d,\alpha}(k^{\rm mm}(\omega),\omega) \equiv
L_{d,\alpha+1}(k^{\rm mm}(\omega),\omega), \nonumber 
\end{eqnarray} 
Regarding $\omega$ as a sort of ``time'', one can therefore say that 
the $U(1)$ magnetic charge in 3-dimensional 
dual space is a conserved quantity.   
Corresponding to this conservation, we might 
as well introduce the $U(1)$ magnetic   
monopole {\it current}; 
\begin{eqnarray}
j^{\rm mm}_{\alpha,\mu}(k,\omega) 
\equiv{\rm sign}\big\{\det V^{\alpha}\big\}  
\frac{d k^{\rm mm}_{\mu}}{d\omega}  
\delta^{3}(k - k^{\rm mm}(\omega)), \label{9-1}
\end{eqnarray}
so that its spatial  divergence is 
balanced by the temporal derivative of 
the magnetic monopole  density;  
\begin{eqnarray}
\sum_{\mu=x,y,z}\partial_{k_{\mu}}j^{\rm mm}_{\alpha,\mu} 
+ \partial_{\omega}\rho_{\alpha}(k,\omega) \equiv 0. 
\nonumber 
\end{eqnarray} 
Then, in analogy to the Maxwell equation in a real space,  
we can introduce the dual version of ``Ampere law''  
for the $U(1)$ electric field.  
Namely, the temporal derivative of  
the dual magnetic field and spatial rotation 
of the dual electric field is originated from this  
magnetic monopole current; 
\begin{eqnarray}
\nabla_{k} \times \bar{\cal E}^{\alpha} 
+ \partial_{\omega} \bar{\cal B}^{\alpha} 
&=& - j^{\rm mm}_{\alpha}. \label{9-2} 
\end{eqnarray}

To see this law in a primitive way, one has only to 
consider the Gauss law in {\it another} 3-dimensional space, such 
as $k_x$ = constant plane, which {\it crosses} the 
doubly degeneracy line mentioned above (see Fig.6(b)).  
Namely, our $\omega$-$k_y$-$k_z$ space, say 
$k_x = k^{\rm mm}_x(\omega_0)$ plane,  contain 
a doubly degeneracy {\it point} at 
\begin{eqnarray}
(\omega,k_y,k_z) = (\omega_0,k^{\rm mm}_y(\omega_0),k^{\rm mm}_z(\omega_0)). 
\nonumber 
\end{eqnarray}
Therefore, generically, the doubly 
degeneracy at $k=k^{\rm mm}(\omega_0)$ also becomes a source of 
the ``magnetic field'' defined in this 3-dimensional space, 
which is now $(\bar{\cal B}^{\alpha}_x,
-\bar{\cal E}^{\alpha}_{z},\bar{\cal E}^{\alpha}_{y})$. 
To determine the {\it sign} of the corresponding 
``magnetic charge'' {\it viewed from this 
$\omega$-$k_y$-$k_z$ space}, expand eq.~(\ref{9-0-a})  
with respect to $\omega$    
around $\omega=\omega_0$, 
\begin{widetext}
\begin{eqnarray}
&&\hspace{-0.8cm} \sum_{\mu,\nu=x,y,z} \big\{k_{\mu}-
k^{\rm mm}_{\mu}(\omega)\big\}\ V^{\alpha}_{\mu\nu}(\omega) 
\ \hat{\sigma}_{\nu} 
\nonumber \\
&&\hspace{0.7cm} \simeq \ 
\left[\begin{array}{ccc}
\omega - \omega_0 & k_y-k^{\rm mm}_y(\omega_0) & k_z-k^{\rm mm}_z(\omega_0)
\end{array}\right]\cdot\left[\begin{array}{ccc}
-\frac{\partial k^{\rm mm}_x}{\partial \omega} & 
-\frac{\partial k^{\rm mm}_y}{\partial \omega} & 
-\frac{\partial k^{\rm mm}_z}{\partial \omega} \\
0 & 1& 0 \\
0 &  0& 1 \\
\end{array}
\right]_{|\omega=\omega_0}\cdot \big[V^{\alpha}(\omega_0)\big]\cdot\left[
\begin{array}{c}
\hat{\sigma}_x \\
\hat{\sigma}_y \\
\hat{\sigma}_z \\
\end{array}\right]. \nonumber 
\end{eqnarray}
\end{widetext}
where we imposed $k_x=k^{\rm mm}_x(\omega_0)$ in the right hand side, 
only to discuss the Gauss law in $k_x$ = constant plane.  
Namely, comparing this with eqs.~(\ref{9-0},\ref{9-0-a}), we obtain 
the {\it Gauss law in this $\omega$-$k_y$-$k_z$ space};  
\begin{eqnarray}
&&\hspace{-0.5cm}\partial_{\omega} \bar{\cal B}^{\alpha}_{x} 
+ \partial_{k_y}(-\bar{\cal E}^{\alpha}_z) 
+ \partial_{k_z}\bar{\cal E}^{\alpha}_y = - {\rm sign}\Big\{ \det V^{\alpha}  
\frac{\partial k^{\rm mm}_x}{\partial \omega}\Big\}_{|\omega=\omega_0}  
\nonumber \\ 
&&\hspace{0.5cm}
\times\delta(\omega - \omega_0)
\delta
\big(k_{y}-k^{\rm mm}_{y}(\omega_0)\big)
\delta\big(k_z-k^{\rm mm}_z(\omega_0)\big). 
\label{9-5a} 
\end{eqnarray}
By solving $\omega_0$ in favor of $k_x$, 
we can actually see that this equation is nothing 
but the $x$-component of eq.~(\ref{9-2}).  

\section{Examples \hspace{6.3cm} -- electron-phonon coupling systems --}

In this appendix, based on a specific model calculations, 
we will present the crude estimate of 
the  dual version of $U(1)$ electric field. 
Basically, the electric component arises, in  
a perturbative calculation, from the frequency dependence 
of the self-energy.  Moreover, we are interested in the 
hermitian part of the self-energy. This is analytically 
more difficult to extract than well-studied 
anti-hermitian part. Namely, the latter involves an
energy-conserving delta function, while 
the former takes over the corresponding principal 
integrals. As a result, a consideration for the
case of the Coulomb interaction becomes 
quite complex, the frequency
dependence arising only at two loops.  
We therefore consider instead the 
self-energy due to an electron-phonon 
interaction, for which we already have a  
non-trivial frequency dependence at one loop.  
Though we did not discuss the electron-phonon interaction explicitly 
in our derivations, the
considerations there still apply to this problem. Namely, 
our formulations are clearly free from the specific origin 
of the self-energy, provided that its anti-hermitian part 
can be negligible at low frequency region.     

As a simplest model, we considered the 2D 
Rashba model defined on a square lattice, subjected to
an external magnetic field along the $z$-direction; 
\begin{eqnarray} 
&&{\cal H}_{0} = \frac{t}{2}\sum_{\langle jm\rangle,\alpha}
{\psi}^{\dagger}_{j,\alpha}{\psi}_{m,\alpha}  + h_z\sum_{j,\alpha,\beta}
{\psi}^{\dagger}_{j,\alpha}[\hat{\tau}_z]_{\alpha,\beta}
{\psi}_{j,\beta}  \nonumber \\
&&\hspace{0.8cm} - a\sum_{j}\sum_{\mu,\nu=x,y} \frac{i\epsilon_{\mu\nu}}{2}
\Big({\psi}^{\dagger}_{j+e_{\mu},\alpha}\ 
[\hat{\tau}_{\nu}]_{\alpha,\beta}\ {\psi}_{j,\beta}-{\rm h.c.}\Big). \nonumber 
\end{eqnarray} 
We employ the {\it lattice-regularized} model,  
since the hermitian part of the self-energy 
usually depends on the {\it ultraviolet cut-off}. 
Namely, $\hat{\sigma}$ obtained from the Matsubara Green function 
is composed of a {\it principal integral} with respect to  
momentum, which shows a ultraviolet logarithmic divergence in a 2D model    
(see for example eq.(\ref{10-2})). 
The above Hamiltonian, however, can be referred as 
a ``Rashba'' model, in a sense 
that, when Fourier transformed, it reproduces the 
same energy dispersion as the standard 
(i.e. continuum) Rashba model, around the $\Gamma$ point; 
\begin{eqnarray}
&& \hspace{0.42cm} {\cal H}_{0} \equiv   \sum_{\alpha,\beta} \sum_{k}
{\psi}^{\dagger}_{\alpha}(k) \  
[\hat{H}_{0}(k)]_{\alpha,\beta}\ {\psi}_{\beta}(k), \nonumber \\
&& \hat{H}_{0}(k) 
\equiv - t(\cos k_x + \cos k_y-2)\ \hat{1}   \nonumber \\
&& \hspace{1.7cm} + h_z\hat{\tau}_z + 
a\sin k_x  \hat{\tau}_y - a\sin k_y  \hat{\tau}_x, \nonumber 
\end{eqnarray}
with $\psi^{\dagger}_{\alpha}(k)\equiv 
\frac{1}{\sqrt{N}}\sum_{j}e^{ik\cdot j}\  
\psi^{\dagger}_{\alpha}(j)$. 

A finite magnetic field $h_z$ lifts the band degeneracy at $k=(0,0)$, $(\pi,0)$, 
$(0,\pi)$ and $(\pi,\pi)$; 
\begin{eqnarray}
{\sf e}_{\pm,k}= -t\big(\cos k_x + \cos k_y-2\big) \pm |\lambda_k|. \nonumber 
\end{eqnarray}   
where $|\lambda_{k}|\equiv \sqrt{h^2_{z} 
+ a^2(\sin^2 k_x + \sin^2 k_y)}$. 
Concomitantly, the magnetic Bloch wavefunctions for these two 
energy bands also acquire finite $z$-components of spin in the presence of $h_z$;
\begin{eqnarray}
|{\sf u}^{+}_{k}\rangle = \left[ 
\begin{array}{c}
\cos\big(\frac{\theta_k}{2}\big) \\
\sin\big(\frac{\theta_k}{2}\big)e^{i\varphi_k}\\ 
\end{array} 
\right], \  
|{\sf u}^{-}_{k}\rangle = \left[ 
\begin{array}{c}
-\sin\big(\frac{\theta_k}{2}\big)e^{-i\varphi_k} \\
\cos\big(\frac{\theta_k}{2}\big)\\
\end{array} 
\right], \nonumber 
\end{eqnarray}
where $\theta_k$ and $\varphi_k$ are defined as follows;  
\begin{eqnarray}
(\cos\theta_k,\sin\theta_k)&\equiv&\frac{1}{|\lambda_k|} 
(h_z,a\sqrt{\sin^2 k_x + \sin^2 k_y}) \nonumber \\ 
(\cos\varphi_k,\sin\varphi_k)&\equiv& 
\frac{1}{\sqrt{\sin^2 k_x + \sin^2 k_y}}
(\sin k_x,\sin k_y). \nonumber 
\end{eqnarray}
We have introduced the 
magnetic field, because, in a time-reversal symmetric system, the 
a.e.f.-contribution to the anomalous Hall effect (AHE) 
as well as that of the dual magnetic field 
(a.m.f.) always vanishes. Namely, the electric field  
is time-reversally ``even'' in a $T$-invariant system;
\begin{eqnarray}
{\cal E}^{\alpha}(k) = {\cal E}^{\alpha}(-k), \nonumber 
\end{eqnarray}
while the quasi-particle velocity is $T$-reversal ``odd'';
\begin{eqnarray}
{\bf v}_{\alpha}(k) = - {\bf v}_{\alpha}(-k). \nonumber 
\end{eqnarray} 
Thus, ${\cal E}^{\alpha}(k) \times {\bf v}_{\alpha}(k)$  
would vanish after the $k$-integration in a $T$-invariant system.

As for ``many-body'' effects, 
we consider an electron-phonon 
coupled Hamiltonian, in which the Einstein  
phonon having constant mass $\omega_0>0$ 
interacts with this Rashba fermion; 
\begin{eqnarray}
&&\hspace{-0.5cm}
{\cal H}_{\rm ph} + {\cal H}_{\rm ep} = \nonumber \\
&&\hspace{-0.3cm}
\sum_{q}{\omega}_{0}\ b^{\dagger}_{q}b_{q}  
+ \sum_{k,q,\alpha,\beta}
{\psi}^{\dagger}_{\alpha}(k+q)
[\hat{g}_q]_{\alpha,\beta}{\psi}_{\beta}(k)
\cdot \big(b_{q} + b^{\dagger}_{-q}\big).  \nonumber 
\end{eqnarray}
The hermiteness clearly requires 
$\hat{g}^{\dagger}_q=\hat{g}_{-q}$ in general. 
We take the simplest possible forms 
for this e-p coupling constant: 
$\hat{g}_q \simeq \hat{g}_{q=0}$, and  
$\hat{g}_{q=0}\simeq g\cdot \hat{1}$.  
Even employing this over-simplified 
coupling form, we already find a 
substantial magnitude of the electric 
field, as shown below.  
       
The lowest order (imaginary-time) self-energy associated with 
this electron-phonon system is 2nd order in $g$;   
\begin{eqnarray}
\hat{\Sigma}(k,i\omega_m) &=& \frac{g^2}{N}\sum_q
\sum_{\gamma=\pm}  |{\sf u}_{k+q}^{\gamma}\rangle 
\langle {\sf u}_{k+q}^{\gamma}| \nonumber \\ 
&&\hspace{-2.5cm}  \times  \Big\{\frac{n_{\rm b}({\omega}_0) 
+ n_{\rm f}({\sf e}_{\gamma,k+q})}
{i\omega_m + {\omega}_0 -{\sf e}_{\gamma,k+q}} \ +\  
\cdot\frac{n_{\rm b}({\omega}_0) + 1 - n_{\rm f}({\sf e}_{\gamma,k+q})}
{i\omega_m - {\omega}_0 -{\sf e}_{\gamma,k+q}}\Big\}, \nonumber
\end{eqnarray}
where $n_{\rm b}(\omega)=\frac{1}{e^{\beta\omega}-1}$ denotes the 
Bose distribution function. 

At equilibrium, the life-time part of the self-energy 
and hermitian part of  the self-energy $\hat{\sigma}$ 
are analytically continued 
from this imaginary-time one 
(see also eqs.~(\ref{1-6a},\ref{1-6b})); 
\begin{eqnarray}
\Gamma_{\alpha\beta}(k,\omega) &\equiv& 
i\Sigma_{\alpha\beta}(k,i\omega_m=\omega+i|\delta|)  \nonumber \\
&& \hspace{0.6cm} -i\Sigma_{\alpha\beta}(k,i\omega_m=\omega-i|\delta|),  \nonumber \\ 
\sigma_{\alpha\beta}(k,\omega) &\equiv& \frac{1}{2} 
\big\{\Sigma_{\alpha\beta}(k,i\omega_m=\omega+i|\delta|)  \nonumber \\
&&\hspace{0.6cm}  +\Sigma_{\alpha\beta}(k,i\omega_m=\omega-i|\delta|)\big\}. \nonumber 
%
\end{eqnarray}
The life-time term clearly becomes zero at $T=0$ 
and $\omega\simeq \mu$. Namely its integral region over 
the internal line is exponentially small: 
\begin{eqnarray}
\big(n_{\rm b}(\omega_0) 
+ n_{\rm f}({\sf e})\big)\cdot
\delta(\omega+\omega_0-{\sf e})
&\simeq& 0, \nonumber \\
\big(n_{\rm b}(\omega_0)
 + 1-n_{\rm f}({\sf e})\big)
\cdot \delta(\omega-\omega_0-{\sf e}) 
&\simeq& 0,  \nonumber  
\end{eqnarray} 
when $ |\omega-\mu| \ll \omega_0$ at $T\rightarrow 0$. 
On the other hand,  the 
hermitian part of the self-energy, which is made up of  
the principal integral, remains finite even at $T=0$;   
\begin{eqnarray}
&&\hspace{-0.8cm}\hat{\sigma}(k,\omega) = g^2\sum_{\gamma=\pm}     
\Big\{\int_{{\sf e}_{\gamma,k'}\le \mu} {\cal P}\  
\frac{|{\sf u}_{k'}^{\gamma}\rangle 
\langle {\sf u}_{k'}^{\gamma}|}
{\omega + {\omega}_0 -{\sf e}_{\gamma,k'}}\ \nonumber \\
&& \hspace{1.2cm} + \int_{{\sf e}_{\gamma,k'}\ge \mu} 
{\cal P}\ \frac{|{\sf u}_{k'}^{\gamma}\rangle 
\langle {\sf u}_{k'}^{\gamma}|}  
{\omega - {\omega}_0 -{\sf e}_{\gamma,k'}}\ \Big\} 
\frac{dk'}{(2\pi)^2}, \label{10-2}
\end{eqnarray} 
where we introduced the new integral variable $k'\equiv k+q$. 

When integrated over $k'$, 
the off-diagonal elements of the 2 by 2 matrix 
$|{\sf u}_{k'}^{\gamma}\rangle
\langle {\sf u}_{k'}^{\gamma}|$  
vanish. Namely,  
they are always odd functions of $k'_{x}$ or  
$k'_{y}$, while ${\sf e}_{\gamma,k'}$ is even; 
\begin{eqnarray}
|{\sf u}_{k'}^{\gamma}\rangle\langle {\sf u}_{k'}^{\gamma}|
&=&\frac{1}{2}\ \hat{1} + \frac{{\rm sign}\gamma}{2} 
\big( \cos\theta_{k'}\hat{\tau}_z 
\nonumber \\
&& \hspace{-0.5cm} +  \sin\theta_{k'}\cos\varphi_{k'}\hat{\tau}_x 
- \sin\theta_{k'}\sin\varphi_{k'}\hat{\tau}_y \big). \nonumber  
\end{eqnarray}
Accordingly, eq.(\ref{10-2}) becomes diagonal; 
\begin{eqnarray}
&& \hat{\sigma}(k,\omega) = \big\{\cdots\big\}\cdot \hat{1}\ 
+\ \frac{g^2}{2} \sum_{\gamma=\pm} \nonumber \\ 
&&\  \  \ \bigg\{\int_{\epsilon_{k',\gamma}\le \mu} {\cal P}\cdot 
\frac{{\rm sign}\gamma \cos\theta_{k'}} 
{\omega + {\omega}_0 -\epsilon_{k',\gamma}}  \frac{dk'}{(2\pi)^2}  \nonumber \\
&&\  \  \ \ \ + \int_{\epsilon_{k',\gamma}\ge \mu} {\cal P}\cdot 
 \frac{{\rm sign}\gamma \cos\theta_{k'}} 
{\omega - {\omega}_0 -\epsilon_{k',\gamma}}\frac{dk'}{(2\pi)^2}  \bigg\}
\ \hat{\tau}_z, \nonumber \\ 
&&\hspace{1.06cm} \equiv S_{0}(\omega)\cdot\hat{1} + 
S_{1}(\omega)\ \hat{\tau}_z. \label{10-3} 
\end{eqnarray} 
The scalar function $S_{1}(\omega)$ is  
always a smooth function at $|\omega-\mu|\ll \omega_0$. 
When this self-energy diagonalized in combination with $\hat{H}_{0}(k)$, 
a non-zero derivative of $S_{1}(\omega)$ with respect 
to $\omega$ becomes indispensable for a finite a.e.f., 
which we will see below.  

In a 2D model, we generally have 
2-component electric field and 
a single-component magnetic field;
\begin{eqnarray}
&&(-{\cal E}^{\gamma}_{y},{\cal E}^{\gamma}_{x}, 
{\cal B}^{\gamma}_{z})\equiv  
\nabla\times {\cal A}^{\gamma}, \nonumber \\ 
&&{\cal A}^{\gamma}_{\mu} \equiv i\langle u_{k,\omega}^{\gamma}|
\partial_{\mu}u_{k,\omega}^{\gamma}\rangle\, 
\nonumber 
\end{eqnarray}
with $\mu = (k_x,k_y,\omega)$. Note that       
$|u_{k,\omega}^{\gamma}\rangle$ diagonalizes 
the $\hat{\sf L}(k,\omega)=\hat{H}_{0}(k) + \hat{\sigma}(\omega)$. 
For any 2-band model, 
the Lagrangian can be decomposed in terms of the Pauli matrices;  
\begin{eqnarray}
\hat{\sf L}(k,\omega)=\sum_{\mu=x,y,z}
N_{\mu}(k,\omega)\cdot\hat{\tau}_{\mu} + {\rm const.},  \label{10-4}  
\end{eqnarray}  
where $N$ now reads  
$(-a\sin k_y,a\sin k_x, h_z + S_1(\omega))$. Then, one readily 
see that the dual fields are identified as 
the solid angle subtended by this normalized 
vector $\hat{N}\equiv N/|N|$. For example, its 
spatial component reads  
\begin{eqnarray}
{\cal B}_{z}^{\gamma} &=& -\frac{{\rm sign}\gamma}{2}
\Big(\nabla_{k_x}\hat{N}\times \nabla_{k_y}\hat{N}\Big)\cdot\hat{N}, \nonumber \\
&=& \frac{{\rm sign}\gamma\cdot  N_z}{2|N|^3}\cdot
\frac{\partial N_y}{\partial k_x}\frac{\partial N_x}{\partial k_y} \nonumber \\
&=& - \frac{{\rm sign}\gamma\cdot a^2 N_z}{2|N|^3}\cos k_x \cos k_y,  
\label{10-5}   
\end{eqnarray}
where the band index $\gamma = \pm$. 
Similarly, the temporal components are given as follows; 
\begin{eqnarray}
\big({\cal E}_{x}^{\gamma},{\cal E}_{y}^{\gamma}\big)&=& 
-\frac{{\rm sign}\gamma}{2|N|^3}
\frac{\partial S_1}{\partial \omega}\   
\hat{z}\cdot \Big(\frac{\partial N}{\partial k_x}\times N,
\frac{\partial N}{\partial k_y}\times N\Big)
\nonumber \\
&=&-\frac{{\rm sign}\gamma \cdot a^2}{2|N|^3}\frac{\partial S_1}{\partial \omega} 
\big(\cos k_x\sin k_y,-\cos k_y \sin k_x\big),  \nonumber \\ 
&\equiv &{\rm sign}\gamma \cdot |E_{k}|
\big(\cos k_x\sin k_y,-\cos k_y \sin k_x\big), \nonumber  
\end{eqnarray} 
which clearly indicates that a  
finite $\frac{\partial S_{1}}{\partial \omega}$ 
is the essential origin of the electric fields. 
With the lattice constant ${\rm a}_{\rm lattice}$ being explicit, 
the magnitude of the dual electric field estimated on a Fermi 
surface, i.e. $\omega=\mu$, is given as follows; 
\begin{eqnarray}
 && \hspace{-0.4cm} |E_{k}|_{|\omega=\mu} = \frac{{\rm a}_{\rm lattice}}{t}
\cdot \frac{\bar{a}^2 \bar{g}^2 \bar{h}_z}{2|\bar{N}_{k}|^3} 
\times f, \nonumber \\ 
&& \hspace{-0.4cm}f= \sum_{\gamma=\pm}\Big\{\int_{\bar{\sf e}_{\gamma,q}\le 0} 
\frac{{\rm sign}\gamma}{\bar{\Delta}_{q}\cdot 
(\bar{\omega}_0-\bar{\sf e}_{\gamma,q})^2} \nonumber \\ 
&& \hspace{0.5cm} \hspace{-0.8cm} + \ \int_{\bar{\sf e}_{\gamma,q}\ge 0}
\frac{{\rm sign}\gamma}{\bar{\Delta}_{q} \cdot 
(\bar{\omega}_0+\bar{\sf e}_{\gamma,q})^2}
\Big\}\frac{dq}{(2\pi)^2}, \label{10-7} 
\end{eqnarray}
where $\bar{\Delta}_q$ denotes the direct band gap 
at $q$-point, {\it measured with the transfer 
integral ``$t$'' being an energy unit}; 
$\bar{\Delta}_q\equiv 2|\lambda_q|/t$. In a same 
sense, we also made it dimensionless, 
the Rashba coupling energy, e-p coupling energy, 
Zeeman energy, phonon energy and
band dispersion, like $\bar{a}$ , $\bar{g}$ , $\bar{h}_z$ ,  
 $\bar{\omega}_0$ and 
$\bar{\sf e}_{\gamma,q}\equiv ({\sf e}_{\gamma,q}-\mu)/t$ 
respectively. In terms of these, the 
other dimensionless function of $k$, i.e. $|\bar{N}_{k}|$,  
reads;  
\begin{eqnarray} 
|\bar{N}_k|&=&\sqrt{\bar{a}^2(\sin^2 k_y + \sin^2 k_x) + 
{\bar{h}_z}^2 (1+ \bar{S}_{1})^2}, \nonumber \\
\bar{S}_{1} &=& \bar{g}^2 \sum_{\gamma=\pm}\Big\{
\int_{\bar{\sf e}_{\gamma,q}\le 0} 
\frac{{\rm sign}\gamma}{\bar{\Delta}_{q}\cdot 
(\bar{\omega}_0-\bar{\sf e}_{\gamma,q})}  \nonumber \\
&&\ \ -\ \int_{\bar{\sf e}_{\gamma,q}\ge 0}
\frac{{\rm sign}\gamma}{\bar{\Delta}_{q} \cdot 
(\bar{\omega}_0+\bar{\sf e}_{\gamma,q})}
\Big\}\frac{dq}{(2\pi)^2}.  
\nonumber 
\end{eqnarray}

\begin{figure}
\begin{center}
\includegraphics[width=0.45\textwidth]{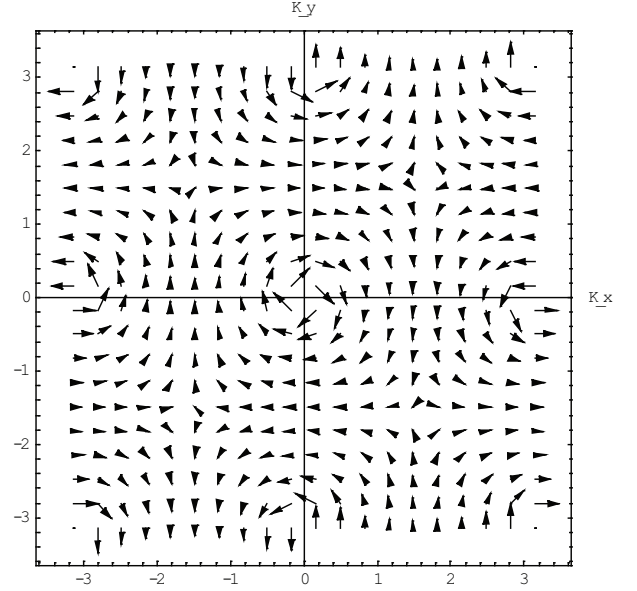}
\caption{A distribution of $({\cal E}^{-}_{x},{\cal E}^{-}_{y})$  
estimated on a Fermi surface, i.e. $\omega=\mu$.   
$\bar{a}=1.0,\bar{g}=2.0,\bar{h}_z=0.5,\bar{\omega}_0=0.5$ and $\mu=0$.}
\label{fig7}
\end{center}
\end{figure}

The expression eq.~(\ref{10-7}) clearly demonstrates 
that the magnitude of the electric field is finite only in the presence of 
the applied magnetic field $h_z$ and Rashba coupling $a$. 
It is also proportional to the dimensionless factor, i.e. $f$, 
whose value depends on a specific shape of 
the upper band and lower band in $k$-space. From its form, however, 
there is no reason that 
this real-valued factor always has to reduce 
identically to zero. 
In fact, a simple numerical estimation shows that $f=-0.31$ and 
$\bar{S}_1=-0.24$, 
in the case of $\bar{a}=1.0, \bar{g}=2.0, \bar{h}_z=0.5, \mu=0$ 
with $\bar{\omega}_0 = 0.5$. 
Observing these numerical factors, the magnitude of the a.e.f. is   
estimated as follows;    
\begin{eqnarray}
|{\cal E}^{\alpha}|
\simeq 10^{-1}\times \frac{{\rm a}_{\rm lattice}}{t}. 
\label{10-8}
\end{eqnarray} 
The distribution of the a.e.f. has a  
4 vortex structure within the unit cell $[-\pi,\pi]\times [-\pi,\pi]$ 
(see Fig.~\ref{fig7}). 
These vortices 
reflect the 4 band-crossing points located at $(k_x,k_y,\omega)=
(0,0,\omega_1),(\pm \pi,\pm \pi,\omega_1),(0,\pm \pi,\omega_1)$ and 
$(\pm \pi,0,\omega_1)$ with $\omega_1$ defined as follows;
\begin{eqnarray}
h_z + S_{1}(\omega_1)=0. \nonumber 
\end{eqnarray}
Namely, 
$(-{\cal E}^{-}_{y},{\cal E}^{-}_{x},{\cal B}^{-}_{z})$ 
has sources ($2\pi$ ``charges'') 
from the former two points, while 
it has  sinks ($-2\pi$ ``charges'') at
the latter two.

The many-body correction to the anomalous velocity is given by the outer
product between the a.e.f. and the quasi-particle velocity, i.e.,
$\big({\cal E}_{k}^{\alpha}\times {\bf v}_{\alpha}\big)_{z}$. Thus,
taking $|{\bf v}_{\alpha}|$ to be ${\rm a}_{\rm lattice}\cdot t$, we find it of
the order of $10^{-1}\times {\rm a}^2_{\rm lattice}.$
The bare contribution of the 
anomalous velocity on a Fermi surface can be directly 
estimated from eq.(\ref{10-5});  
${\cal B}^{\alpha}_{z} \simeq 1.0 \times {\rm a}^{2}_{\rm lattice}$. 
Observing these two quantities, we can then insist that   
the electric-field contribution becomes almost at the same order  
of this bare contribution and not negligible even in this oversimplified 
model.


\begin{thebibliography}{99}
\bibitem{note1} Note that we have neglected the following terms which is 
of the order of ${\cal O}(\lambda^3)$;  
\begin{eqnarray}
\big[\big[\hat{\epsilon}_{\alpha},\hat{\Omega}^{\alpha}_{X_jQ_j}
\big]_{+},\hat{A}_{\alpha}\big]_{-}  
\simeq {\cal O}(\lambda^2_1\lambda_2,\lambda_1\lambda^2_2). 
\end{eqnarray}
Namely, $\hat{A}_{\alpha}$ being unit matrix at equilibrium, 
this commutator is 3rd order in $\lambda$.
\bibitem{note3} 
Note that the $\sigma$,$\sf b$ and $G_0^{-1}-\hat{\Sigma}^{\rm HF}$ 
observe the followings; 
\begin{eqnarray}
&&\hspace{-0.5cm}\sigma(1,1')^{*} \equiv \sigma(1',1)\ , \ 
{\sf b}(1,1')^{*} \equiv {\sf b}(1',1), \nonumber \\
&&\hspace{-0.5cm}{(G_{0}^{-1} - \hat{\Sigma}^{\rm HF})(1,1')}^{*} 
\equiv (G_{0}^{-1} -\hat{\Sigma}^{\rm HF})(1',1), \nonumber 
\end{eqnarray} 
while 
\begin{eqnarray} 
&&\hspace{-0.5cm}  
{\sf g}^{<(>)}(1,1')^{*} \equiv - {\sf g}^{<(>)}(1',1), \nonumber \\
&&\hspace{-0.5cm}
{\Sigma}_{\rm c}^{<(>)}(1,1')^{*} \equiv 
-{\Sigma}_{\rm c}^{<(>)}(1',1). \nonumber 
\end{eqnarray}
\bibitem{note4} As in the standard literature~\cite{KB},   
we took the following conventions of the Fourier transformation 
of lesser/greater green function/collisional self energy with 
respect to the frequency;
\begin{eqnarray}
&&\hspace{-0.1cm} 
{\sf g}^{<}(\omega) = -i\int_{-\infty}^{\infty} e^{i\omega t}
{\sf g}^{<}(t)\ dt,  \nonumber \\ 
&&\hspace{-0.1cm}{\sf g}^{>}(\omega) = i\int_{-\infty}^{\infty} 
e^{i\omega t} {\sf g}^{>}(t)\ dt, \nonumber \\
&&\hspace{-0.1cm} \Sigma_{\rm c}^{<}(\omega) = -i
\int_{-\infty}^{\infty} e^{i\omega t} \Sigma_{\rm c}^{<}(t)
\ dt, \nonumber \\
&&\hspace{-0.1cm}
\Sigma_{\rm c}^{>}(\omega) = 
i\int_{-\infty}^{\infty} e^{i\omega t} 
\Sigma_{\rm c}^{>}(t)\ dt, \nonumber 
\end{eqnarray}  
only to obtain the relation between 
the spectral function and green function given below;
\begin{eqnarray}
{\sf A}(\omega) \equiv \int_{-\infty}^{\infty} e^{i\omega t} 
{\sf A}(t)\ dt = {\sf g}^{>}(\omega) + {\sf g}^{<}(\omega), 
\end{eqnarray}
(see also eq.~(\ref{A-def}))
\end{thebibliography}
\end{document}